\documentclass[aps,twocolumn,floatfix]{revtex4}
\usepackage{amsmath}
\usepackage{amssymb}
\usepackage{graphicx}% Include figure files
\usepackage{subfigure}
\usepackage{dcolumn}% Align table columns on decimal point
\usepackage{bm}% bold math
\begin{document}

\title{Scattering function of semiflexible polymer chains under good solvent conditions}

\author{Hsiao-Ping Hsu$^a$, Wolfgang Paul$^b$, and Kurt Binder$^a$}

\affiliation{$^a$Institut f\"ur Physik, Johannes Gutenberg-Universit\"at Mainz,\\
 Staudinger Weg 7, D-55099 Mainz, Germany \\
$^b$Theoretische Physik, Martin Luther Universit\"at \\
Halle-Wittenberg, von Seckendorffplatz 1, 06120 Halle, Germany}

\begin{abstract}
Using the pruned-enriched Rosenbluth Monte Carlo algorithm, the scattering 
functions of semiflexible macromolecules in dilute solution under good 
solvent conditions are estimated both in $d=2$ and $d=3$ dimensions, 
considering also the effect of stretching forces. Using self-avoiding walks 
of up to $N = 25600$ steps on the square and simple cubic lattices, variable 
chain stiffness is modeled by introducing an energy penalty $\epsilon_b$ 
for chain bending; varying $q_b=\exp (- \epsilon_b/k_BT)$ from $q_b=1$ 
(completely flexible chains) to $q_b = 0.005$, the persistence length can 
be varied over two orders of magnitude. For unstretched semiflexible chains 
we test the applicability of the Kratky-Porod worm-like chain model to 
describe the scattering function, and discuss methods for extracting 
persistence length estimates from scattering. While in $d=2$ the direct 
crossover from rod-like chains to self-avoiding walks invalidates the 
Kratky-Porod description, it holds in $d=3$ for stiff chains if the number of Kuhn 
segments $n_K$ does not exceed a limiting value $n^*_K$ (which depends on the persistence 
length). For stretched chains, the Pincus blob size enters as a further 
characteristic length scale. The anisotropy of the scattering is well 
described by the modified Debye function, if the actual observed chain 
extension $\langle X \rangle$ (end-to-end distance in the direction of the 
force) as well as the corresponding longitudinal and transverse linear 
dimensions $\langle X^2 \rangle - \langle X \rangle^2$, 
$\langle R_{g,\bot}^2 \rangle$ are used.
\end{abstract}
\maketitle

\section{Introduction}
Small angle (neutron) scattering from polymers in dilute solution is the 
method of choice to obtain a complete picture of the conformations of long 
flexible or semiflexible macromolecules, from the length scale of the 
monomeric units to the gyration radius of the chain molecules~\cite{1,2,3,4}. 
Classical experiments have shown that the gyration radius $R_g$ of long 
flexible chains in dense melts (and also in dilute solution under Theta 
conditions) scales with chain length N according to the classical random 
walk picture~\cite{5,6,7}, $R_g \propto N^{1/2}$, while in dilute 
solution under good solvent conditions flexible polymers form swollen 
coils~\cite{1,2,3,4,8}, $R_g \propto N ^\nu$, with~\cite{9} 
$\nu \approx 0.588$ (in $d=3$ dimensions) or $\nu=3/4\; (d=2)$.
It was also shown that on intermediate length scales the dependence of 
the scattering intensity $S(q)$ on wave number $q$ reflects the self-similar 
fractal structure of the polymer~\cite{8}, $S(q) \propto q^{-1/\nu}$ 
(under good solvent conditions) or $S(q) \propto q^{-2}$ (Theta conditions 
or melts, respectively). Also the crossovers between these regimes when 
either the temperature distance from the Theta point or the concentration 
of the solution are varied have been investigated~\cite{4}, and the length 
scales ruling these crossovers (i.e., diameter of ``thermal blobs'' or 
``concentration blobs'', respectively) have been identified~\cite{1,2,3,4}.

The behavior gets more complicated, however, when chain stiffness plays a 
prominent role: only when chain stiffness is essentially due to chain 
thickness, i.e. when the effective persistence length $\ell_p$ scales 
proportional to the local chain diameter $D$ the problem can still be 
reduced to a rescaled self-avoiding walk problem~\cite{10,11}. However, 
when $\ell_p \gg D$, one finds (in $d=3$) a double crossover, since then 
short chains behave like rigid rods (i.e., $R_g \propto N$ as long as 
$R_g < \ell_p$), and then a crossover to Gaussian random-walk like coils 
occurs, while for $R_g \approx R_0$ ($R_0$ will be discussed below) a second
crossover to swollen coils  
$(R_g \propto N^\nu)$ starts. Also this behavior has been established in 
beautiful classical experiments~\cite{12}, as well as in recent 
simulations~\cite{11,13}. The standard model for semiflexible worm-like 
chains, the Kratky-Porod model~\cite{14}, can only describe the first 
crossover (from rods to Gaussian coils) but fails to account for the second 
crossover to swollen coils, due to its complete neglect of excluded volume 
effects. It turns out that this second crossover still is incompletely 
understood: while in the early experiments~\cite{12} it was suggested that 
this crossover occurs for $n_K=n_K^*$ with $n_K^*=50$ Kuhn segments, 
independent of the persistence length, a Flory-type argument~\cite{15,16} 
suggests that the crossover occurs for a polymer radius 
$R_0 \propto \ell_p^2 /D$ (corresponding to $n^*_K \propto (\ell_p/D)^2)$, 
while the simulations rather find~\cite{13} $n^*_K \propto \ell_p^{1.5}$. 
Remember that a worm-like chain can be described as 
an equivalent freely jointed chain of $n_k$ Kuhn segments of length 
$2 \ell_p $ each~\cite{5,6}. In view of an apparent conflict of the estimate~\cite{12} 
$n^*_K=50$ with the Yamakawa-Stockmayer-Shimata 
theory~\cite{16,161,162,163} for worm-like chains the result 
$n^*_K=50$ was considered as a very fundamental 
problem~\cite{164,165}. However, Tsuboi et al~\cite{165} confirmed this 
estimate for another stiff polymer and concluded that the result $n^*_K=50$ 
is compatible with the theory. We also note that in $d=2$ there occurs a 
single crossover from rods to self-avoiding walks, any regime of 
Gaussian-like behavior is completely absent~\cite{13,17}. We emphasize 
however, that the results of~\cite{11,13,15,16} imply that a universal 
number $n_K^*$ (independent of $\ell_p$) up to which the Kratky-Porod model 
holds in $d=3$ does not exist.

It then is interesting to ask how chain stiffness shows up on intermediate 
length scales, that can be probed via the scattering function $S(q)$. 
If one disregards excluded volume~\cite{1} and bases the treatment on the 
Kratky-Porod model, one can show~\cite{18} that the rod-like behavior at 
large $q$ leads to a scattering law proportional to $q^{-1}$, 
i.e. (in $d=3$ dimensions)
\begin{equation}\label{eq1}
N \ell_b qS(q) = \pi + \frac 2 3 (q \ell_p) ^{-1}\, , \qquad 
N \rightarrow \infty\,,
\end{equation}
where we have assumed that the chain has a contour length $L=N\ell_b$, 
where $\ell_b$ is the bond length while the persistence length~\cite{6} is 
$\ell_p=(N/n_K)\ell_b/2$. While for Gaussian ``phantom chains'' 
(i.e., excluded volume interactions are completely neglected) 
the structure factor $S(q)$ is readily found~\cite{1,2,3,4,5,6} in terms of 
the Debye function, and the only length that enters is the gyration 
radius $\langle R_g ^2 \rangle ^{1/2}$, choosing a normalization 
where $S(q=0)=1$
\begin{equation}\label{eq2}
S(q) = 2 [\exp (-\zeta) -1+\zeta]/\zeta^2 \, ,\, \zeta = 
q^2 \langle R_g^2\rangle = q^2N\ell_b^2/6 \, ,
\end{equation}
for semiflexible polymers the calculation of $S(q)$ for chains with $N$ 
finite is a formidable 
problem~\cite{19,20,21,22,23,231,24,25,26,27,271,28,281,29,291,292},
even in the absence of excluded volume effects. However, including excluded 
volume effects in the description of scattering of semiflexible chains 
is even more an unsolved problem: existing phenomenological approaches 
require the adjustment of many empirical parameters~\cite{29}. It will be 
one of the tasks that will be addressed in the present paper, to investigate 
$S(q)$ for semiflexible chains numerically in the presence of excluded 
volume interactions between the effective monomers, varying $\ell_p$ over 
a wide range.

In recent years also the behavior of macromolecules under the influence 
of stretching forces has found enormous interest 
(e.g.~\cite{30,31,32,33,34,35,36,37,38,39,40,41,42,43,44,45}), 
in particular for the study of bio-macromolecules. Experimentally this 
can be realized e.g. by pulling at one end of a chain, that is anchored 
at a substrate with the other chain end, by the tip of an atomic force 
microscope~\cite{32,36,38,39,42,45} but it is also conceivable to stretch 
polymers by the forces occurring when a polymer solution is exposed to 
strong shear flow~\cite{46,461,462} or elongational flow~\cite{463}. 
Of course, it is not obvious that it will be possible to carry out 
scattering experiments on such stretched chains and measure the structure 
factor (which then is anisotropic and has two relevant parts 
$S_{||}(q_{||})$, $S_\bot(q_\bot)$ since the direction of the 
scattering vector $\vec{q}$ relative to the stretch direction, 
either parallel, $q_{||}$, or perpendicular, $q_\bot$, matters). 
But nevertheless a theoretical investigation of 
$S_{||}(q_{||}), S_\bot(q_\bot)$ is worthwhile, because it gives detailed 
insight into the local structure of stretched chains, including also chains 
under cylindrical confinement~\cite{47,48,49} and this may help to understand
problems such as transport of semiflexible polymers through porous materials, 
or channels in nanofluidic devices~\cite{47}, etc. Thus we shall also 
investigate the structure factor of stretched semiflexible chains, extending 
previous work on flexible chains~\cite{30,34}.

The outline of our paper is as follows: in Sec. II, we give a summary of the 
theoretical background, and in Sec. III we define our model and briefly recall 
the simulation methodology. In Sec. IV we present our results for the structure 
factor $S(q)$ of semiflexible chains, for both $d=2$ and $d=3$ dimensions, 
in the absence of stretching forces. Sec. V describes the modifications of the
structure factor due to stretching, while Sec. VI summarizes our conclusions.
The calculation of the scattering function of random walk chains under 
constant pulling forces can be carried out analytically and is presented in 
an appendix.

\section{Theoretical Background}
\subsection{Definitions}

We consider here the scattering from a single polymer chain, assuming that 
the chain can be described by a sequence of $N+1$ (effective) monomers at 
positions $\vec{r}_j$, $j=1,2,\ldots$, $N+1$, so that we can define $N$ 
bond vectors $\vec{a}_j=\vec{r}_{j+1}-\vec{r}_j$. In the absence of 
stretching forces, the structure factor $S(\vec{q})$ does not depend on 
the direction of the scattering vector $\vec{q}$, and can be defined as
\begin{widetext}
\begin{eqnarray}\label{eq3}
S(q) &=& \frac{1}{(N+1)^{2}} \left \langle \sum \limits _{j=1} ^{N+1} \;
\sum \limits _{k=1} ^{N+1} \exp [i \vec{q} 
\cdot (\vec{r}_j - \vec{r}_k)] \right \rangle \nonumber \\
&=& \frac{1}{(N+1)^{2}} \left\{\left \langle \left [\sum \limits _{j=1} ^{N+1} 
\sin (\vec{q} \cdot \vec{r}_j) \right]^2 \right \rangle + 
\left \langle \left [\sum \limits _{j=1} ^{N+1}  
\cos (\vec{q} \cdot \vec{r}_j) \right]^2 \right \rangle \right \}\,.
\end{eqnarray}
\end{widetext}
Note that we have chosen here a normalization for which 
$S(q\rightarrow 0)=1$. When a stretching force is applied to one chain 
end in the +x-direction, the structure factor becomes anisotropic. 
In $d=3$ dimensions, the conformations of chains still have 
axis-symmetric geometries, and we must distinguish between 
$S_{||}(q_{||})$, where $\vec{q}$ is oriented in the x-direction parallel 
to the force, and $S_\bot(\vec{q}_\bot)$, where $\vec{q}$ is oriented 
perpendicular to it. So we define 
$\vec{r}_j = (x_j,y_j,z_j)= (x_j, \vec{\rho}_j)$ to obtain
\begin{widetext}
\begin{equation}\label{eq4}
S_{||}(q_{||}) =\frac{1}{(N+1)^2}\left\{\left \langle \left [
\sum \limits_{j=1}^{N+1} 
\sin (q_{||}x_j) \right ]^2 \right \rangle + \left \langle 
\left [\sum \limits _{j=1}^{N+1} 
\cos (q_{||}x_j) \right]^2 \right \rangle \right\}\, ,
\end{equation}
\begin{equation}\label{eq5}
S_\bot (q_\bot) =\frac{1}{(N+1)^2}\left\{ \left \langle 
\left [\sum \limits_{j=1}^{N+1} 
\sin (\vec{q}_\bot \cdot \vec{\rho}_j)\right]^2 \right \rangle + \left \langle 
\left[\sum \limits _{j=1}^{N+1} \cos (\vec{q}_\bot \cdot \vec{\rho}_j)\right]^2 
\right \rangle \right\} \, .
\end{equation}
\end{widetext}
Note that in $d=2$ dimensions we have $\vec{r}_j=(x_j , y_j)$ and then 
$\vec{q}_\bot \cdot \vec{\rho}_j$ in Eq.~(\ref{eq5}) needs to be replaced 
simply by $q_\bot y_j$, of course.

We also stress that in this paper we are not at all concerned with effects 
due to the local structure of (effective) monomers, such as e.g. chemical 
side groups, etc.; such effects show up at large $q$ when one considers 
the scattering from real chains~\cite{231}. We next define our notation 
for characteristic lengths of the chain. Assuming a rigidly fixed bond 
length $\ell_b$ between neighboring monomers along the chain, the contour 
length $L$ is
\begin{equation}\label{eq6}
L =N \ell_b
\end{equation}
The mean square end-to-end distance (in the absence of stretching 
forces) simply is
\begin{equation}\label{eq7}
\langle R^2 \rangle = \left \langle \left( \sum 
\limits _{j=1}^N \vec{a}_j \right)^2 \right \rangle
\end{equation}
with $\vec{a}_j = \vec{r}_{j+1}-\vec{r}_j$ being the j-th bond vector,
and the mean square gyration radius is given by
\begin{eqnarray}\label{eq8}
\langle R_g^2 \rangle & =& \frac {1}{N+1} \left \langle 
\sum \limits _{j=1}^{N+1}(\vec{r}_j - \vec{r}_{CM})^2 \right \rangle  \nonumber \\
&=& \frac {1}{(N+1)^2} \left \langle \sum \limits _{j=1} ^{N+1} 
\sum \limits _{k=j+1}^{N+1} (\vec{r}_j - \vec{r}_k )^2 \right \rangle \, ,
\end{eqnarray}
where $\vec{r}_{CM} = \sum \limits _{j=1}^{N+1} \vec{r}_j/(N+1)$ is 
the center of mass position of the polymer.

In the presence of stretching forces, the chain takes a mean extension 
$\langle X \rangle$ and mean square extensions also become anisotropic,
\begin{equation}\label{eq9}
\langle X \rangle = \left \langle \sum \limits _{j=1}^N a _{jx}
\right \rangle \,, 
\qquad \left \langle X^2 \right \rangle = \left \langle 
\left( \sum \limits ^N _{j=1} a _{jx} \right)^2 \right \rangle \,,
\end{equation}
\begin{equation}\label{eq10}
\langle R^2 _\bot \rangle = \left \langle \left( \sum \limits _{j=1}^N a_{jy}\right)^2 
\right \rangle + \left \langle 
\left( \sum \limits _{j=1}^N a_{jz}\right)^2 \right \rangle \;.
\end{equation}
Eq.~(\ref{eq10}) refers to the three-dimensional case, for $d=2$ the 
second term in the right hand side needs to be omitted. A related 
anisotropy can then be stated for the gyration radius square as well, namely
\begin{equation}\label{eq11}
\langle R_{g,||}^2\rangle =\frac{1}{(N+1)^2} \left \langle 
\sum \limits _{j=1}^{N+1} \sum \limits _{k=j+1}^{N+1} (x_j-x_k)^2
\right \rangle \, ,
\end{equation}
\begin{equation}\label{eq12}
\langle R_{g, \bot}^2 \rangle = \frac{1}{(N+1)^{2}} \left \langle 
\sum \limits _{j=1} ^{N+1} 
\sum \limits _{k=j+1}^{N+1} \left[(y_j - y_k)^2 +(z_j-z_k)^2
\right] \right\rangle \,,
\end{equation}
for $d=3$, again the term $(z_j-z_k)^2$ simply is omitted in the case $d=2$.

We also recall that the mean square gyration radii describe the 
scattering functions at small $\vec{q}$. So in the absence of 
stretching forces we have
\begin{equation}\label{eq13}
S(q)=1-q^2 \langle R_g^2\rangle /d \, , \qquad q \rightarrow 0\,,
\end{equation}
while if stretching forces are present, one finds instead
\begin{equation}\label{eq14}
S_{||}(q_{||}) = 1 - q^2_{||} \langle R_{g,||}^2\rangle \, , 
\qquad q \rightarrow 0 \,,
\end{equation}
\begin{equation}\label{eq15}
S_\bot(q_\bot)=1-q_\bot^2 \langle R_{g, \bot}^2 \rangle /(d-1) \, , \qquad 
q \rightarrow 0 \,.
\end{equation}
In addition to the limit $q\rightarrow 0$, also the limiting behavior 
of $S(q\rightarrow \infty)$ is trivially known: then all interference 
terms in Eq.~(\ref{eq3}) average to zero, and only the terms $j=k$ 
contribute to the double sum. Hence we obtain
\begin{eqnarray}\label{eq16}
S(q \rightarrow \infty) &=&\frac{1}{N+1} \, , \nonumber \\
S_{||}(q_{||}\rightarrow \infty) &=& S_\bot(q_\bot\rightarrow \infty) 
=\frac{1}{N+1} \, ,
\end{eqnarray}
irrespective of the value of the persistence length $\ell_p$, the value of 
an applied force $f$, etc.

At this point we emphasize, however, that it is not possible to write 
down a general definition for the persistence length $\ell_p$ that would 
be both universally valid and practically useful~\cite{10,11,13}. The 
definition often found in textbooks~\cite{6,50} in terms of the asymptotic 
decay of the bond vector orientational function (of a very long chain, 
$N \rightarrow \infty)$
\begin{equation}\label{eq17}
\langle \cos \theta (s) \rangle \equiv \langle \vec{a}_j 
\cdot \vec{a}_{j+s}\rangle /\langle \vec{a}_j \cdot \vec{a}_j \rangle \propto 
\exp (-s \ell_b/\ell_p) \, , \, s \rightarrow \infty
\end{equation}
makes sense only for Gaussian PHANTOM chains, and is not applicable to real
polymers under ANY CIRCUMSTANCES, since the asymptotic decay of 
$\langle \cos \theta (s) \rangle $ with the ``chemical distance'' 
$s \ell_b$ along the chain always is a power-law decay. In fact, 
for $\ell_p \ll s \ell_b \ll L$ one has~\cite{10,11,13,51,52,53,54}
\begin{equation}\label{eq18}
\langle \cos\theta (s) \rangle \propto s^{-\beta}
\end{equation}
where $\beta = 3/2$ both in melts~\cite{51,52} and for chains in 
dilute solution under Theta conditions~\cite{10,53}. For the case 
of good solvent conditions, which is the problem of interest for 
the present paper, one rather finds the scaling law~\cite{10,54}
\begin{equation}\label{eq19}
\beta =2(1-\nu)
\end{equation}
which yields $\beta = 0.825 \; (d=3)$ and $\beta = 1/2 \; (d=2)$, 
respectively. In simple cases, such as the semiflexible extension 
of the self-avoiding walk model studied in~\cite{13} and further 
investigated in the present work, one can rather use an analog of 
Eq.~(\ref{eq17}) but for short chemical distances,
\begin{equation}\label{eq20}
\langle \cos \theta (s) \rangle = \exp (- s\ell_b/\ell_p) \, ,
 \qquad 0 \leq s \leq \ell_p/\ell_b
\end{equation}
Eq.~(\ref{eq20}) is useful for the simple model that will be studied 
in the present paper, namely the self-avoiding walk (SAW) on square 
and simple cubic lattices with an energy penalty $\epsilon_b$ for 
``bond bending'' (i.e., kinks of the SAW by 90 degrees), which then 
serves as a convenient parameter to control the persistence length. 
Since for $\epsilon_b/k_BT <2$ the power law, Eq.~(\ref{eq18}), 
already starts to set in even for small $s$ of order unity already, 
we use in practice an alternative definition,
\begin{equation}\label{eq21}
\ell_p/\ell_b= - 1/\ln (\langle \cos \theta (1) \rangle ) \,,
\end{equation}
which is equivalent to Eq.~(\ref{eq20}), if Eq.~(\ref{eq20}) holds 
over a more extended range of $s$. An alternative method considers 
the distribution function $P(n_{str})$ of $n_{str}$ successive bond 
vectors $\vec{a}_i$ having the same orientation without any kink, 
which is found to behave as~\cite{13} $P(n_{str}) = a_p \exp(-n_{str}/n_p)$, 
with $a_p$, $n_p$ being constants. Both $\langle n_{str}\rangle$ 
and $n_p$ can be taken as alternative estimates of a persistence length, 
and for large $n_p$ these estimates agree with the result from 
Eq.~(\ref{eq21}) to within a relative accuracy of a few percent 
(or better), both in $d=2$ and in $d=3$ dimensions~\cite{13}. 
Unfortunately, Eq.~(\ref{eq21}) is not straightforwardly applicable 
for chemically realistic models (such as alkane chains when $\ell_b$ 
means a bond between two successive carbon atoms, but the all-trans 
state corresponds to a zig-zag configuration with a nonzero bond 
angle $\theta (1)$). It also is not useful for coarse-grained models 
of polymers with complex architecture, such as bottle-brush 
polymers~\cite{10,11}. Thus we emphasize that for our model Eq.~(\ref{eq21}) 
is a practically useful definition, while for real polymers studied 
experimentally the estimation of $\ell_p$ is a delicate problem. 
The same caveat applies for the Kuhn length $\ell_K$, which is 
$\ell_K=2 \ell_p$ for worm-like chains, 
but the latter does not apply in solutions, as stated above. 
In dense melts, $\ell_K/\ell_b=6 \langle R_g^2\rangle /(\ell_b^2N)$ is 
supposed to hold, but due to local interactions with neighboring monomers 
in a dense environment it is not obvious that $\ell_K$ for a melt is a 
relevant parameter for a chain under good solvent conditions.

Thus, it is a clear advantage of our model calculations that via 
Eqs.~(\ref{eq20}), (\ref{eq21}) accurate direct estimates of $\ell_p$ are 
possible, unlike in experiment. These estimates for $\ell_p$ are tabulated 
in Ref.~\cite{13}.

\subsection{Theoretical Predictions for the Scattering Function of 
Single Polymers in Good Solvents in the Absence of Stretching Forces}

The classical result for the scattering from Gaussian chains is the well-known 
Debye function~\cite{1,2,3,4}
\begin{equation}\label{eq22}
S_{\rm Debye} (q) = \frac {2} {q^2\langle R_g^2\rangle} 
\left\{1-\frac {1}{q^2\langle R_g^2\rangle} 
[1 - \exp (-q^2\langle R^2_g\rangle )]\right\}
\end{equation}
Note that for large $q$ this reduces to 
$S_{\rm Debye}(q) \approx 2/(q^2\langle R_g^2\rangle)$, reflecting the 
random-walk like fractal structure of a Gaussian coil, 
$S_{\rm Debye} (q) \propto q^{-1/\nu_{MF}} $ with $\nu_{MF}= 1/2$. 
Eq.~(\ref{eq22}) does not tell how large $q$ can be in 
order for this power law to be still observable. For semiflexible 
Gaussian chains, the contour length $L=N\ell_b$ can be written as 
$L=n_p\ell_p$ and the mean square end-to-end distance and gyration 
radius are~\cite{14,55}
\begin{equation}\label{eq23}
\frac {\langle R^2\rangle}{2 \ell_p L} =1 - \frac {1} {n_p} [1-\exp (-n_p)] \,,
\end{equation}
\begin{equation}\label{eq24}
\frac{6 \langle R_g^2\rangle } {2 \ell_p L} = 1 - 
\frac {3}{n_p} + \frac {6}{n_p^2} - \frac {6} {n_p^3} [1-\exp(-n_p)]
\end{equation}
From Eqs.~(\ref{eq23}), (\ref{eq24}) one can clearly recognize that 
Gaussian behavior of the radii is only seen if the number $n_p$ of 
persistence lengths that fit to a given contour length of the chain 
is large, $n_p \gg 1$ (for $n_p$ of order unity, a crossover to rod-like 
behavior occurs). Since $q^{-1}$ also is a length scale, one concludes 
that the Gaussian coil behavior reflected in Eq.~(\ref{eq22}) also 
implies that a scale $q^{-1}$ requires that a subchain with this 
gyration radius contains many persistence lengths as well, 
i.e. Eq.~(\ref{eq22}) can only hold for
\begin{equation}\label{eq25}
q \ell_p \ll 1 \,.
\end{equation}
In the regime
\begin{equation}\label{eq26}
\ell_p^{-1} \ll q \ll \ell_b^{-1}
\end{equation}
we expect that the scattering function will resemble the scattering 
function of a rigid rod of length $L_{\rm rod}$~\cite{56}
\begin{equation}\label{eq27}
S_{\rm rod} (q) = \frac {2} {qL_{\rm rod}} \left[\int_0 ^{qL_{\rm rod}} dx 
\frac {\sin x}{x} - \frac {1-\cos (qL_{\rm rod})} {qL_{\rm rod}} \right ]\, ,
\end{equation}
which for large $q$ varies like
\begin{equation}\label{eq28}
S_{\rm rod} (q\rightarrow \infty) = \pi /(qL_{\rm rod}) \,.
\end{equation}
Note that Eqs.~(\ref{eq27}), (\ref{eq28}) refer to a rigid rod on 
which the scattering centers are uniformly and continuously distributed. 
In the lattice model that is studied here, the scattering centers are 
just the subsequent lattice sites along the rod, and since for a rod of 
length $L_{\rm rod}$ there are then $L_{\rm rod} +1$ such scattering centers, 
one has~\cite{57}
\begin{widetext}
\begin{equation}\label{eq29}
S_{\rm rod}(q) = \frac{1}{L_{\rm rod}+1}
\left[-1 + \frac{2}{L_{\rm rod}+1} \sum \limits _{k=0}^{L_{\rm rod}} 
(L_{\rm rod} +1-k) \frac {\sin q k}{q k} \right] \,, \qquad q <\pi \,.
\end{equation}
\end{widetext}
Both Eqs.~(\ref{eq27}), (\ref{eq29}) have a simple smooth crossover 
from $S_{\rm rod}(q) = 1-q^2 \langle R_g^2\rangle _{\rm rod}/3$ with 
$\langle R_g^2\rangle _{\rm rod} = L_{\rm rod}^2/12$ to the $1/q$ power law 
(Eq.~(\ref{eq28})). Of course, on the lattice consideration of $q >\pi$ 
does not make sense, since distances of the order of a lattice spacing 
and less are not meaningful.

While for a Gaussian coil no direction of $\vec{q}$ is singled out, for 
a rod it makes sense to consider also the special case where the wave 
vector $\vec{q}$ is oriented along the rod; rather than considering the 
case when all orientations of $q$ are averaged over, as done in
Eqs.~(\ref{eq27}) - (\ref{eq29}). Then one rather obtains, $q_{||}$ being the
component of $\vec{q}$ parallel to the axis of the rod~\cite{46}
\begin{equation}\label{eq30}
S_{\rm rod}(q_{||}) = \frac {2}{(q_{||} L_{\rm rod})^2} 
[1-\cos (q_{||} L_{\rm rod})] \,.
\end{equation}
Note that Eq.~(\ref{eq30}) leads to an oscillatory decay since 
$\cos (q_{||}L_{\rm rod})=1 $ for 
$q_{||}^{(k)}=k(2\pi/L_{\rm rod})$, $k = 0,1,2,\;\ldots ,\;L_{\rm rod}$, 
and $S_{\rm rod} (q_{||})$ hence has zeros for all 
$q_{||}^{(k)}$, $k = 0,1,2,\; \ldots ,\; L_{\rm rod}$.

When we consider the scattering from semiflexible Gaussian chains, we now 
expect a smooth crossover between the Debye function, 
$S_{\rm Debye}(q)$ and the 
rod scattering, Eq.~(\ref{eq27}), similar to the smooth crossovers from rods 
to Gaussian coils, as described for the radii by Eqs.~(\ref{eq23}) and 
(\ref{eq24}). It turns out that this is a formidable problem, and no simple 
explicit formula exists, despite the fact that excluded volume effects still 
are neglected~\cite{161,162,163,164,165,18,19,20,21,22,23,231,24,25,26,27,28,29,291,292}. Kholodenko~\cite{26} 
derived an interpolation formula which 
describes the two limiting cases of Gaussian coils and rigid rods exactly, 
and which is expected to show only small deviations from the exact result 
in the intermediate crossover regime. His result can be cast in the form
\begin{equation}\label{eq31}
S(q) = \frac 2 x [I_1(x)-\frac 1 x I_2 (x)], \quad x = 3L/2 \ell_p
\end{equation}
where
\begin{equation}\label{eq32}
I_n(x) = \int \limits _0^x dzz^{n-1}f(z),
\end{equation}
and the function $f(z)$ is given by
\begin{eqnarray}\label{eq33}
f(z) =  \left\{ \begin{array}{r@{\, , \qquad}l}
\frac{1}{E} \frac{\sinh (Ez)}{\sinh z} & q \leq \frac{3}{2 \ell_p}  \, , \\
\frac {1} {\hat{E}} \frac {\sin (\hat{E}z)} {\sinh z} & q >\frac{3}{2\ell_p}\,,
\end{array} \right.
\end{eqnarray}
with
\begin{equation}\label{eq34}
E = \left[1-\left(\frac {2q\ell_p}{3} \right)^2 \right]^{1/2}\, , \, 
\hat{E} = \left[\left(\frac {2q\ell_p}{3}\right)^2 -1\right]^{1/2}\;.
\end{equation}

In addition, Stepanow~\cite{28} has developed a systematic expansion of the 
scattering function in terms of the solution for the quantum rigid rotator 
problem, which converges fast if $L/\ell_p$ is not too large. Note that the 
opposite limit, $L/\ell_p \rightarrow \infty$, has already been considered 
by des Cloizeaux~\cite{18}, and his result has been quoted in the introduction
(Eq.~(\ref{eq1})). This expansion (as well as equivalent representations  
written as continued fractions~\cite{27}) can only be 
evaluated numerically~\cite{58}.

However, a few qualitative statements can be made on the structure factor 
$S(q)$ in the representation of a Kratky plot, $qLS(q)$ plotted vs.~$Lq$. 
While Eqs~(\ref{eq27})-(\ref{eq29}) imply a monotonous increase from the 
straight line $qLS(q\approx 0)= qL$ towards the plateau value 
$qLS(q \gg2 \pi/L)=\pi$ for simple rigid rods, for semiflexible polymers 
this Kratky plot exhibits a maximum, since in the regime of interest
we may crudely approximate $S(q)$ by its leading terms
\begin{equation}\label{eq35}
qLS(q) \approx qL - \frac 1 3 (qL)^3\frac{\langle R_g^2\rangle}{L^2}\, .
\end{equation}
\begin{equation}\label{eq36}
q_{\rm max} =1/\sqrt{\langle R_g^2\rangle} \, , \,
q_{\rm max} LS(q_{\rm max}) = \frac 2 3 L/\sqrt{\langle R_g^2\rangle}\,
\end{equation}
i.e. From Eqs.~(\ref{eq35}), (\ref{eq24}) we immediately find, for $n_p \gg 1$,
\begin{equation}\label{eq37}
q_{\rm max} = \sqrt{3/(\ell_pL)} \, , \, q_{\rm max} LS(q_{\rm max}) = \frac {2}{\sqrt{3}} \sqrt{L/\ell_p} \; .
\end{equation}
Using the full Debye function one finds a different prefactor,
$q_{\rm max} \approx \sqrt{6.4/(\ell_pL)}$, but the general
scaling behavior is the same as given by Eq.~(\ref{eq37}).
Thus, when  $L$ is known (as is the case in simulations) for semiflexible 
Gaussian chains estimation of the coordinates of the maximum in the Kratky 
plot allows a straightforward estimation of the persistence length $\ell_p$.

Considering now the effects of excluded volume, we emphasize that 
Eq.~(\ref{eq36}) still is supposed to be valid, while Eq.~(\ref{eq37}) 
no longer holds. In particular, in Ref.~\cite{13} it was shown that 
in $d=2$ the Kratky-Porod model~\cite{14} of semiflexible chains, 
on which Eqs.~(\ref{eq23}), (\ref{eq24}), and (\ref{eq31})-(\ref{eq34}) 
are based, has no validity whatsoever: rather it was shown that 
around $n_p=1$ a smooth crossover from the rigid rod behavior to the 
behavior of two-dimensional self-avoiding walks occurs. Thus, we expect 
similarly instead of Eq.~(\ref{eq24}) that (recall $n_p=L/\ell_p)$
\begin{equation}\label{eq38}
\frac {\langle R_g^2\rangle }{\ell_p L} = f (n_p), 
\quad \textrm{with} \quad f(n_p<1) = n_p\, ,
\end{equation}
as in Eq.~(\ref{eq24}), but
\begin{equation}\label{eq39}
f(n_p \gg 1) = C_g^sn_p^{2\nu_2-1} \quad , \quad \nu_2=3/4 \, ,
\end{equation}
$C_g^s$ being (for $\ell_p/\ell_b \gg 1$) a non-universal constant. 
Eqs.~(\ref{eq38}), (\ref{eq39}) hence imply in $d=2$ dimensions
\begin{equation}\label{eq40}
\langle R_g^2\rangle = C_g^s \ell_p ^{1/2} L^{3/2}\, , 
\qquad L \rightarrow \infty \,,
\end{equation}
in full analogy to the result for fully flexible chains,
\begin{equation}\label{eq41}
\langle R_g^2\rangle = C_g^f \ell_b^{1/2} L^{3/2}\,, \qquad L 
\rightarrow \infty \,,
\end{equation}
where $C_g^f$ is another non-universal constant. Similar relations were 
found~\cite{13} for the end-to-end distance
\begin{eqnarray}\label{eq42}
\langle R^2 \rangle &=& C_e^f \ell_b^{1/2} L^{3/2} \; (\textrm{flexible})\, ,
\nonumber  \\ 
\langle R^2 \rangle &=& C_e^s \ell_p ^{1/2} L^{3/2} \; 
(\textrm{semiflexible})
\end{eqnarray}
with $C_e^f$ and $C_e^s$ other (non-universal) constants for flexible and 
semiflexible chains, respectively. The ratios $C_e^f/C_g^f$ and 
$C_e^s/C_g^s$ are expected to be universal, however (for Gaussian chains 
$\langle R_e^2 \rangle /\langle R_g^2 \rangle =6$), for both flexible and 
semiflexible chains.

In $d=3$ dimensions, however, the situation in the presence of excluded 
volume is considerably more involved~\cite{13}. For the end-to-end 
distance $\langle R^2\rangle$ of semiflexible chains, two successive 
crossovers were found: for $n_p\approx 1$ a crossover from rods to 
Gaussian coils occur, while excluded volume effects become prominent for
\begin{equation}\label{eq43}
n_p >n_p^* \, , \qquad n_p^* \propto (\ell_p/D)^\zeta
\end{equation}
where we have introduced the chain diameter $D$ as another characteristic 
length that may be needed in general (while in our model $D=\ell_b$, however), 
and $\zeta$ is an exponent that is not yet known precisely. Arguments 
based on Flory theory yield~\cite{13,15,16} $\zeta =2$ while Monte Carlo 
results rather suggested~\cite{13} $\zeta \approx 1.5$. We recall that 
in $d=3$ Flory arguments are not~\cite{1} exact, implying~\cite{3,9} $\nu = 3/5$ 
instead of $\nu=0.588$. A similar double crossover from rods 
to first Gaussian coils and then to $d=3$ self-avoiding walks is 
expected to be visible in $\langle R_g^2\rangle$, too. If we could 
rely on Flory theory, we would predict from these considerations that
\begin{equation}\label{eq44}
\langle R_g^2\rangle /(\ell_pL)=C_g^s(n_p/n^*_p)^{2\nu-1}\, , 
\qquad n_p >n_p^*
\end{equation}
and hence (using the Flory value $\nu=3/5$)
\begin{equation}\label{eq45}
\langle R_g^2\rangle = C_g^sL^{6/5} (\ell_pD)^{2/5}\; .
\end{equation}
From $q_{\rm max}=(\sqrt{C_g^{s}}L^{3/5} (\ell_pD)^{1/5})^{-1}$ 
the persistence length $\ell_p$ 
can be inferred, provided $C_g^s$ has been determined. However, 
if $\ell_p \gg \ell_b$ and $D=\ell_b$, one can study the regime 
$1 < n_p <n_p^*$, where Gaussian statistics for the gyration radius is 
still applicable, and hence Eq.~(\ref{eq37}) applies.

\subsection{The structure factor in the presence of stretching forces}

For Gaussian chains under stretch, where a force $f$ is applied at a chain 
at one end in the +x-direction, the other end being fixed at the 
coordinate origin, the structure factor $S(\vec{q})$ has been derived 
by Benoit et al.~\cite{30} as follows $\{\vec{q}=(q_x,q_y,q_z)$ with 
$q_{||}=q_x$ and $q_\bot = \sqrt{q_y^2+q_z^2}\}$
\begin{equation}\label{eq46}
S_{||}(q_{||}) = 2 Re \left\{ \frac {\exp(-X_{||})-1+X_{||}}{X_{||}^2} 
\right\}\, ,
\end{equation}
\begin{equation}\label{eq47}
S_\bot(q_\bot) = 2  \frac {\exp(-X_\bot)-1+X_\bot}{X_\bot^2} \, ,
\end{equation}
where $X_\bot = q_\bot^2 (\langle R^2 \rangle_0/6 \lambda_\bot^{-2})$, with
$\langle R^2 \rangle_0$ the mean-square end-to-end distance of the chain 
in the absence of any force ($f=0$), and $\lambda_{\bot}$ describes the 
modification of the Gaussian distribution in the transverse directions 
($y$ and $z$-direction, for $d=3$). The quantity $X_{||}$ is complex 
(therefore the real part of Eq.~(\ref{eq46}) is taken) and is given by
\begin{equation}\label{eq48}
   X_{||} = \frac{q_{||}^2\langle R^2 \rangle_0}{6\lambda_{||}^{-2}}
+i \langle X \rangle q_{||}
\end{equation}
where $\lambda_{||}$ describes the modification of the Gaussian distribution
in the $x$-direction (parallel to the force). Benoit et al.~\cite{30}
explicitly state that their result is restricted to deformations of 
small amplitudes, and do not specify how $\lambda_{\bot}$, $\lambda_{||}$
are related to the applied force. However, considering the small $q$
expansion of Eqs.~(\ref{eq46}), (\ref{eq47}) one can relate these
parameters to the mean square gyration radius components of the chain,
since for $X_{\bot} \ll 1$
\begin{equation}
   S(q_{\bot}) = 1-q_{\bot}^2 \frac{\langle R^2 \rangle_0}
{18 \lambda_{\bot}^{-2}} = 1-q_{\bot}^2 \langle R^2_{g,\bot}\rangle /2 \, ,
\end{equation}
where in the last step Eq.~(\ref{eq15}) was used. Hence we conclude 
(note that $\langle R_g^2 \rangle_0=\langle R^2 \rangle_0 /6$ for Gaussian
chains) that $\lambda_{\bot}^2=(3/2) 
\langle R_{g,\bot}^2 \rangle / \langle R_g^2 \rangle_0 = 
(\langle R^2_{gy} \rangle + \langle R^2_{gz} \rangle)/
(\langle R^2_{gy} \rangle_0 + \langle R^2_{gz} \rangle_0)$,
as expected.
Similarly, Eq.~(\ref{eq46}) yields for $q_{||} \rightarrow 0$, using also 
Eq.~(\ref{eq14})
\begin{equation}
S(q_{||})=1-q_{||}^2\left(\frac{\langle R^2 \rangle_0}{18 \lambda_{||}^{-2}}
+\frac{\langle X \rangle^2}{12} \right) = 
1-q_{||}^2 \langle R_{g,||}^2 \rangle \, ,
\end{equation}
and hence we see that $\lambda_{||}$
in $X_{||}$ can be expressed in terms of the gyration radius component
$\langle R^2_{g,||} \rangle$ of the stretched chain and the extension
$\langle X \rangle$. However, since from the work of Benoit et al.~\cite{30}
it is not clear that Eqs.~(\ref{eq46}), (\ref{eq47}) are applicable for
conditions where $\langle X \rangle /L$ is not very small, we hence rederived 
Eqs.~(\ref{eq46}), (\ref{eq47}) by an independent method, which is more 
transparent with respect to the basic assumptions that are made.
This derivation is presented in an Appendix, and it shows that 
$S(q_{||})$ can be cast into the form
\begin{widetext}
\begin{equation}\label{eq51p}
   S(q_{||}) = \frac{1}{(N+1)^2} \sum_{i,j} \exp
\left[ -\frac{1}{2} q_{||}^2 (\langle X^2 \rangle - \langle X \rangle^2)
\frac{\mid i-j \mid}{N} \right] \cos \left(q_{||} 
\frac{\mid i-j \mid \langle X \rangle }{N} \right)
\end{equation}
\end{widetext}
which is equivalent to Eq.~(\ref{eq46}) but with a somewhat different 
expression for $X_{||}$, namely
\begin{equation}\label{eq52p}
  X_{||} = \frac{1}{2} q_{||}^2 \left( \langle X^2 \rangle - \langle
X \rangle^2 \right) + i q_{||} \langle X \rangle \, .
\end{equation}
It is interesting to note that Eqs.~(\ref{eq46}) and (\ref{eq51p}) can be 
given a very simple physical interpretation: with respect to the 
correlation in stretching direction, the stretched polymers is 
equivalent to a harmonic one-dimensional ``crystal" (which at nonzero
temperature lacks long range order, of course) of length
$Na=\langle X \rangle$, $a$ being the ``lattice spacing" of the crystal.

Writing the Hamiltonian of the one dimensional chain as~\cite{Emery1978,Ricci2007}
\begin{equation}
   {\cal H} = \frac{1}{2} \sum_{\ell} \left[ 
\frac{\pi_{\ell}^2}{m} 
+ mc^2 \frac{(x_{\ell+1}-x_{\ell} -a)^2}{a^2} \right]\, ,
\end{equation}
where point particles of mass $m$ have positions $x_{\ell}$ and
conjugate momenta $\pi_{\ell}$, and the spring potential coupling
neighboring particles is written in terms of the sound velocity $c$.
At $T=0$, particles would be localized at positions
$x_n^{(0)}=x_0^{(0)}+na$, $n=0,1,\; \ldots,\; N$. So it makes
sense to consider displacements relative to the ground state,
$u_n=x_n-x_n^{(0)}=x_n-na$, putting $x_0^{(0)}$ at the origin.
Due to the harmonic character of this ``crystal", one can calculate
the mean square displacements easily to find (assuming periodic
boundary conditions) that 
$\langle (u_n-u_0)^2 \rangle=na^2k_BT/(mc^2)=n\delta^2$, where
$\delta$ characterizes the local displacement for two neighboring
particles. Applying the formula also for the end-to-end distance
of a chain without periodic boundary conditions, 
$\langle (u_N-u_0)^2 \rangle = N \delta^2$, one immediately finds that
$S(q_{||})$ for the harmonic chain yields the above expressions of 
$S(q_{||})$, since $X=x_N-x_0=Na+u_N-u_0$, $\langle X \rangle=Na$,
and $\langle X^2 \rangle = N^2a^2+N \delta^2=\langle X \rangle^2
+N \delta^2$. This consideration also emphasizes that a condition
$\langle X \rangle \ll L=N \ell_b$ in fact is not required for the
validity of Eqs.~(\ref{eq46})-(\ref{eq52p}). 

\begin{table*}[!htb]
\caption{Values of persistence lengths
$\ell_p/\ell_b$ for semiflexible chains in $d=2$ and $d=3$,
estimated by Eq.~(\ref{eq21}), and the crossover length $N^*$
between the intermediate Gaussian regime and the SAW regime in $d=3$,
estimated empirically from Fig.~7b of Ref.~\cite{13}
($N^*=N^{\rm rod}=2\ell_p/\ell_b$ in $d=2$) for various values of $q_b$.}
\begin{center}
\begin{tabular}{|c|ccccccccc|}
\hline
$q_b$     & 0.005 & 0.01 & 0.02 & 0.03 & 0.05 & 0.10 & 0.20 & 0.40 & 1.0 \\
\hline
$\ell_{p}/\ell_b (d=2)$ & 118.22 & 59.44 & 30.02 & 20.21 & 12.35 & 6.46 & 3.50 & 2.00 & 1.06 \\

$\ell_{p}/\ell_b (d=3)$ & 51.52 & 26.08 & 13.35 & 9.10 & 5.70 & 3.12 & 1.18 & 1.12 & 0.67  \\
$N^* (d=3)$ & 36000 & 9000 & 1850 & 700 & 180 & 41 & 11 & - & - \\
\hline
\end{tabular}
\end{center}
\label{table1}
\end{table*}

For the unstretched case 
$(\langle X \rangle =0)$ Eqs.~(\ref{eq46})-(\ref{eq52p}) reduce to 
Eq.~(\ref{eq22}), as 
it should be. We recall that according to the Kratky-Porod model 
simple approximations for the extension $\langle X \rangle$ of a 
chain as a function of the force can be derived (see~\cite{13} for a 
review), namely
\begin{equation}\label{eq49}
\frac{f\ell_p}{k_BT} = \frac 3 4 \frac {\langle X\rangle}{L} + 
\frac {1}{8(1-\langle X\rangle/L)^2} - \frac 1 8 \, , \qquad d =2\,,
\end{equation}
and
\begin{equation}\label{eq50}
\frac{f\ell_p}{k_BT} = \frac 3 4 
\frac {\langle X\rangle}{L} + \frac {1}{4(1-\langle X\rangle/L)^2} - 
\frac 1 4 \, , \qquad d =3\,.
\end{equation}
Eqs.~(\ref{eq49}), (\ref{eq50}) imply in the linear response regime,
 where $\langle X\rangle \propto f$, that
\begin{equation}\label{eq51}
\frac{f\ell_p}{k_BT} = \frac d 2 \frac {\langle X\rangle}{L}\, .
\end{equation}
However, from linear response one can show generally that
\begin{equation}\label{eq52}
\langle X \rangle = f \langle X^2 \rangle _0/(k_BT) 
= f \langle R^2\rangle _0 /( dk_BT)\, ,
\end{equation}
where $\langle R^2\rangle_0$ is the mean square end-to-end distance in the 
absence of forces. Eqs.~(\ref{eq51}), (\ref{eq52}) are compatible with 
each other for Gaussian semiflexible chains, for which 
$\langle R^2\rangle _0 = 2 \ell_p L $ \{Eq.~(\ref{eq23})\}, but are 
incompatible in the presence of excluded volume forces. In this case, one 
observes a crossover from the linear response regime, as described by 
Eq.~(\ref{eq52}) together with Eq.~(\ref{eq42}) for $d=2$ and a result 
analogous to Eq.~(\ref{eq50}), namely
\begin{equation}\label{eq53}
\langle R_e^2\rangle = C_e^s L^{6/5} (\ell_pD) ^{2/5} \, , \, d=3 \, 
(\textrm{using} \; \nu \approx 3/5) \,,
\end{equation}
to the so-called ``Pincus blob''~\cite{31} regime, described by a power 
law for the extension versus force relation
\begin{equation}\label{eq54}
\langle X \rangle /L \propto (f\ell_p /k_BT) ^{1/\nu-1}\, .
\end{equation}
While in $d=2$ Eq.~(\ref{eq52}) holds up to $\langle X \rangle /L$ of order
unity, where then saturation effects ($\langle X \rangle /L \rightarrow 1$ 
for large enough $f$) set in, in $d=3$ the regime of validity of 
Eq.~(\ref{eq52}) 
is much more restricted, namely we have to require~\cite{13}
\begin{equation}\label{eq55}
\xi_p \equiv k_BT /f >R_0 \propto \ell_p^2/D
\end{equation}

\begin{figure*}[htb!]
\begin{center}
\includegraphics[scale=0.60,angle=270]{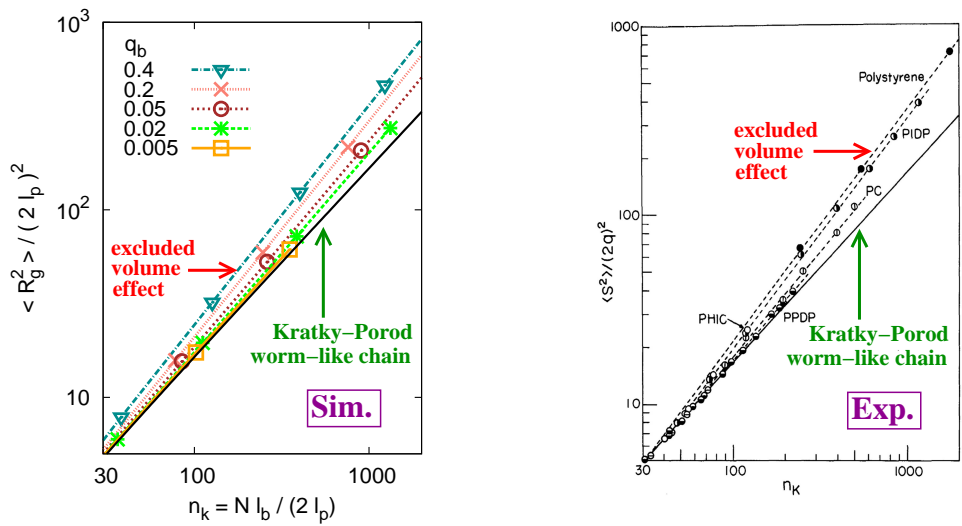}
\caption{\label{fig1}
(a) Log-log plot of normalized gyration radius square
$\langle R_g^2\rangle/(2 \ell_p)^2$ versus the number of Kuhn segments,
for the range $30 \leq n_K \leq 3000$, for chains of widely varying stiffness,
and comparison to corresponding experimental data (b), taken from Norisuye
and Fujita~\cite{12}. The full straight line is the Kratky-Porod model,
Eq.~(\ref{eq24}).}
\end{center}
\end{figure*}

For stronger forces (corresponding to $\xi_p<R_0$) the Kratky-Porod results, 
Eqs.~(\ref{eq49}), (\ref{eq50}), are expected to become valid. In the 
Pincus blob regime, also nontrivial power laws for the fluctuations 
$\langle X ^2\rangle - \langle X\rangle ^2$ and the transverse linear 
dimensions are predicted~\cite{44}
\begin{equation}\label{eq56}
\langle X^2\rangle - \langle X\rangle ^2 \propto 
\langle R_\bot^2\rangle \propto (f\ell_p/k_BT)^{1/\nu-2}\, .
\end{equation}
Since we are not aware of any treatment of the structure factor of the 
Kratky-Porod model under stretch, we shall use 
Eqs.~(\ref{eq46}), (\ref{eq47}) also for semiflexible chains 
(but using the numerical results for $\langle X \rangle$ and 
$\langle X^2\rangle - \langle X \rangle ^2, \; \langle R_{g,\bot}^2 \rangle$, 
rather than theoretical predictions).

\begin{figure*}[htb]
\begin{center}
(a)\includegraphics[scale=0.29,angle=270]{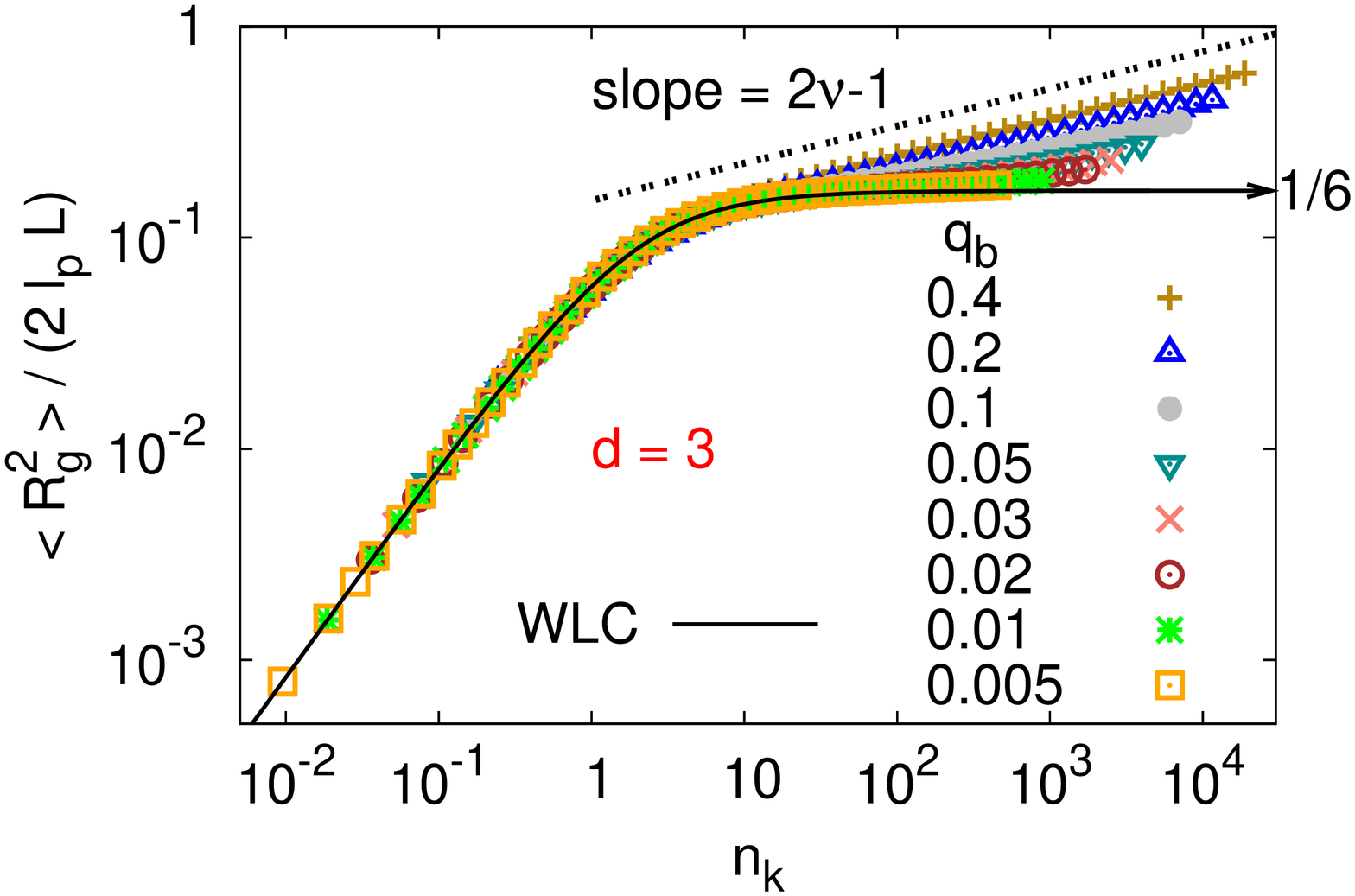}\hspace{0.4cm}
(b)\includegraphics[scale=0.29,angle=270]{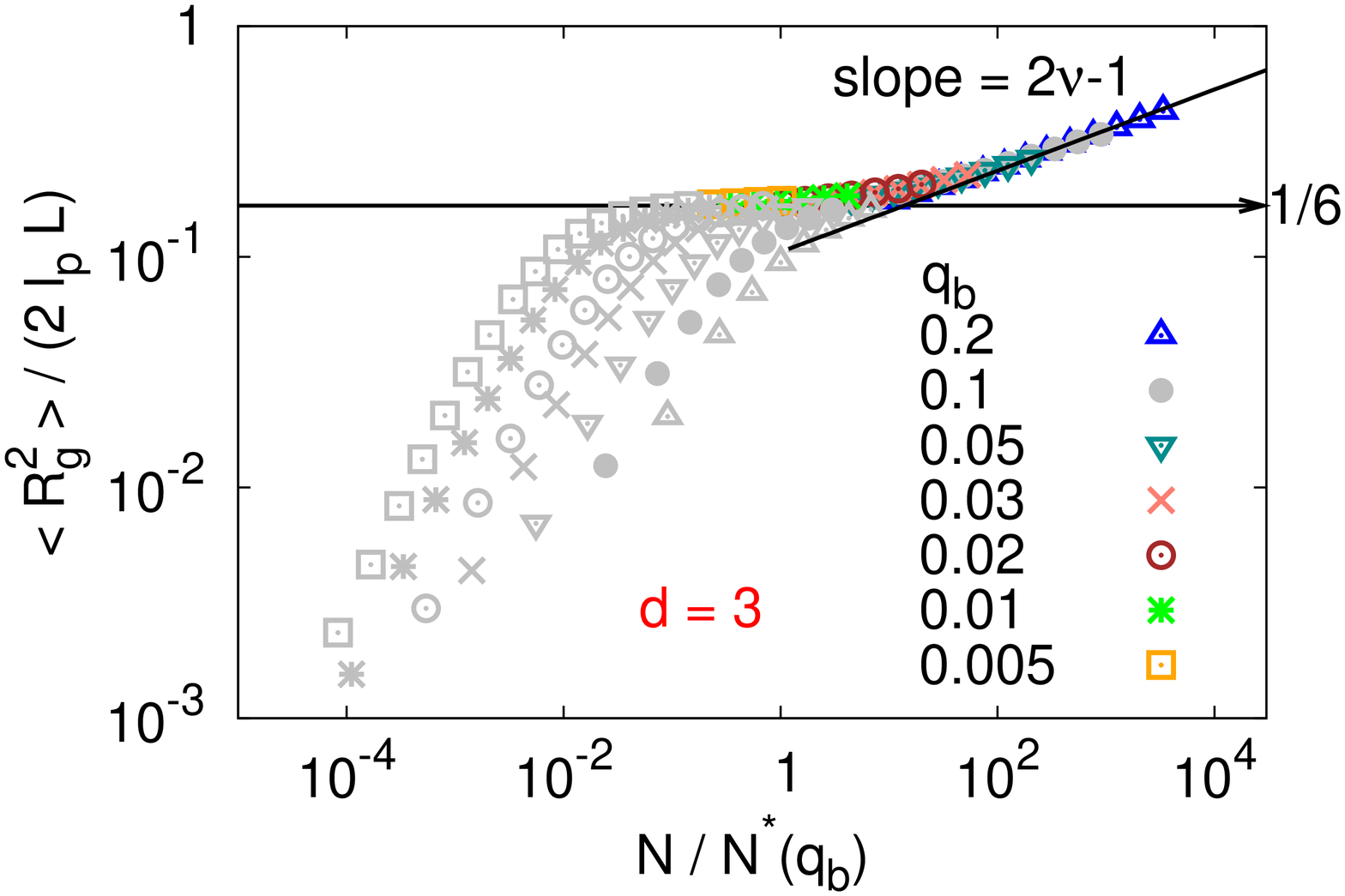}\\
\caption{\label{fig2}
(a) Log-log plot of $\langle R_g^2\rangle /(2 \ell_pL)$ versus $n_K $ for
$d=3$, and including data for widely varying persistence lengths. Note that
the worm-like chain result, WLC (Eq.~(\ref{eq24})), describes correctly the
crossover from rods to coils, but not the onset of excluded volume effects.
(b) Same as (a), but choosing $N/N^*(q_b)$ as an abscissa variable
($N^*(q_b)$ was already estimated for a similar scaling plot for the mean
square end-to-end distance, cf. Fig.~7b of Ref.~\cite{13}). Here $L=N\ell_b$.}
\end{center}
\end{figure*}

\begin{figure}[htb]
\begin{center}
\includegraphics[scale=0.29,angle=270]{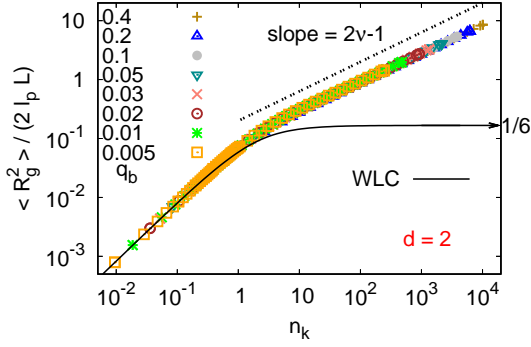}
\caption{\label{fig3}
Log-log plot of $\langle R_g^2\rangle/(2 \ell_pL)$ versus $n_K$ for $d=2$,
including data for widely varying persistence lengths. Note that in $d=2$
there occurs a direct crossover from rods to SAWs, an intermediate Gaussian
regime in absent.}
\end{center}
\end{figure}

\begin{figure}[htb]
\begin{center}
(a)\includegraphics[scale=0.29,angle=270]{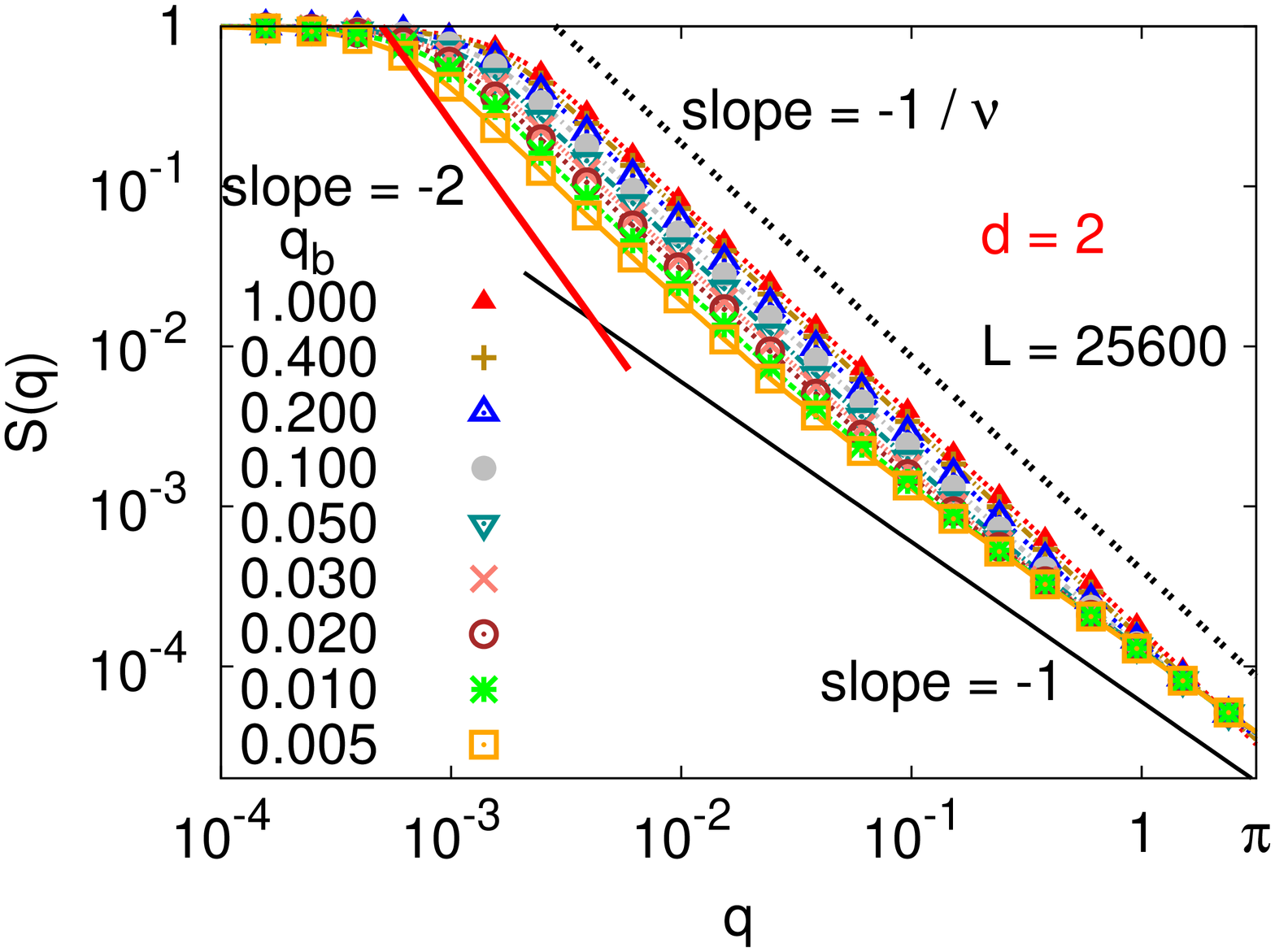}\hspace{0.4cm}
(b)\includegraphics[scale=0.29,angle=270]{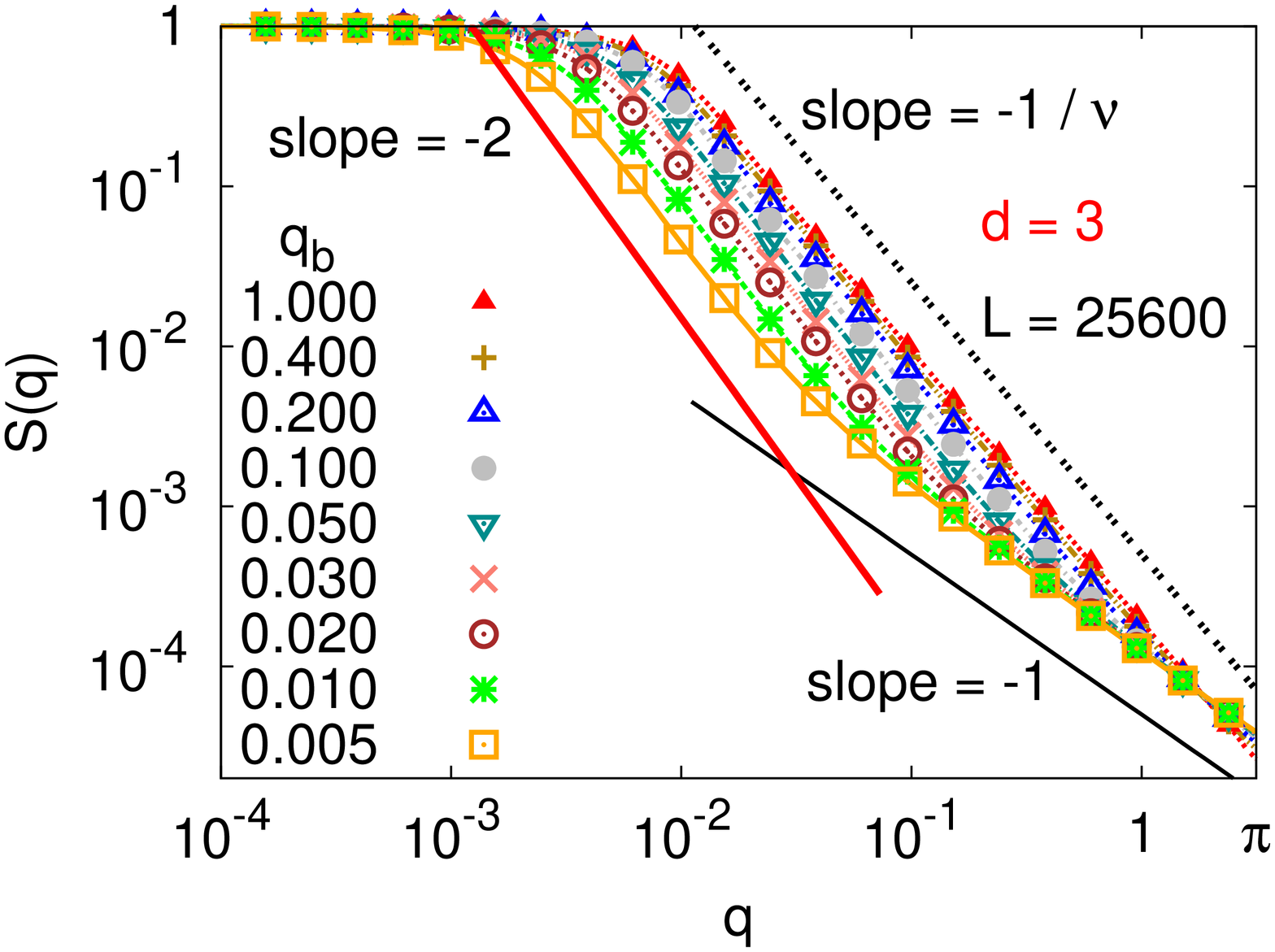}\\
\caption{\label{fig4}
Structure factor $S(q)$ plotted $q$, on log-log- scales, for $d=2$ (a)
and $d=3$ (b); only data for $L=25600$ are shown, but many different
choices of the stiffness parameter $q_b$ are included. The straight lines
indicate the rod-like behavior at large $q$ (slope $= -1$) and
the SAW behavior for flexible chains (slope $=-1/\nu$). Also the slope
expected in the Gaussian regime is included (slope $=-2$). Only data up to
$q=\pi$ have been included (larger $q$ cannot be studied due to the lattice
character of our model).}
\end{center}
\end{figure}

\begin{figure*}[htb]
\begin{center}
(a)\includegraphics[scale=0.29,angle=270]{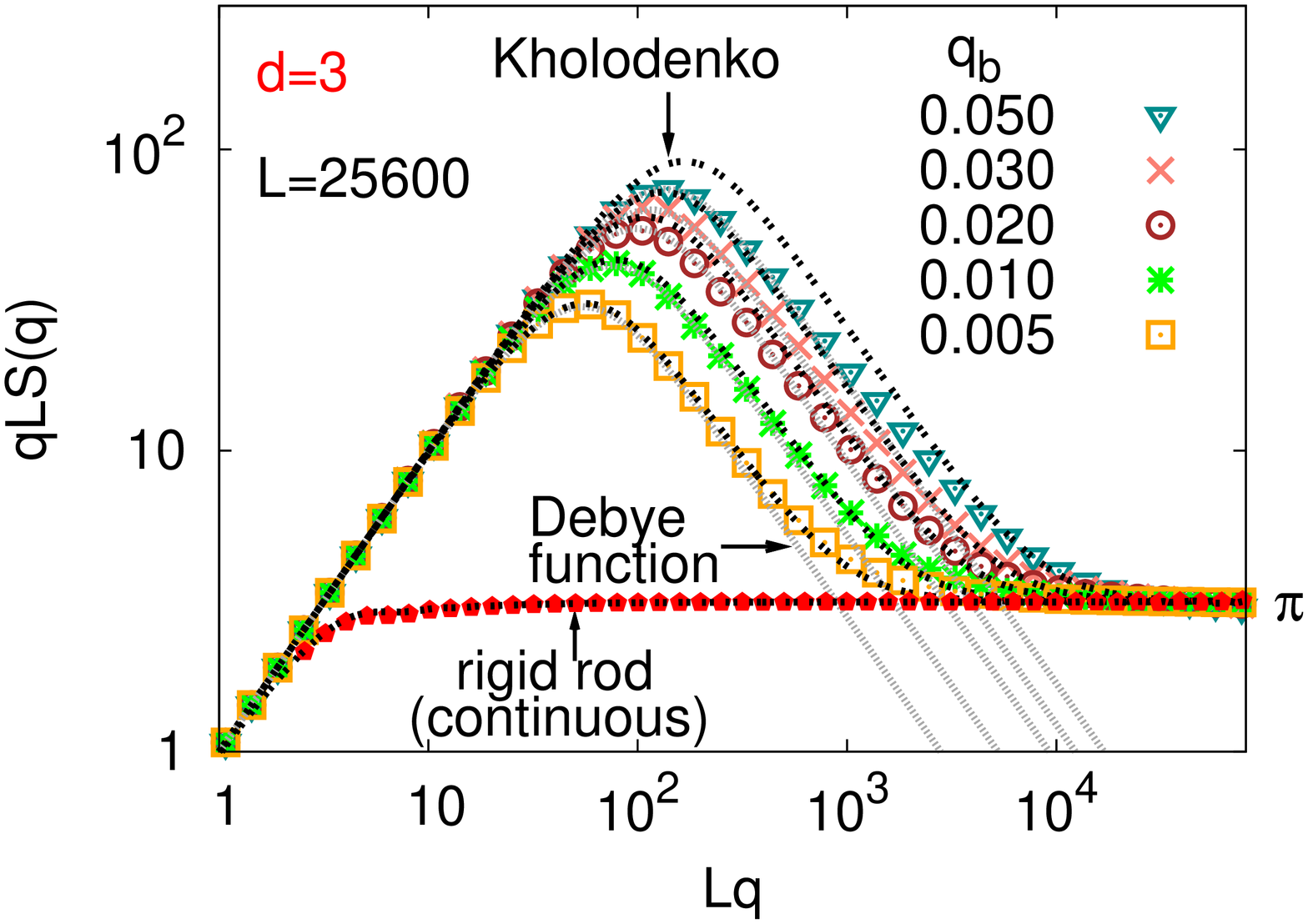}\hspace{0.4cm}
(b)\includegraphics[scale=0.29,angle=270]{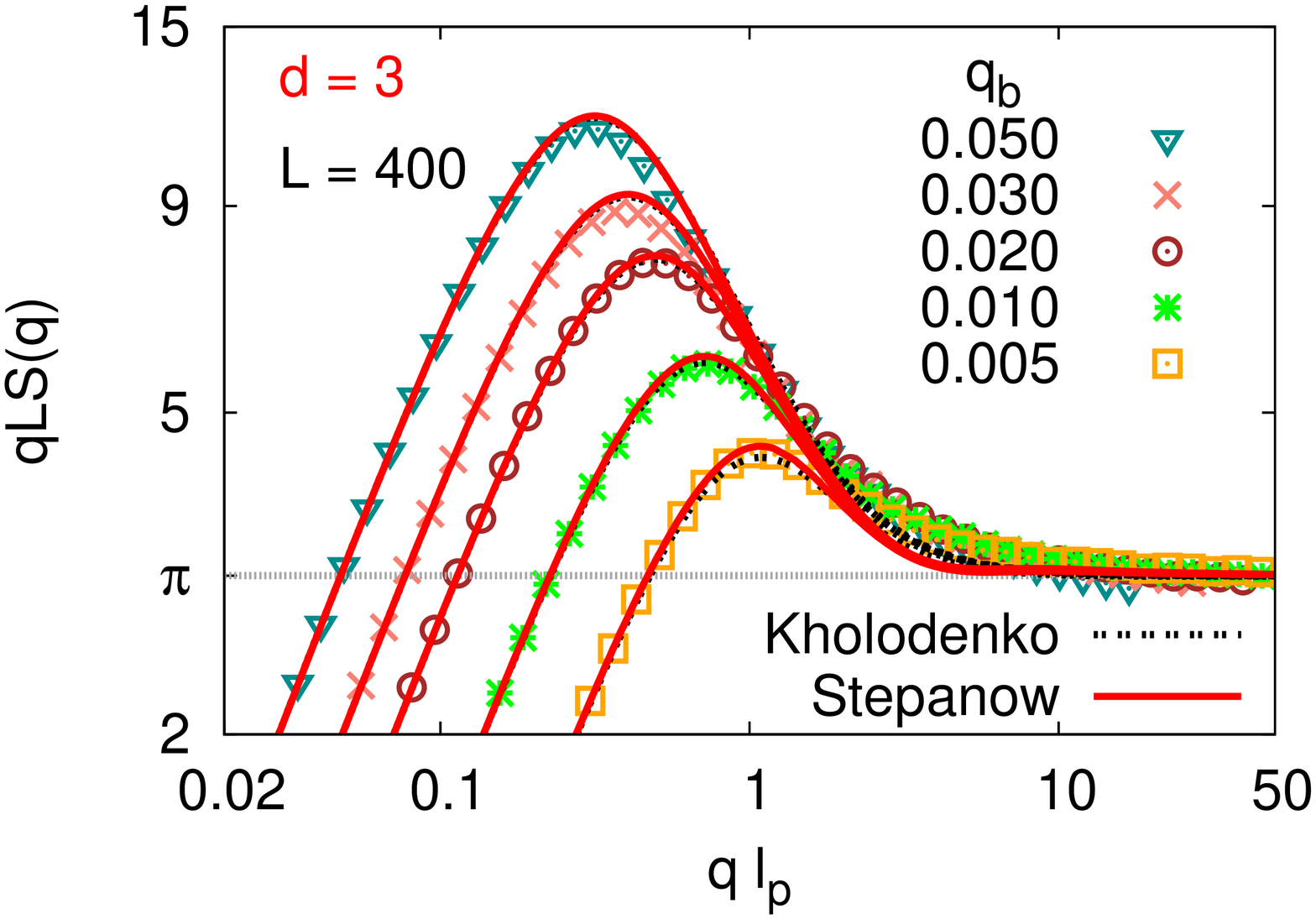}\\
\vspace{0.5cm}
(c)\includegraphics[scale=0.29,angle=270]{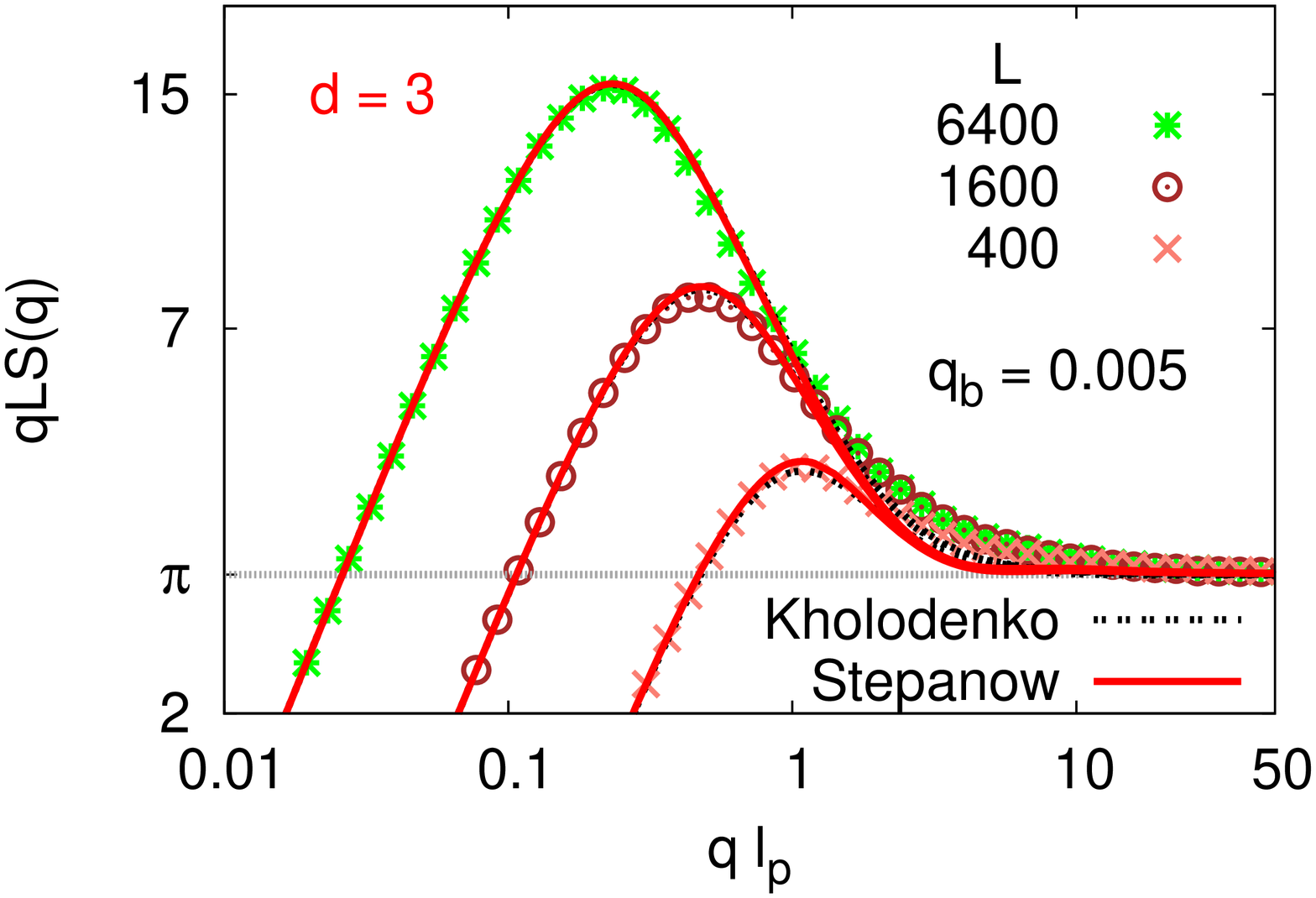}
\caption{\label{fig5}
(a) Rescaled structure factor $qLS(q)$ plotted against $Lq$ for $d=3$
and $L=25600$, including 5 choices of the stiffness. The result for Gaussian
chains (Debye function) and for continuous rigid rods (for which
$qLS(q) \rightarrow \pi$ for large $q$. cf.~Eq.~(\ref{eq28})) are included
for comparison. Also predictions obtained from the formulas proposed by
Kholodenko \{Eqs.~(\ref{eq31})-(\ref{eq34})\} are shown. (b)
Rescaled structure factor $qLS(q)$ plotted against $q\ell_p$ for $d=3$ and
for $L=400$, including both the predictions due to Stepanow~\cite{271} and
Kholodenko \{Eq.~(\ref{eq31}) - (\ref{eq34})\}, which are essentially
indistinguishable on the scale of the figure. (c) Same as (b), but for
3 different choices of $L$ for $q_b=0.005$.}
\end{center}
\end{figure*}

\begin{figure*}[htb]
\begin{center}
(a)\includegraphics[scale=0.29,angle=270]{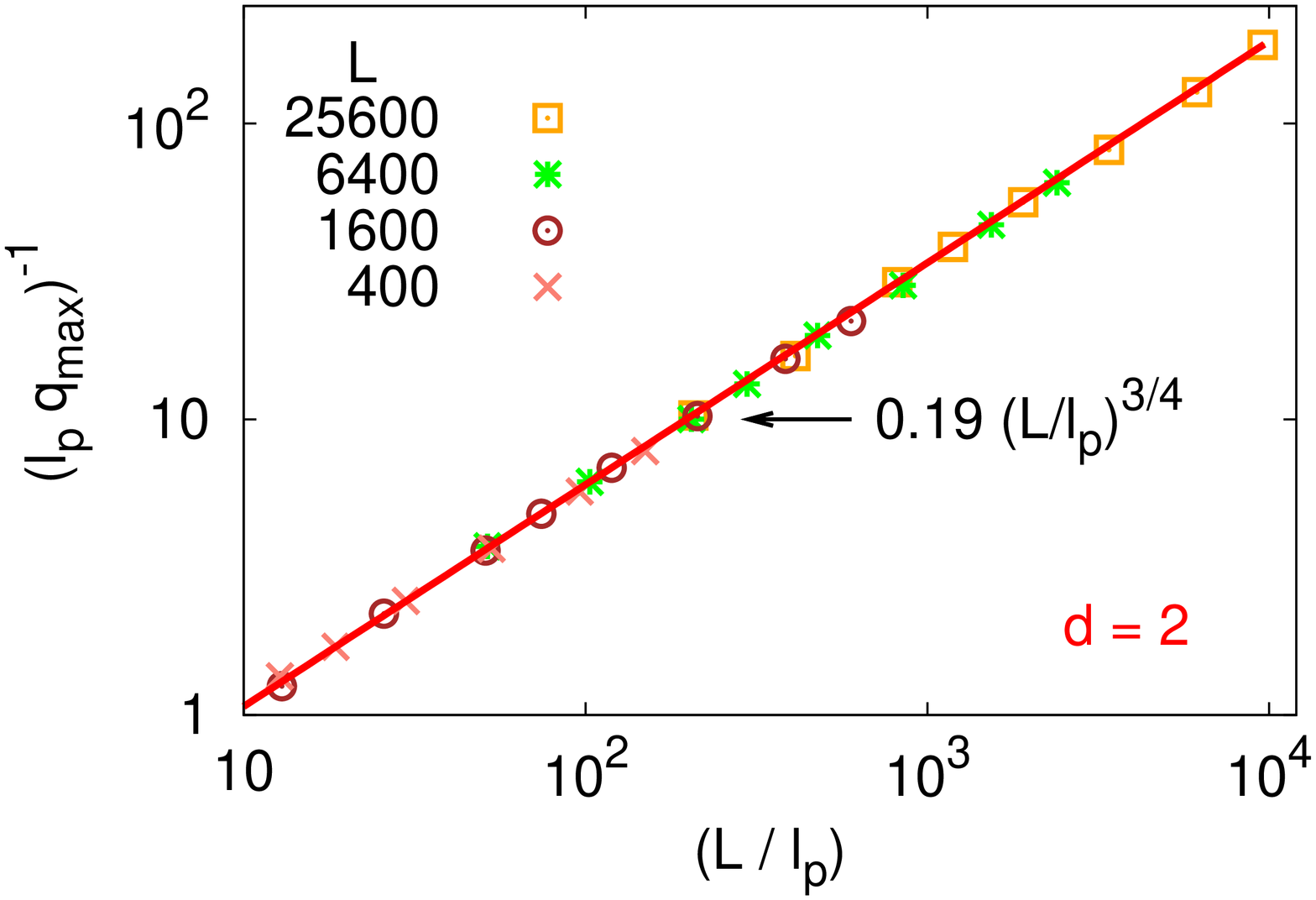}\hspace{0.4cm}
(b)\includegraphics[scale=0.29,angle=270]{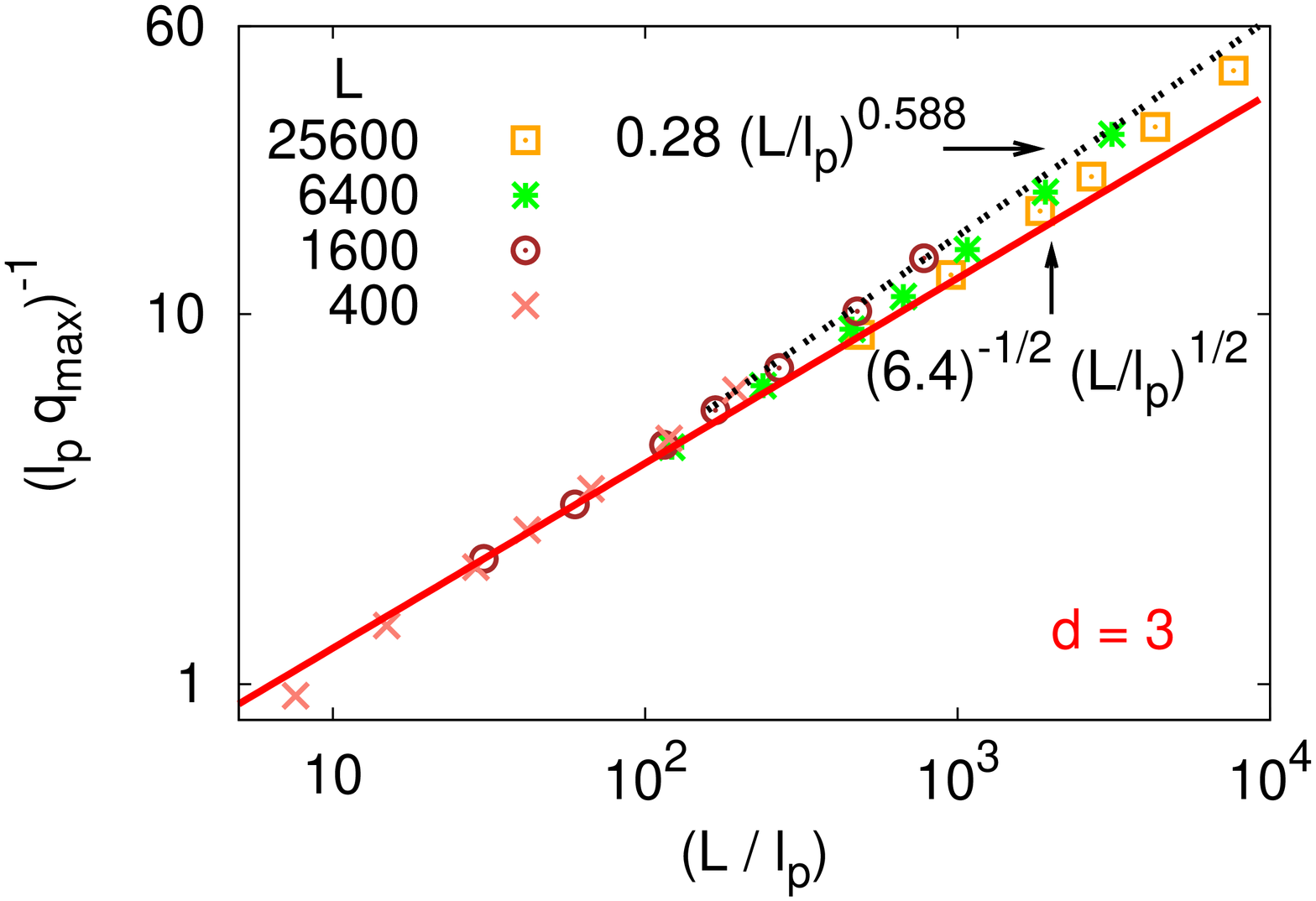}\\
\caption{\label{fig6}
Log-log plot of the inverse position of the maximum in the Kratky plot,
$(\ell_pq_{\rm max})^{-1}$, versus the rescaled contour length $L/\ell_p$,
for $d=2$ (a) and $d=3$ (b). The full straight line included in (a)
is described by the equation $0.19 (L/\ell_p)^{3/4}$.
In case (b), two straight lines
with slope $\nu = 1/2$ and $\nu = 0.588$ are included, to illustrate the
crossover.
In both cases the persistence length $\ell_p$ is varied over a wide range,
and the contour lengths chosen are $L=400$, $1600$, $6400$ and $25600$,
as indicated.}
\end{center}
\end{figure*}

\begin{figure*}[htb]
\begin{center}
(a)\includegraphics[scale=0.29,angle=270]{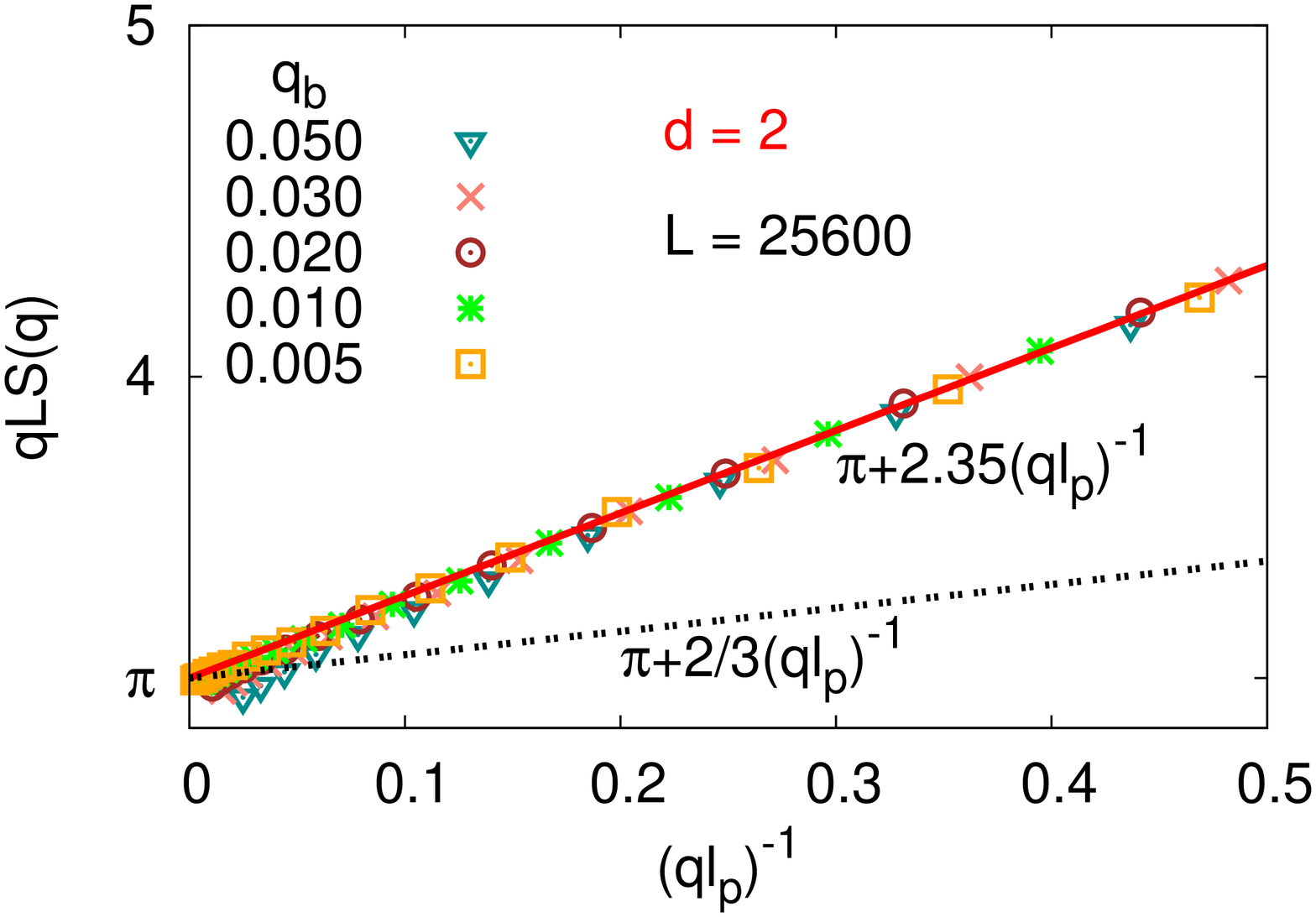}\hspace{0.4cm}
(b)\includegraphics[scale=0.29,angle=270]{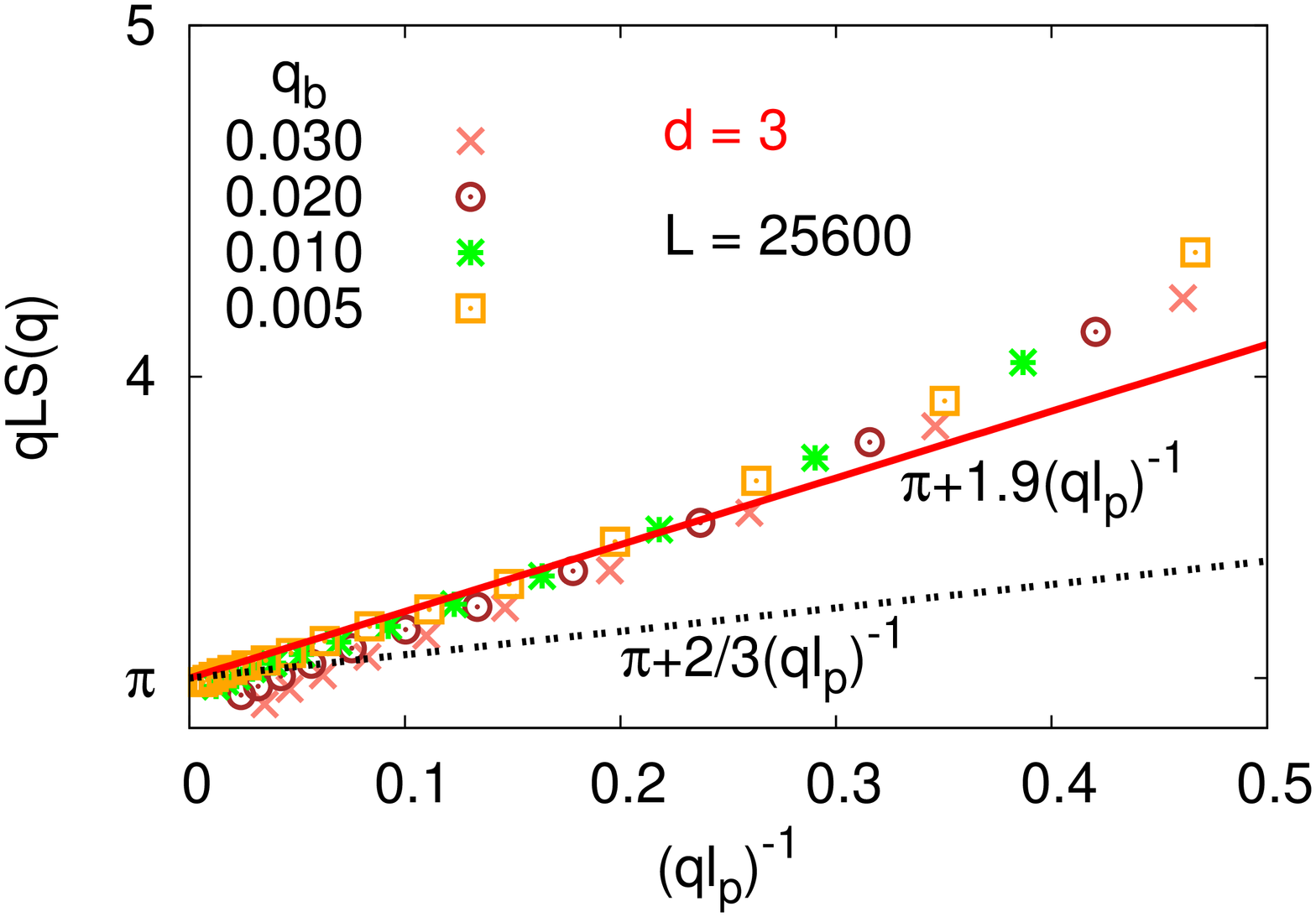}\\
\caption{\label{fig7}
Plot of $qLS(q)$ vs.~$(qL_p)^{-1}$, for $d=2$ (a) and $d=3$ (b).
All data are for $L=25600$ only, and many different choices of $q_b$,
as indicated. In each case, two straight lines are shown: the result of
des Cloizeaux~\cite{18}, Eq.~(\ref{eq1}), $qLS(q)= \pi+(2/3)(q\ell_p)^{-1}$,
and empirical fits, $qLS(q)=\pi+2.35(q\ell_p)^{-1}$ for $d=2$ and
$qLS(q)= \pi + 1.9(q \ell_p)^{-1}$ for $d=3$, respectively.}
\end{center}
\end{figure*}

\section{Model and Simulation Technique}

Our model is the standard self-avoiding walk (SAW) on the square and 
simple cubic lattices, effective monomers being described by occupied 
lattice sites, connected by bonds. Each site can be taken only once, 
and thus we realize the excluded volume interaction. The lattice 
spacing henceforth is our unit of length, $\ell_b=1$. We introduce 
an energy $\epsilon_b$ for any kink the walk takes (by an angle of 
$\pm 90^\circ$). Any such kink introduces hence a factor 
$q_b=\exp(-\epsilon_b/k_BT)$ to the statistical weight of the walk.

In the presence of a force $f$ coupling to the extension $X$ of the 
chain in x-direction, the statistical weight gets another factor 
$b^X$, with $b =\exp (f/k_BT)$. Then the partition function of a SAW 
with $N$ bonds (i.e., $N+1$ effective monomers) and $N_{\rm bend}$ 
local kinks becomes
\begin{equation}\label{eq57}
Z_{N,N_{\rm bend}} (q_b,b) = \sum \limits _{\textrm{config.}} 
C(N,N_{\rm bend},X) q_b^{N_{\rm bend}} b^X
\end{equation}
By the pruned-enriched Rosenbluth method (PERM) it is possible to 
obtain estimates of the partition function and quantities derived 
from it (e.g. $\langle X \rangle, \langle X^2\rangle)$ and additional 
averages such as $S(q)$, using chain lengths up to $N=25600$. 
Both the chain stiffness and the force $f$ have been varied over 
a wide range; for $q_b=1$ one has fully flexible self-avoiding random walks, 
while for $q_b=0.005$ the persistence length (computed from 
Eq.~(\ref{eq21})) is of the order of $120$ in $d=2$ and $52$ in $d=3$ 
(Table I lists our corresponding estimates). For technical details 
on the implementation of the algorithm, we refer to the literature~\cite{13}.

\section{Results for the scattering function of unstretched chains}

We start with our data for the mean square gyration radius 
$\langle R_g^2\rangle$, normalized by the square of the Kuhn length 
$\ell_K=2 \ell_p$, plotted vs.~the number of Kuhn segments 
$n_K=L/\ell_K=N\ell_b/(2 \ell_p)$, Fig.~1a since this was the
representation chosen for the experimental data of Norisuye and 
Fujita~\cite{12}, 
which we reproduce in Fig.~1b. Both diagrams show the same range of 
abscissa $(30 \leq n_K \leq 3000)$ and ordinate 
$(5 < \langle R_g^2\rangle /(2 \ell_p)^2 <1000)$. The qualitative 
similarity between both simulation and experiment is striking. 
Since only the regime of rather large $n_K$ is shown, the crossover 
from rods to Gaussian chains is not included (the full straight line 
represents the Gaussian chain behavior, as described by Eq.~(\ref{eq24}) 
for $n_K= (1/2) n_p \gg 1$). One can see that Eq.~(\ref{eq24}) works for very
stiff chains and not too large $n_K$, 
while for large $n_K$ systematic
deviations occur, which can be attributed  
to excluded volume effects. Both the simulation and the experiment include 
data for widely varying persistence lengths (in the experiment, this could 
only be achieved by combining data for chemically different polymers in 
this plot). From their results (see Fig.~1b) the experimentalists concluded that the excluded volume 
effects set in for $n_K>50$, irrespective of the precise value of the 
persistence length.

However, this latter conclusion needs to be questioned: in fact, for 
large $n_K$ the data do not superimpose in this representation for 
different choices of $\ell_p$, indicating that the behavior is more 
complicated. To elucidate this, we take out the leading power law in 
the Gaussian coil regime, plotting $\langle R_g^2\rangle/(\ell_pL)$ 
versus $n_K$ over the full range (Fig.~2a). One sees that nice scaling 
behavior occurs with respect to the crossover from rigid rods to 
Gaussian coils; in this regime, Eq.~(\ref{eq24}) works in $d=3$. 
However, now one can see rather clearly that the crossover from 
Gaussian coils to SAWs does not scale in this representation: 
rather for large $n_K$ the curves ``splay out'', the larger $\ell_p$ 
the longer the data follow Eq.~(\ref{eq24}), before an onset of excluded volume 
effects can be seen. This behavior has already been studied in Ref.~\cite{13} 
with respect to the end-to-end distance. Empirically, it was found that 
scaling $N$ with $N^*$, where $N^* \propto \ell_p^{2.5}$ rather 
than $N^*= \ell_p$. Fig.~2b shows that a master curve results as an 
envelope of the curves for individual $\ell_p$. We also recall, that 
Flory arguments predict $N^*\propto \ell_p^3$, cf. Eq.~(\ref{eq43}) 
and the subsequent discussion. The data in Fig.~2 are fully analogous 
to our data on the end-to-end distance that were discussed recently 
elsewhere~\cite{13}. In $d=2$, however, the behavior is clearly simpler 
(Fig.~3): there occurs a single crossover from rods to $d=2$ SAWs, and 
a regime where the Kratky-Porod worm-like chain model presents a faithful 
description of the data is absent. These results for $\langle R_g^2\rangle$ 
confirm our earlier analogous findings~\cite{13} for $\langle R_e^2\rangle$.

Fig.~4 shows some of our raw data for the structure factor $S(q)$. 
For small $q$, one recognizes the Guinier regime, 
$S(q) \approx \exp (-q^2\langle R_g^2\rangle/3) 
\approx 1-q^2\langle R_g^2\rangle/3$, 
and then a crossover occurs to the power law of SAWs or of Gaussian chains 
(the latter is seen clearly only for $d=3$ and very stiff chains). 
At large $q$ and stiff chains, the expected $q^{-1}$ behavior is in 
fact compatible with the data.

It turns out that an analysis of $S(q)$ in the form of Kratky
plots (Eq.~(\ref{eq35})) is more illuminating, cf. Fig.~5: The location of 
the maximum in the Kratky plot, as discussed in Eqs.~(\ref{eq36}), (\ref{eq37}), 
is easily identified, and it shows the expected scaling with $L/\ell_p$ both 
in $d=3$ and in $d=2$ (Fig.~6). In $d=3$, one notes that with increasing 
$L/\ell_p$ a crossover from Gaussian behavior to SAW behavior occurs. 
Our data also confirm that for rather stiff chains in $d=3$ both the 
Kholodenko and the Stepanow theories describe $S(q)$ very accurately.
Note that Fig.~\ref{fig5}b refers to rather short chains, for which strong
effects due to excluded volume interactions are not yet expected,
and hence the good agreement with the theories is not surprising.

Another issue of interest is the behavior $qLS(q)$ in the rigid rod limit, 
where one clearly notes the approach to $\pi$ (Fig.~5). Can we then use 
$S(q)$ in this region, applying the des Cloizeaux formula \{Eq.~(\ref{eq1})\} 
to extract quantitatively reliable estimates for the persistence length $\ell_p$ 
from a plot of $qLS(q)$ versus $1/q$ ? Fig.~7 suggests that although a 
regime occurs where the variation is linear in $(q\ell_p)^{-1}$, the coefficient of 
this linear variation is inconsistent with the des Cloizeaux result. We have no 
final answer to offer to explain this discrepancy; we suspect that in the regime 
where $q^{-1}$ and $\ell_p$ are of the same order, the discreteness of our 
lattice model (opposed to the Kratky-Porod continuum model) might matter.

\section{Results for the scattering function of chains under stretch}

While for unstretched chains it is only $\langle R_g^2 \rangle$ as a measure 
of the linear dimension of the whole chain which is relevant (Figs.~1-3), 
for chains under the action of stretching forces 
anisotropy of the conformation of the chain comes into play. However,
from the small $q$ expansion of 
Eqs.~(\ref{eq4}), (\ref{eq5}) one can show straightforwardly that 
$S_{||}(q_{||})$ yields information on $\langle R_{g,||}^2\rangle$ and 
$S_\bot(q_\bot)$ on $\langle R_{g,\bot}^2 \rangle$ cf. Eqs.~(\ref{eq14}) 
and (\ref{eq15}). Since for larger $q_{||}$ where also the extension 
$\langle X\rangle$ of the chain along the direction of the force enters 
the description of the scattering, we begin with a discussion of these 
linear dimensions and describe their variation as a function of the 
force $f$ (for a more detailed discussion and related results, we refer 
the reader to Ref.~\cite{13}).

\begin{figure*}[htb]
\begin{center}
(a)\includegraphics[scale=0.29,angle=270]{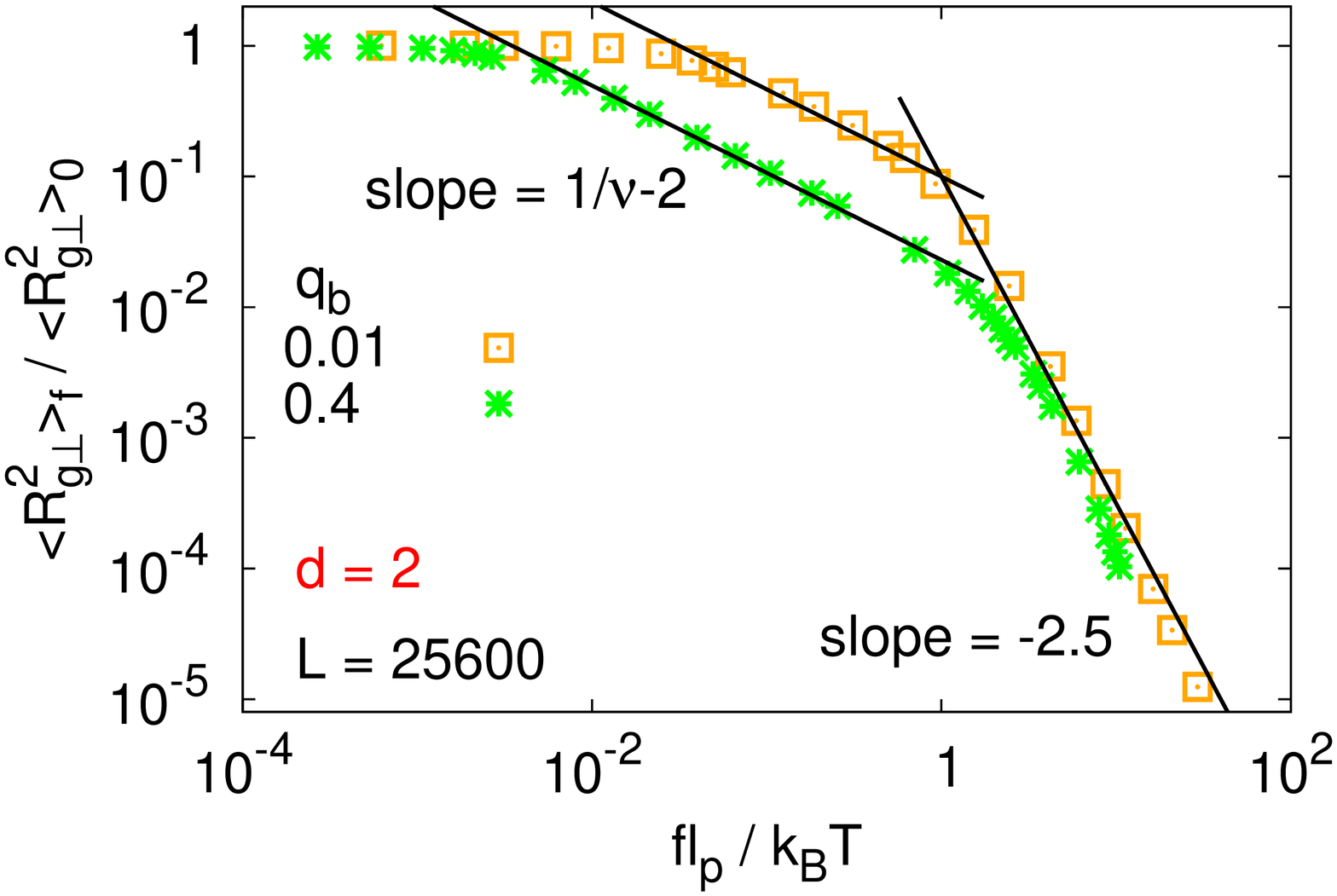}\hspace{0.4cm}
(b)\includegraphics[scale=0.29,angle=270]{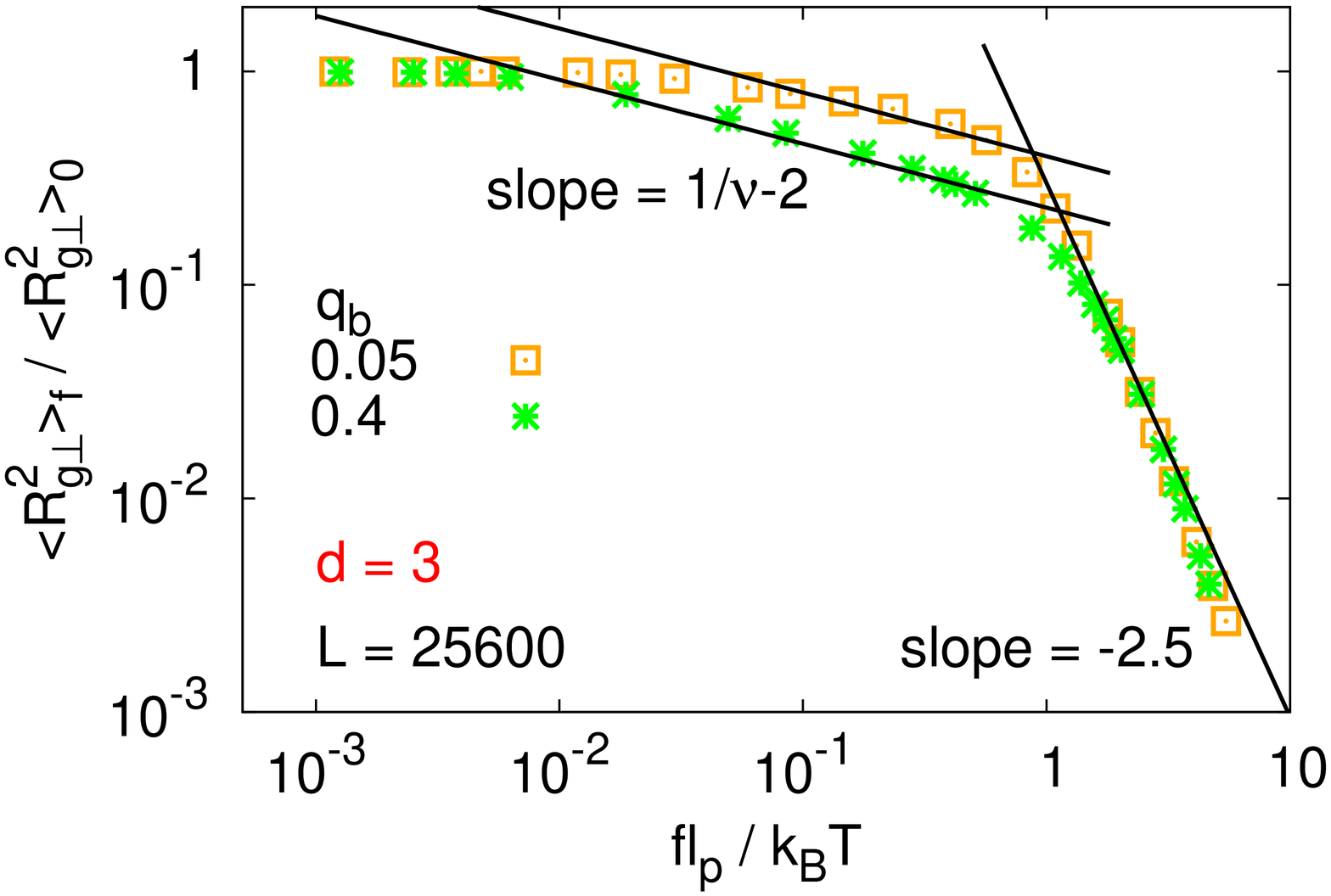}\\
(c)\includegraphics[scale=0.29,angle=270]{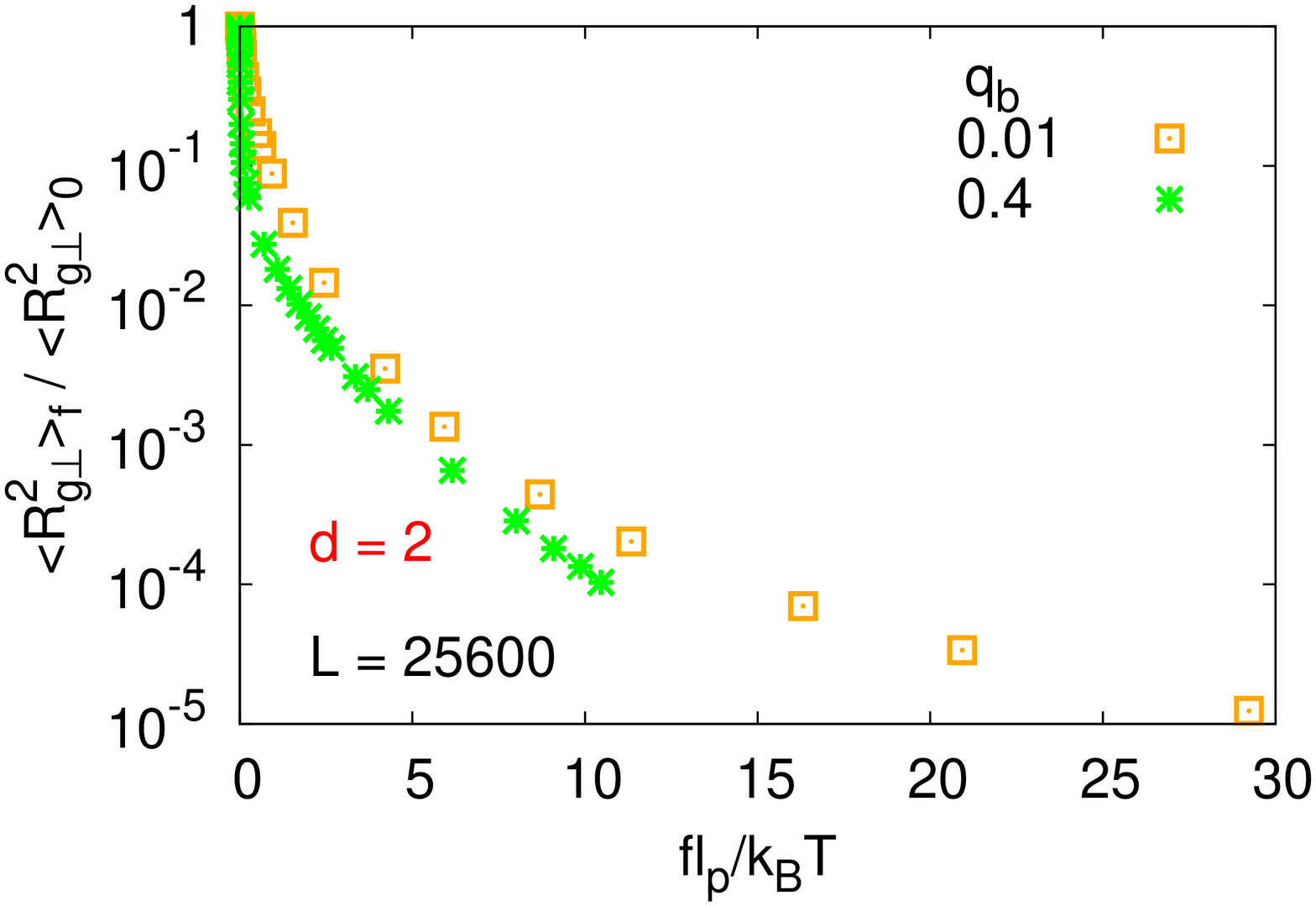}\hspace{0.4cm}
(d)\includegraphics[scale=0.29,angle=270]{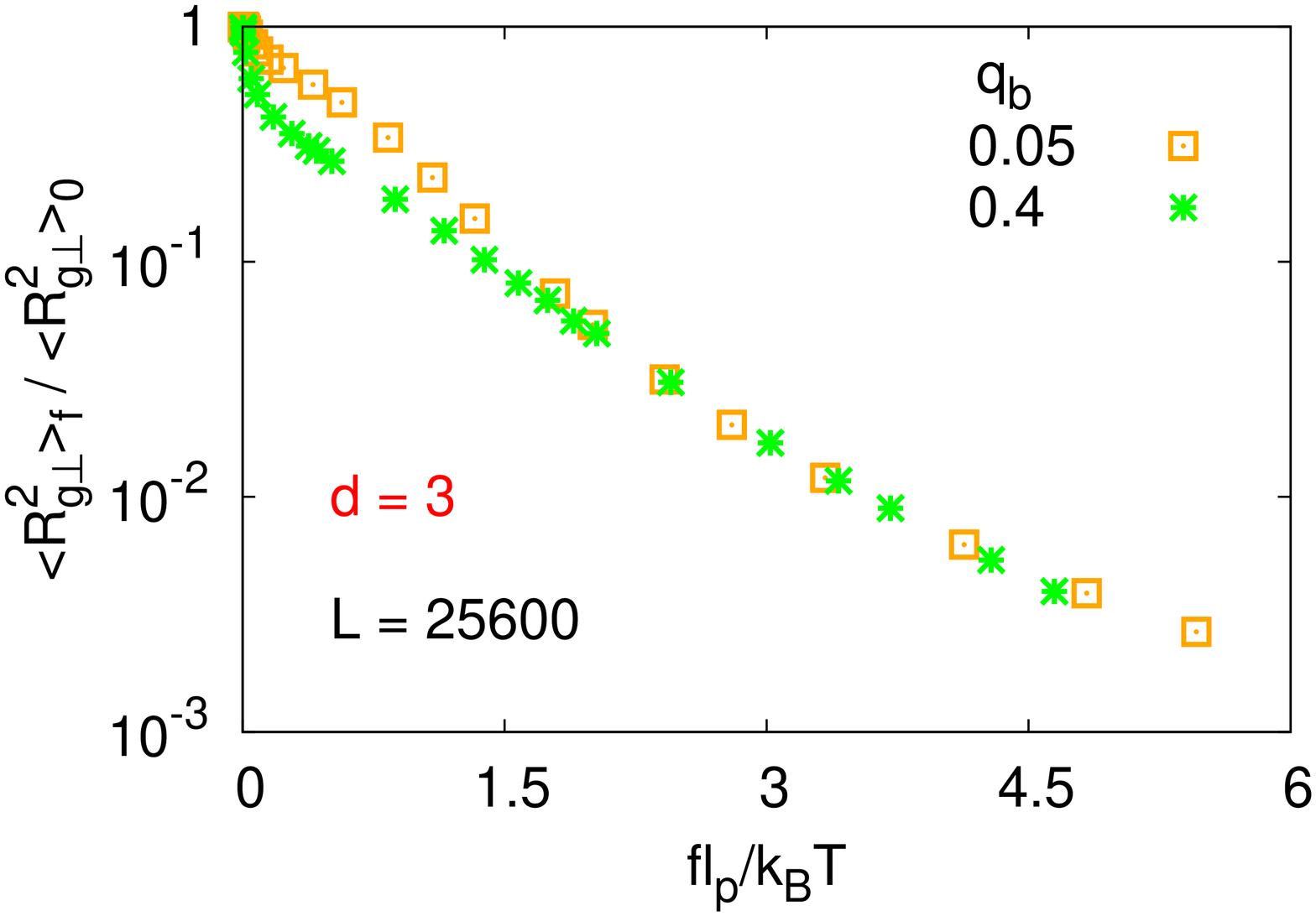}
\caption{\label{fig8} Log-log plot of
$\langle R_{g,\bot}^2\rangle /\langle R_{g, \bot}^2\rangle _0$
vs.~$f\ell_p/k_BT$, for $d=2$ (a) and for $d=3$ (b),
using $L=N\ell_b=25600$ in both cases, for two choices of $q_b$.
In parts (c,d) the same data as in (a,b) for large $f\ell_p/k_BT$ are
replotted choosing a linear scale for $f\ell_p/k_BT$, to show
that the behavior is
compatible with an exponential decay in $d=3$.}
\end{center}
\end{figure*}

Fig.~8 shows typical data of $\langle R_{g,\bot}^2 \rangle$ versus 
$f \ell_p/k_BT$, both for rather flexible chains $(q_b=0.4)$ and for 
rather stiff chains $(q_b = 0.01$ in $d=2$, $q_b=0.05$ in $d=3$, respectively). 
We recognize three regimes: for very small forces 
$\langle R_{g,\bot}^2\rangle \approx \langle R_{g,\bot}^2 \rangle_0$, 
the unperturbed value in the absence of forces. In this linear response regime, 
the force orients the coil without deforming it. Then we recognize a regime 
where $\langle R_{g,\bot}^2 \rangle$ decreases according to a power law,
namely Eq.~(\ref{eq56}). This power law holds in the regime where the radius
of the  Pincus blob, $\xi_p = k_BT/f$, is smaller than the unperturbed radius, but much 
larger than the persistence length $\ell_p$ itself. Thus this is the analog of 
the ``Pincus blob'' law that yields another power law for the extension 
vs.~force curve, Eq.~(\ref{eq54}). 
The physical picture invoked here for the chain 
is an elastic string of Pincus blobs, $\langle R_{g,\bot}^2\rangle$ describing 
the mean square displacement of this string in transverse directions. 
As one can see from Fig.~\ref{fig8}, the data indeed are compatible with 
the predicted power law in $d=2$, and in $d=3$ at least for the flexible chains. 
For stiff chains in $d=3$, the regime where Eq.~(\ref{eq56}) holds is more 
restricted, since the Kratky-Porod regime has a more extended regime of 
validity, effects due to Pincus blobs can only be detected in a regime
$\langle R_{g,\bot}^2 \rangle ^{1/2}_0 > \xi_p > R^* \propto \ell_p^2/D$, 
cf. Eq.~(\ref{eq55})~\cite{13}. 
Therefore we have not included very stiff chains in Fig.~\ref{fig8}b 
(for $q_b=0.005$, leading to $\ell_p \approx 52 \ell_b$, excluded volume 
effects, which also are responsible for the existence of Pincus blobs, 
could even for chains as long as $N= 25600$ hardly be detected in 
the chain linear dimensions in the absence of a force, cf. Figs.~1-3). 
So this failure to detect Pincus blobs for very stiff long chains in 
$d=3$ dimensions is hardly surprising (although Eq.~(\ref{eq56}) 
ultimatively will become valid as $N \rightarrow \infty$, 
irrespective how large $\ell_p$ is).

Another interesting behavior is the apparent power law, 
$\langle R_{g,\bot}^2\rangle \propto (f\ell_p/k_BT)^{-2.5}$, 
seen for $f\ell_p/k_BT >1$. However, we warn the reader to take this 
seriously: a closer look reveals a slight but systematic curvature, 
and a plot versus $f\ell_p/k_BT$ on a linear rather than a logarithmic scale reveals 
that this apparent power law is nothing but the onset of an exponential 
decay (Figs.~\ref{fig8}c,d): indeed, already from the partition function, 
Eq.~(\ref{eq57}) we recognize that for large forces the chains will be 
stretched out almost completely like rigid rods, and the few remaining 
kinks are suppressed exponentially when $f\ell_p/k_BT \gg 1$.

\begin{figure*}[htb]
\begin{center}
(a)\includegraphics[scale=0.29,angle=270]{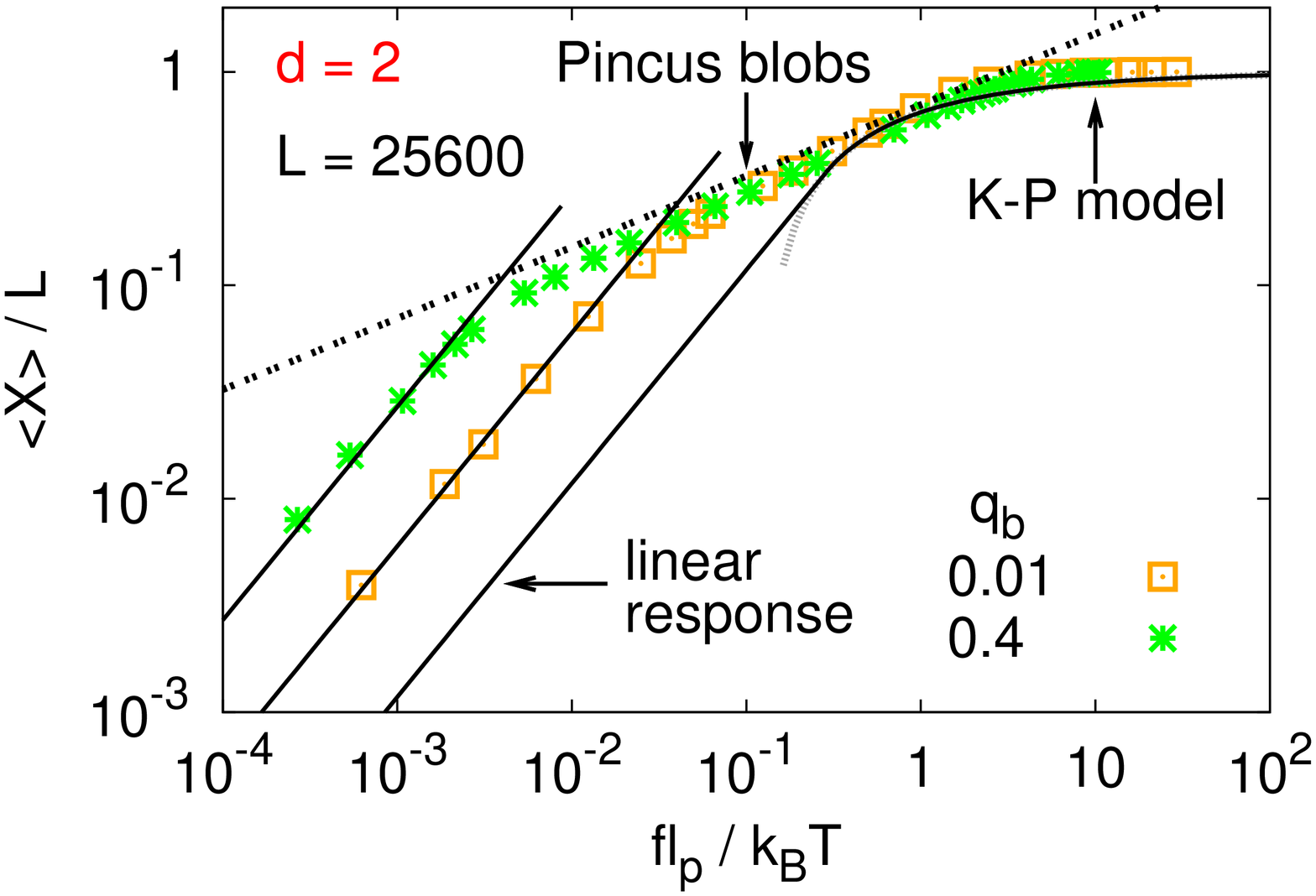}\hspace{0.4cm}
(b)\includegraphics[scale=0.29,angle=270]{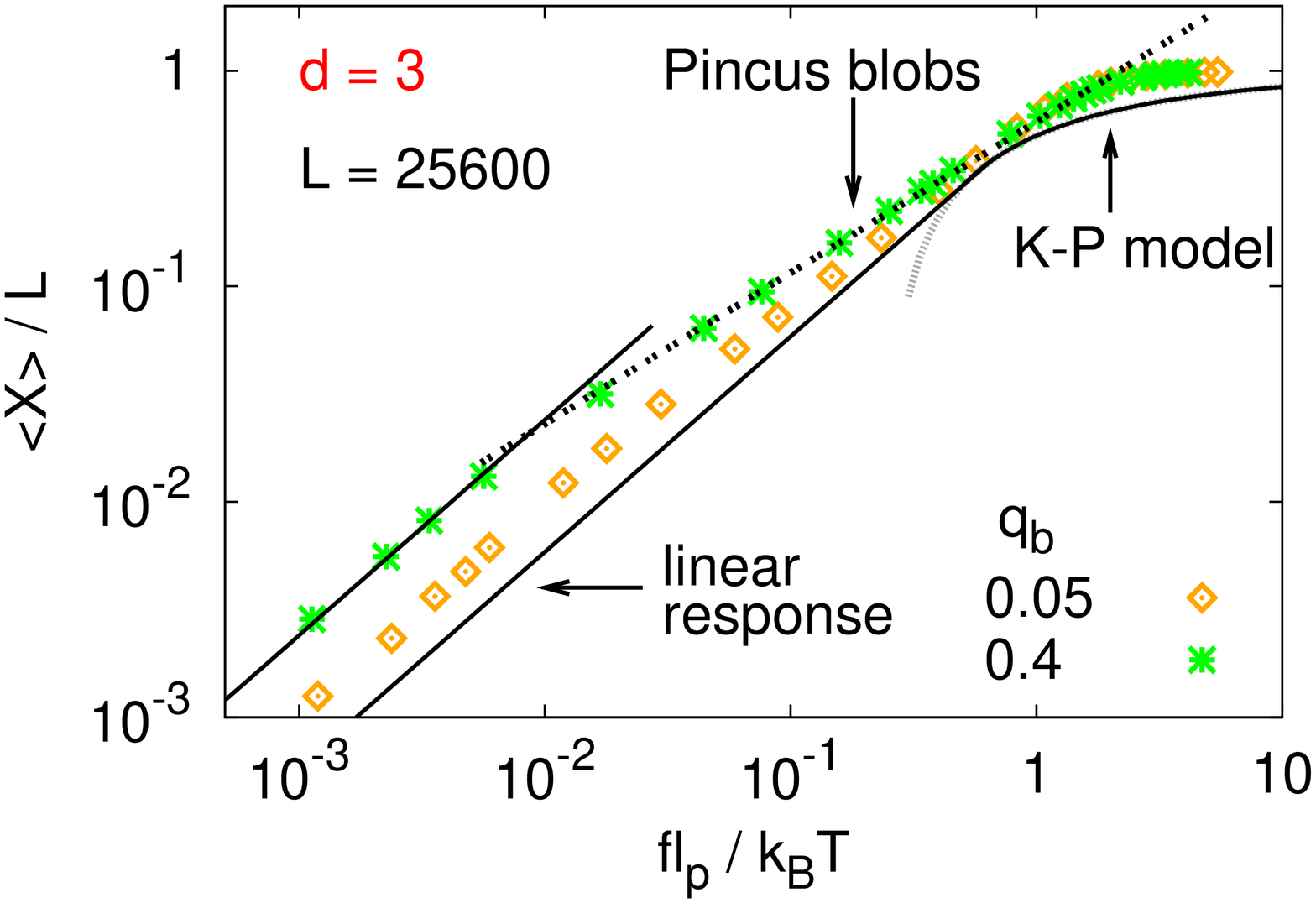}\\
\caption{\label{fig9}
Log-log plot of the relative extension $\langle X\rangle/L $ versus the
scaled force $f\ell_p/k_BT$ for $L=25600$ in $d=2$ (a) and $d=3$, for the
same choices of $q_b$ as shown in Fig.~\ref{fig8}. The straight lines
indicate the linear response regime and the Pincus-blob regime,
respectively \{Eq.~(\ref{eq54})\}. In neither case has the Kratky-Porod model
(curves labeled as K-P-model, cf.~Eqs.~(\ref{eq49}), (\ref{eq50})) a well-defined regime of
validity (this is expected since for large $f\ell_p/k_BT$ the approach to
saturation, $\langle X \rangle /L\rightarrow 1$, is exponentially fast
rather than proportional to $(k_BT/f\ell_p)^{1/2}$, cf. Fig.~\ref{fig8}).}
\end{center}
\end{figure*}

\begin{figure*}[htb]
\begin{center}
(a)\includegraphics[scale=0.29,angle=270]{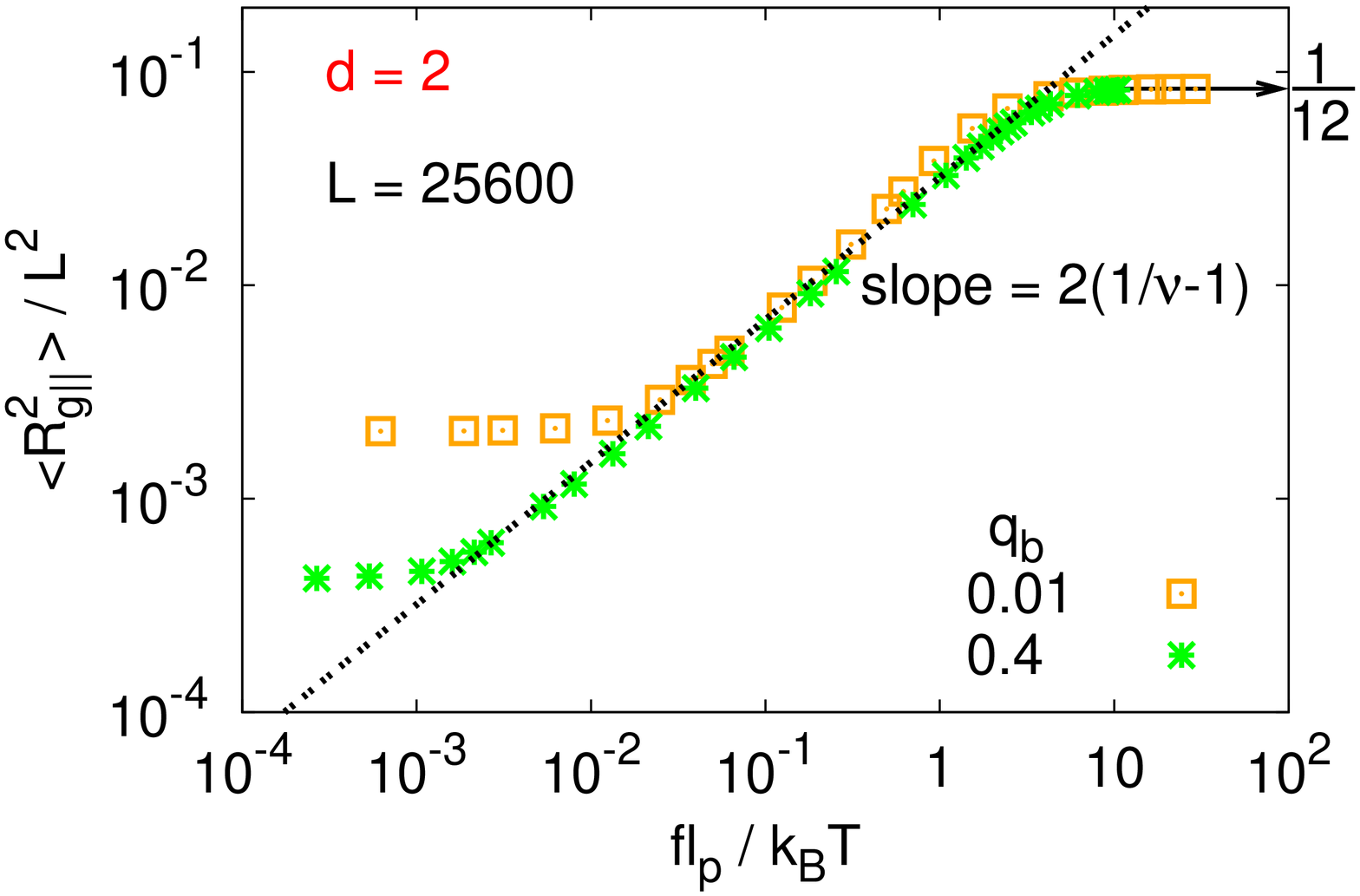}\hspace{0.4cm}
(b)\includegraphics[scale=0.29,angle=270]{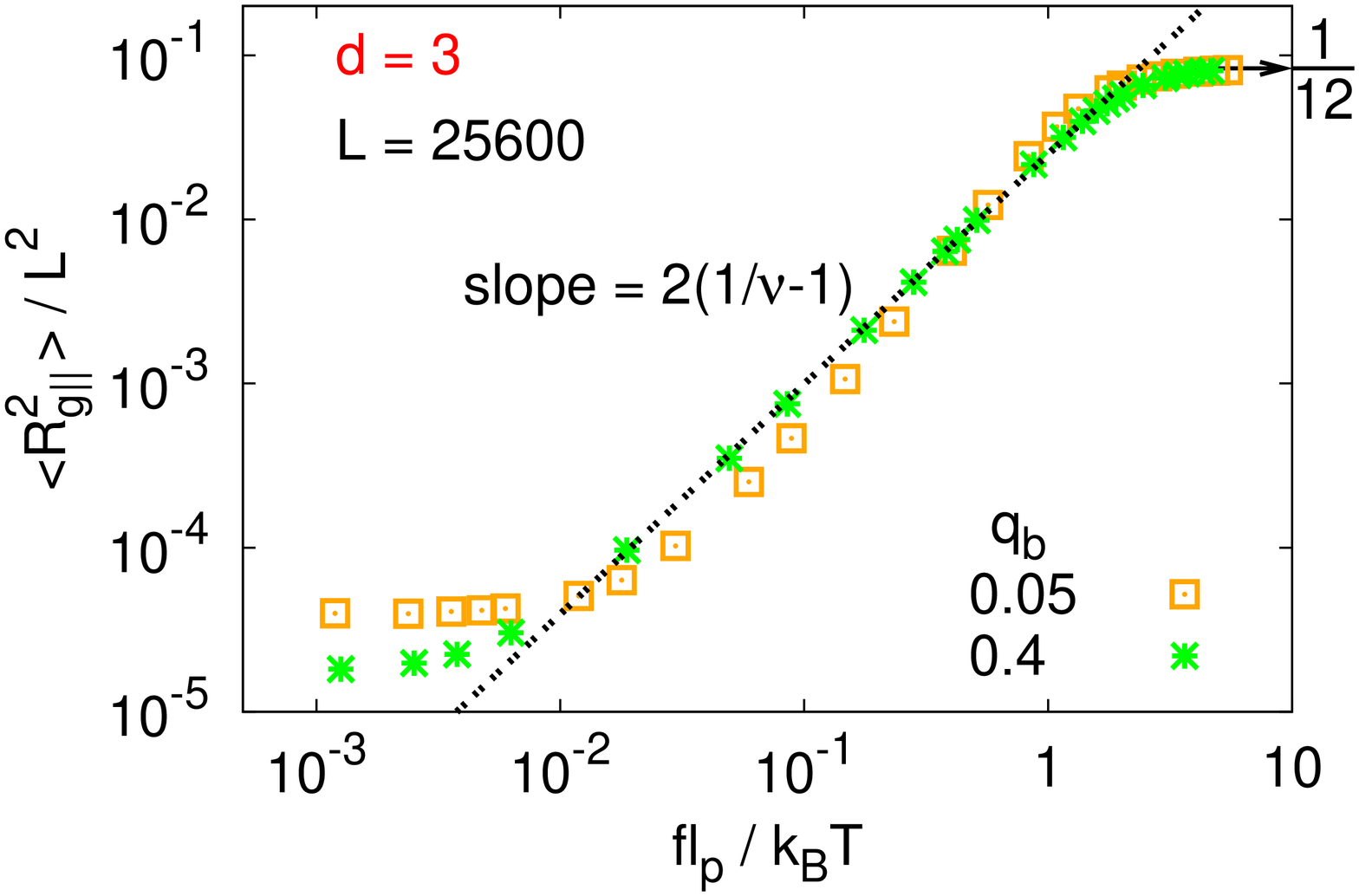}\\
\caption{\label{fig10}
Log-log plot of the parallel component $\langle R_{g,||}^2 \rangle /L^2$
vs.~$f\ell_p/k_BT$, for the same parameters as in Figs.~\ref{fig8}, \ref{fig9}.
Straight lines indicate the Pincus blob regime. Part (a) refers to $d=2$,
part (b) to $d=3$.}
\end{center}
\end{figure*}

For completeness, we show the corresponding simulation data for the 
relative extension $\langle X\rangle/L$ vs.~$f\ell_p/k_BT$ in Fig.~\ref{fig9} 
(related more extensive data for other values of $L$ and $q_b$ can be found 
in Ref.~\cite{13}), and in Fig.~\ref{fig10} we present the corresponding data 
for the longitudinal component $\langle R_{g,||} ^2\rangle/L^2$ of the 
gyration radius in the direction of the force. While for very small 
forces one expects a nonzero plateau (unlike $\langle X \rangle /L$, 
which vanishes as $f \rightarrow 0$), corresponding to the gyration 
radius square component of an unstretched chain, for large $f$ another 
plateau means that the chain has been stretched out fully to a rod of 
length $L$. In between these two plateaus, the Pincus blob behavior 
is seen rather clearly, for the flexible chains.

\begin{figure*}[htb]
\begin{center}
(a)\includegraphics[scale=0.29,angle=270]{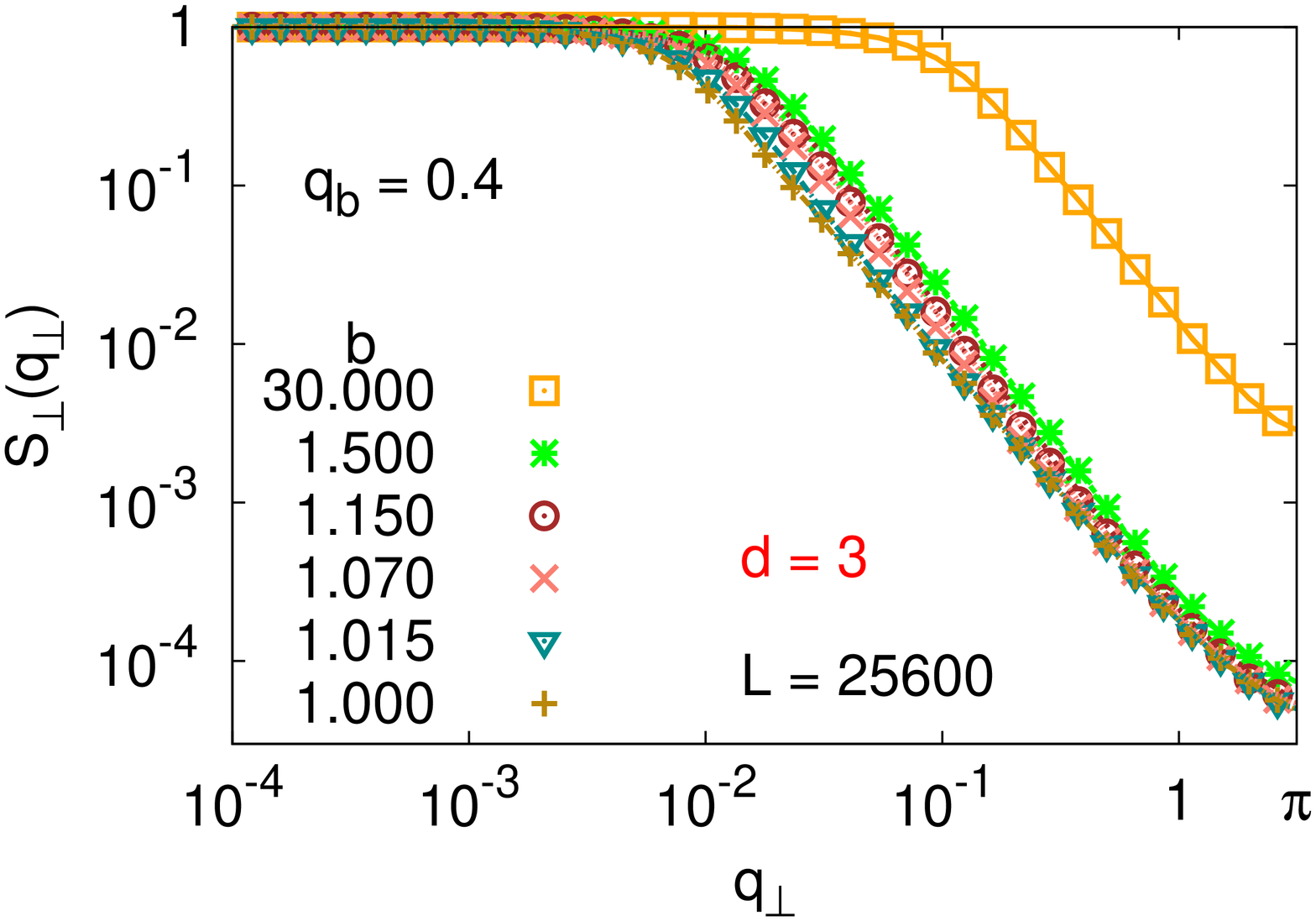}\hspace{0.4cm}
(b)\includegraphics[scale=0.29,angle=270]{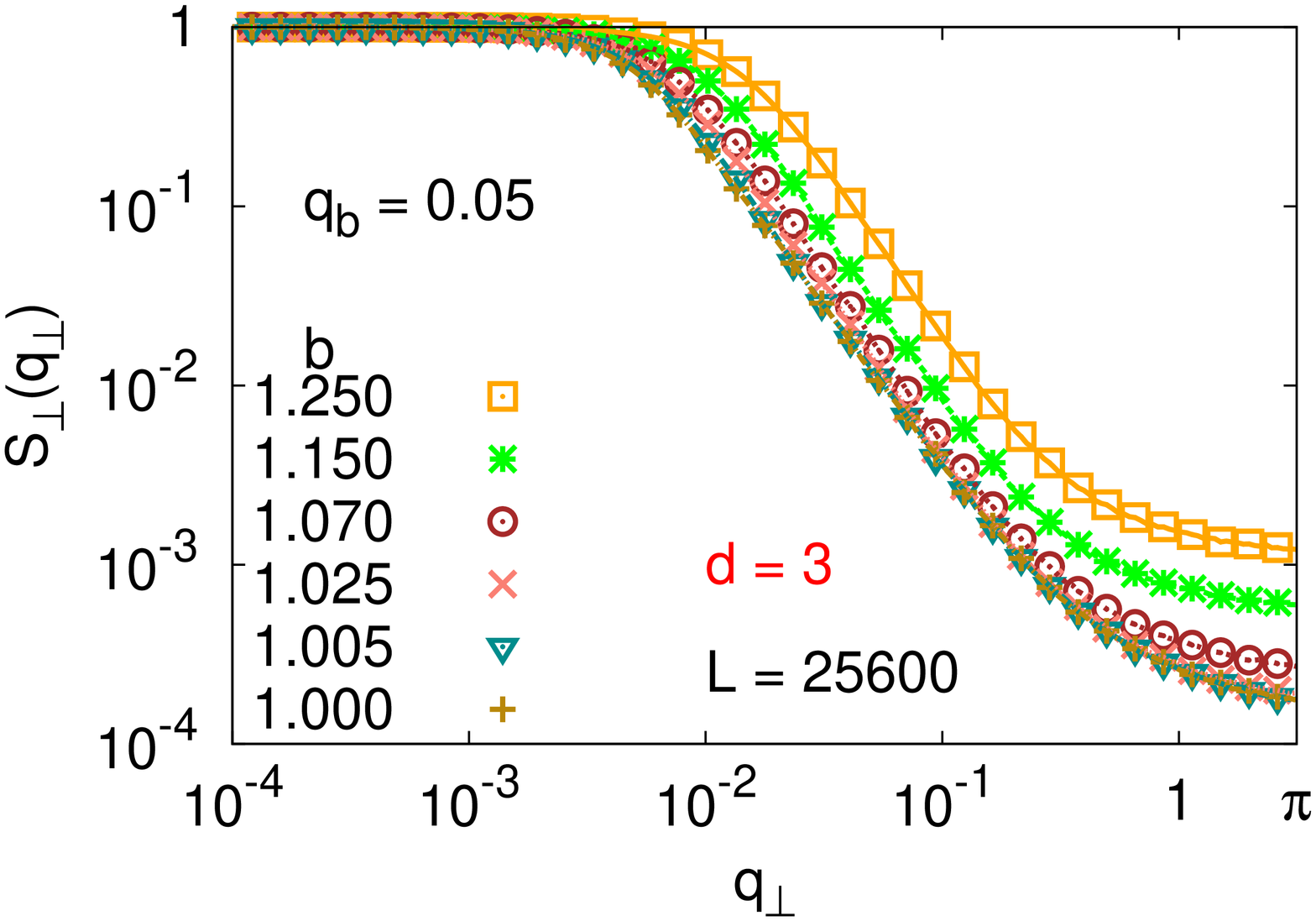}\\
(c)\includegraphics[scale=0.29,angle=270]{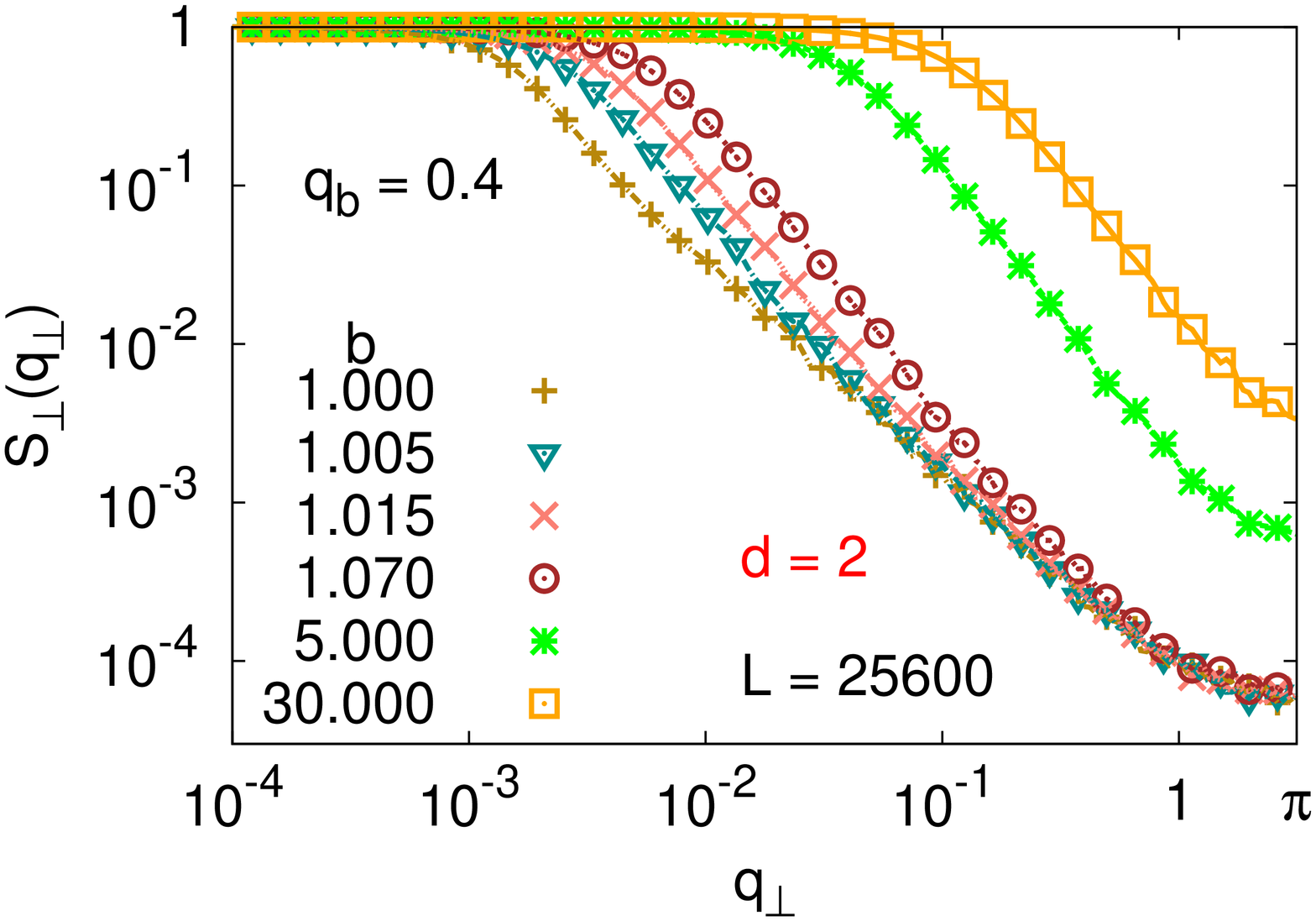}\hspace{0.4cm}
(d)\includegraphics[scale=0.29,angle=270]{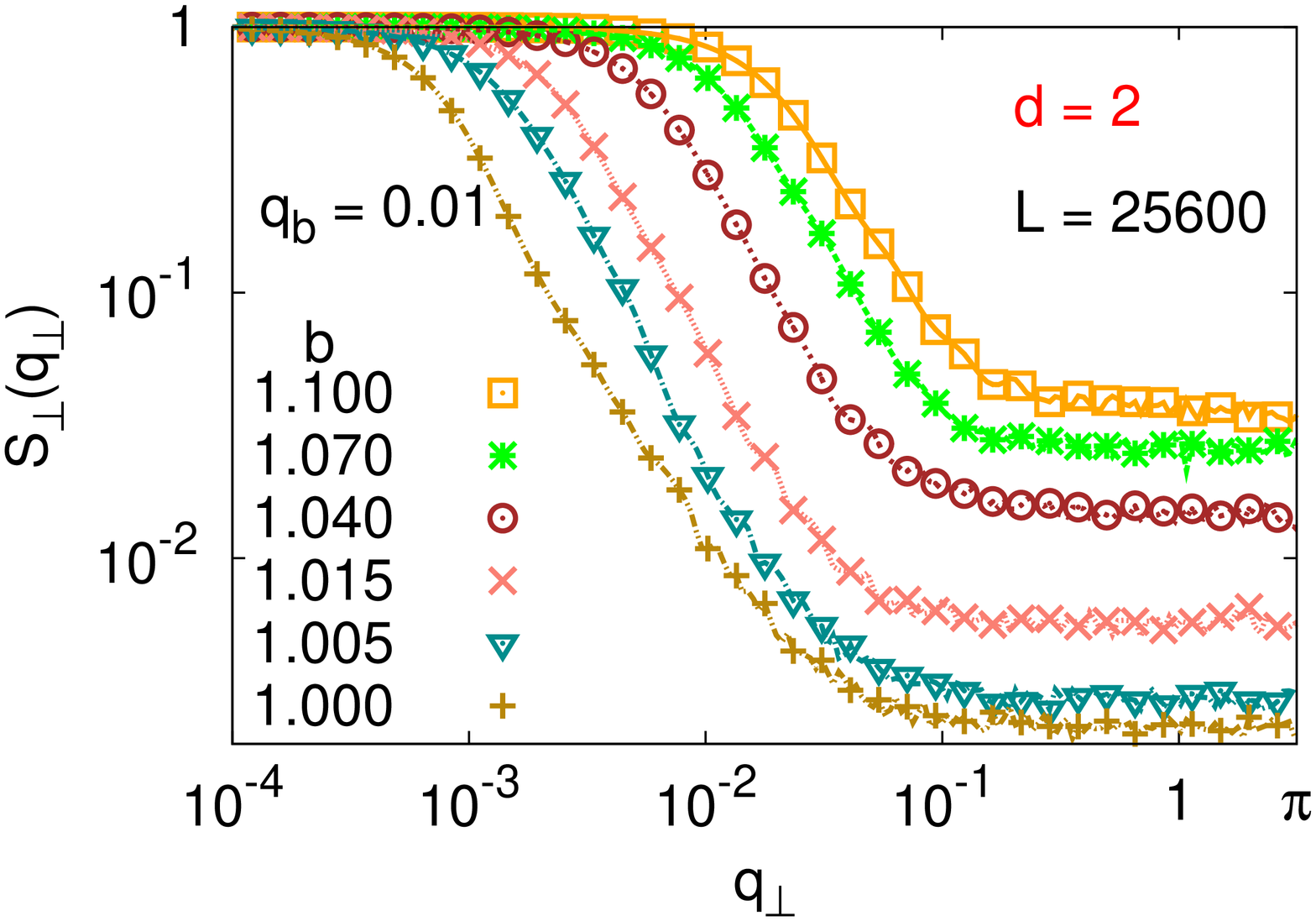}\\
\caption{\label{fig11} Log-log plot of the structure factor $S_\bot(q_\bot)$
vs.~$q_\bot$ for $d=3$ and two choices of the stiffness parameter $q_b$,
$q_b = 0.4$ (a) and $q_b=0.05$ (b), and also for $d=2$ and two choices
of $q_b$, namely $q_b=0.4$ (c) and $q_b=0.01$. In each case several
choices of $b = \exp (f/k_BT)$ are included, as indicated. Chain length
is $L = 25600$ throughout.}
\end{center}
\end{figure*}

\begin{figure*}[htb]
\begin{center}
(a)\includegraphics[scale=0.29,angle=270]{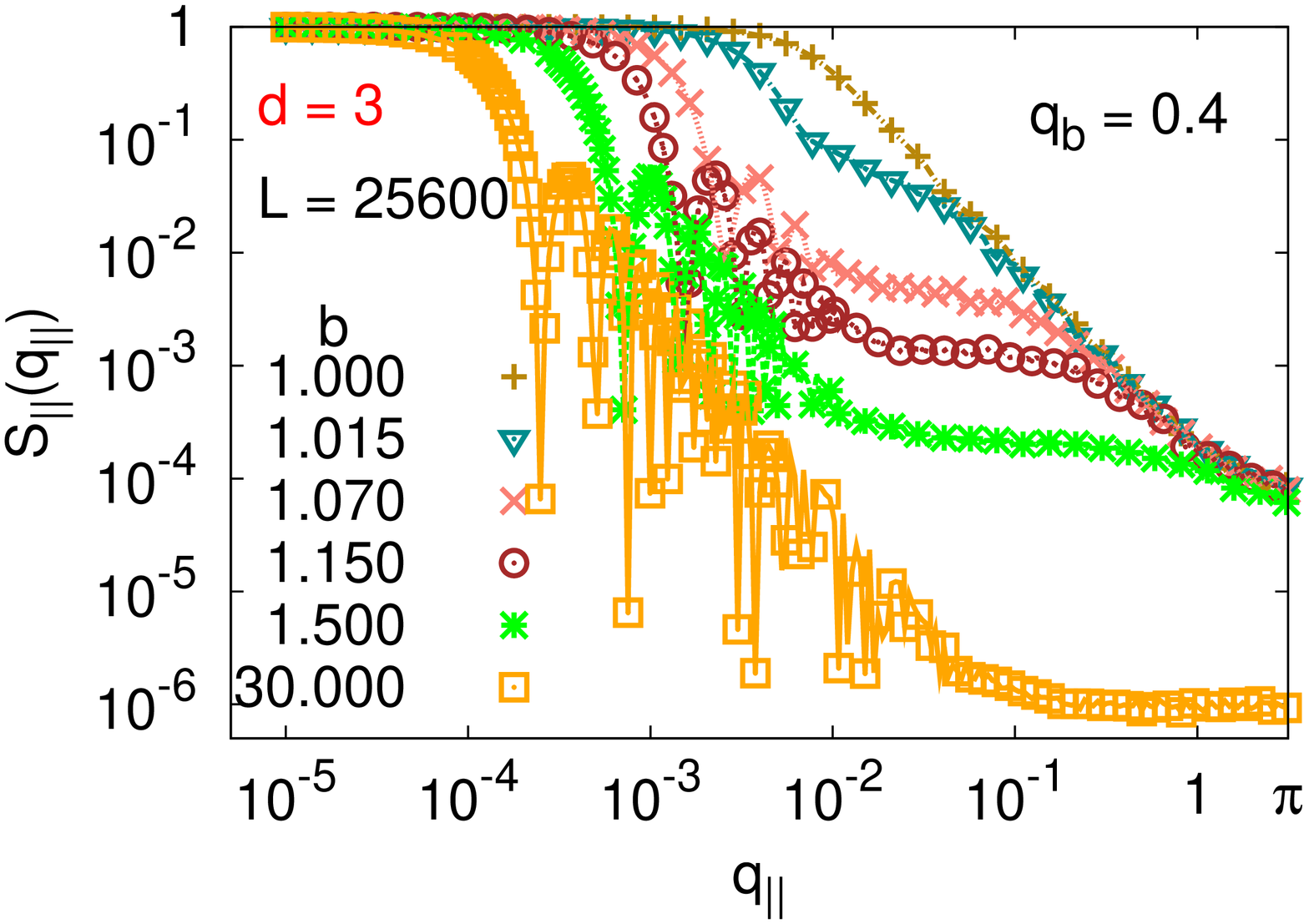}\hspace{0.4cm}
(b)\includegraphics[scale=0.29,angle=270]{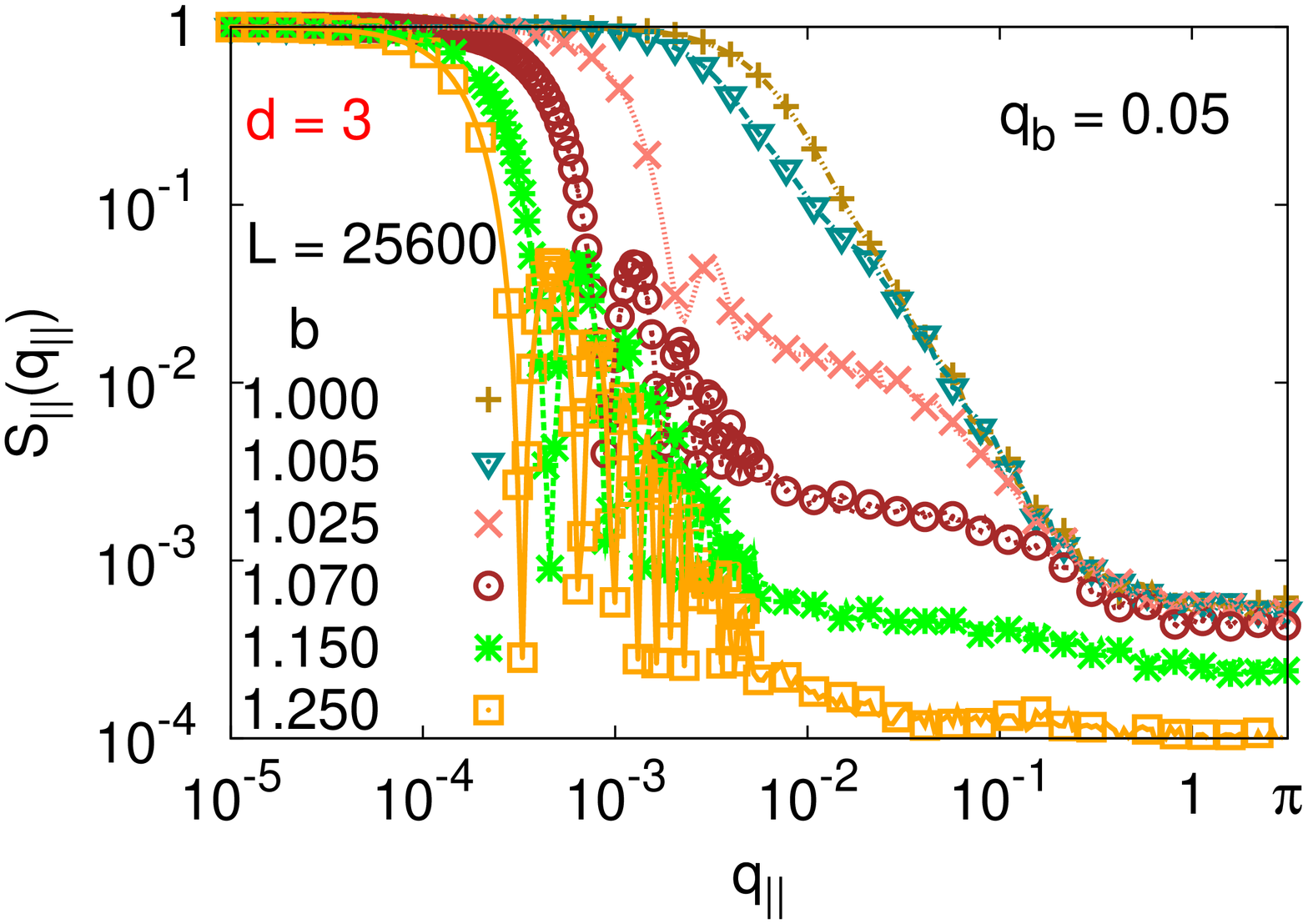}\\
(c)\includegraphics[scale=0.29,angle=270]{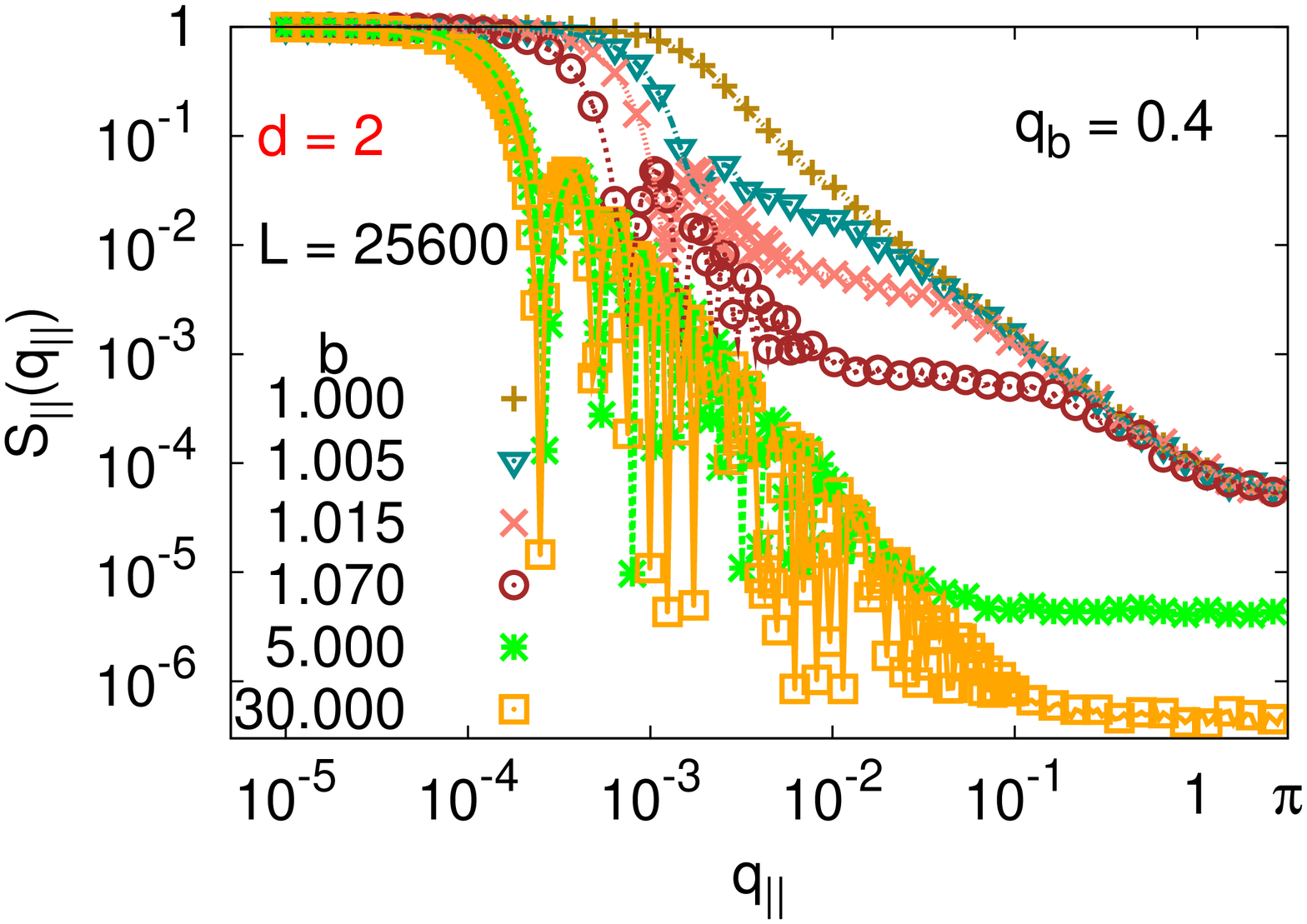}\hspace{0.4cm}
(d)\includegraphics[scale=0.29,angle=270]{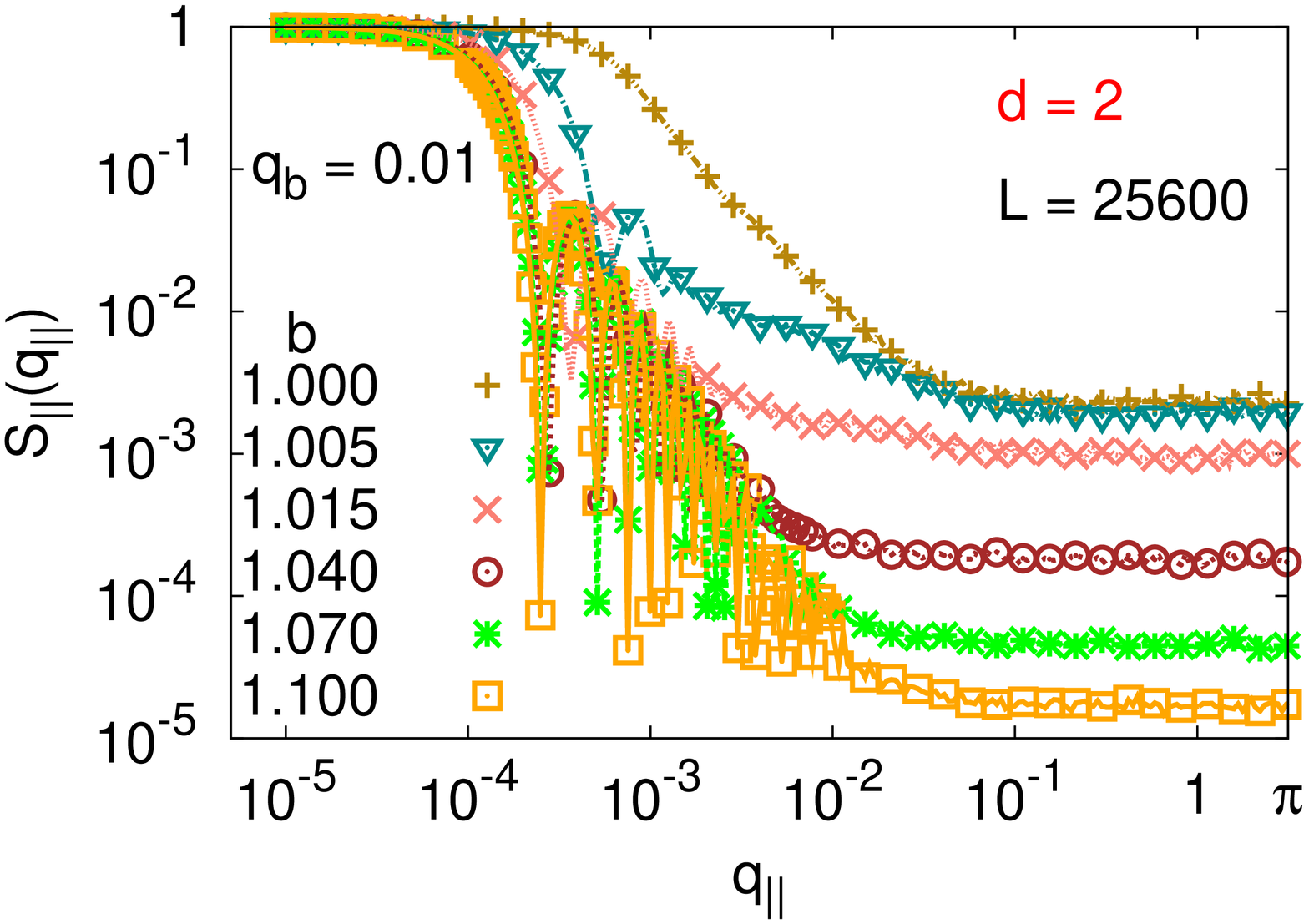}\\
\caption{\label{fig12}
Log-log plot of the structure factor $S_{||}(q_{||})$ vs.~$q_{||}$ for $d=3$ and 
two choices of the stiffness parameter $q_b$, $q_b=0.4$ (a)
and $q_b = 0.05$ (b), and also for $d=2$ and two choices of $q_b$,
namely $q_b=0.4$ (c) and $q_b=0.01$ (d). In each case several choices
of $b = \exp (f/k_BT)$ are included. Chain length is $L=25600$ throughout.}
\end{center}
\end{figure*}

Fig.~11 now shows typical data for $S_\bot(q_\bot)$ vs.~$q_\bot$ and
Fig.~12 the corresponding data for $S_{||}(q_{||})$ vs.~$q_{||}$, 
focusing again on those selected values of $q_b$ that were used already 
in Figs.~8-10. As expected from Eq.~(\ref{eq47}), the perpendicular 
structure factor is similar to the case without stretching force; the 
plateau at small $q$, where $S_\bot(q_\bot)$ deviates only very little 
from unity, gets more extended with increasing $f$, reflecting the 
decrease of $\langle R_{g, \bot}^2\rangle $ with $f$ (Fig.~8). This 
decrease, of course, is more pronounced for stiff chains than for 
flexible chains at the same value of $f$, since we have seen 
(see Fig.~8 and the discussions in Ref.~\cite{13}) that the proper 
control variable is not $f/k_BT$ but rather $f\ell_p/k_BT$. 
A further remarkable feature is the fact, that the power law-like 
decay of $S_\bot(q_\bot)$ with $q_\bot$, that for flexible chains can 
be observed until $S_\bot(q_\bot)$ has decayed up to 
$S_\bot(q_\bot)\approx 10^{-4}$ as $b<1.5$, 
for stiff chains extends only to $S_\bot(q_\bot)\approx 10^{-3}$ in $d=3$ and 
$S_\bot(q_\bot)\approx 10^{-2}$ in $d=2$, respectively. As expected, the theory 
of Benoit et al.~\cite{30} which extended the description of scattering 
from Gaussian chains to elastic stretching deformations, can only be applied 
if $q_\bot \ell_p\ll 1$, and when $\langle R_{g,\bot}^2\rangle$ exceeds 
$\ell_p^2$ only by few orders of magnitudes, the applicability of 
Eq.~(\ref{eq47}) is correspondingly restricted. In fact, noting that, 
for $q_b=0.05$, $\ell_p/\ell_b \approx 5.9$, we conclude that $b=1.5$ means 
$f \ell_p/k_BT \approx 2.4$, and Fig.~\ref{fig8} shows that in this case 
indeed $\langle R_{g, \bot}^2\rangle$ is about an order of magnitude smaller 
than for $f=0$. Assuming that one can represent a stiff chain as a sequence 
of rods of length $\ell_p$ such that $n_p\ell_p=N\ell_b$, and stating 
in the spirit of Eq.~(\ref{eq16}) that at large $q_\bot$ interference
effects of different rods can be neglected, one would expect that 
for $q_\bot\ell_p \approx 1$ 
one obtains a scattering of the order of $S_\bot(q_\bot) \approx n_p^{-1}$, 
independent of $q_\bot$. This (admittedly rough) argument would qualitatively 
explain the systematic increase of the plateau $S(q)$ in Fig.~5 
and $S_\bot(q_\bot)$ 
in Fig.~11 with increasing chain stiffness.

Even more interesting is the behavior of $S_{||}(q_{||})$, Fig.~12. The rapid 
increase of $\langle R_{g,||}^2 \rangle$ with increasing stretching force has 
the consequence that $S_{||}(q_{||})$ deviates from unity for smaller and 
smaller $q_{||}$. While for small $f$ just a shoulder develops, before 
(at large $q_{||}$) the behavior is similar to that of $S_\bot(q_\bot)$, 
for large $f$ pronounced oscillations develop. As pointed out already by 
Pierleoni et al.~\cite{34} for the case of fully flexible chains under 
stretch, this behavior can be attributed to the fact that the chain behaves 
like an elastically stretched string. We shall discuss this behavior in more 
detail below. 
Here we only note that the maxima of these oscillations decay according 
to a power law, which is similar to the power law of $S_\bot(q_\bot)$ in the 
intermediate range of $q_\bot$. The minima of $S_{||}(q_{||})$, as well as 
$S_{||}(q_{||})$ itself at larger values of $q_{||}$ where the oscillations 
of $q_{||}$ have decayed, show a slow further decrease with $q_{||}$. 
While for very large $q_{||}=q_\bot$ but not very strong stretching ($b<1.5$)
we have $S_{||}(q_{||})=S_\bot (q_\bot)$, 
if the chains are flexible ($q_b=1,\, q_b=0.4)$, 
this is not the case for 
stiff chains:
$S_{||}(q_{||}) \ll S_{\bot} (q_{\bot})$ for $q_{||}=q_\bot$ then.

\begin{figure*}[htb]
\begin{center}
(a)\includegraphics[scale=0.29,angle=270]{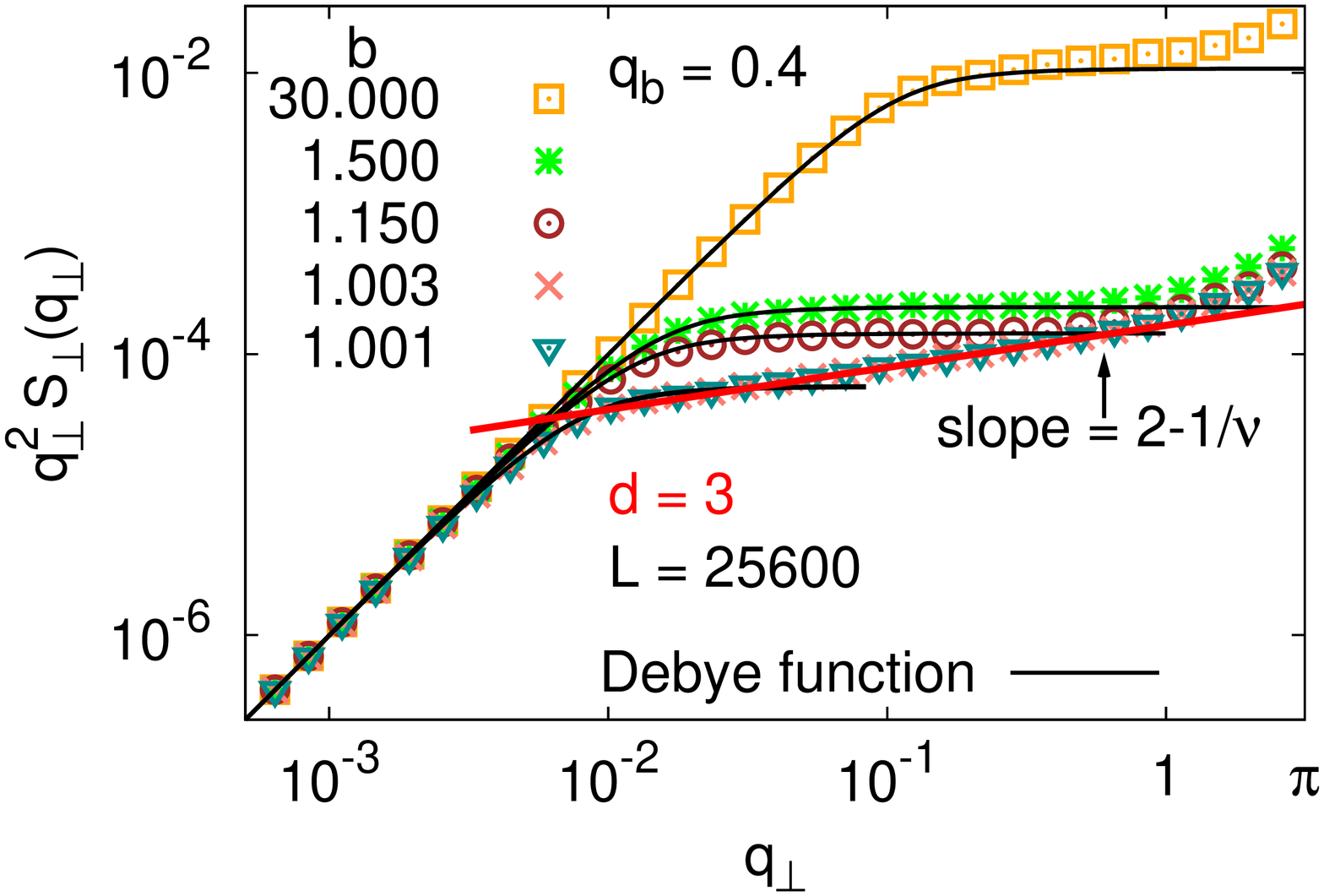}\hspace{0.4cm}
(b)\includegraphics[scale=0.29,angle=270]{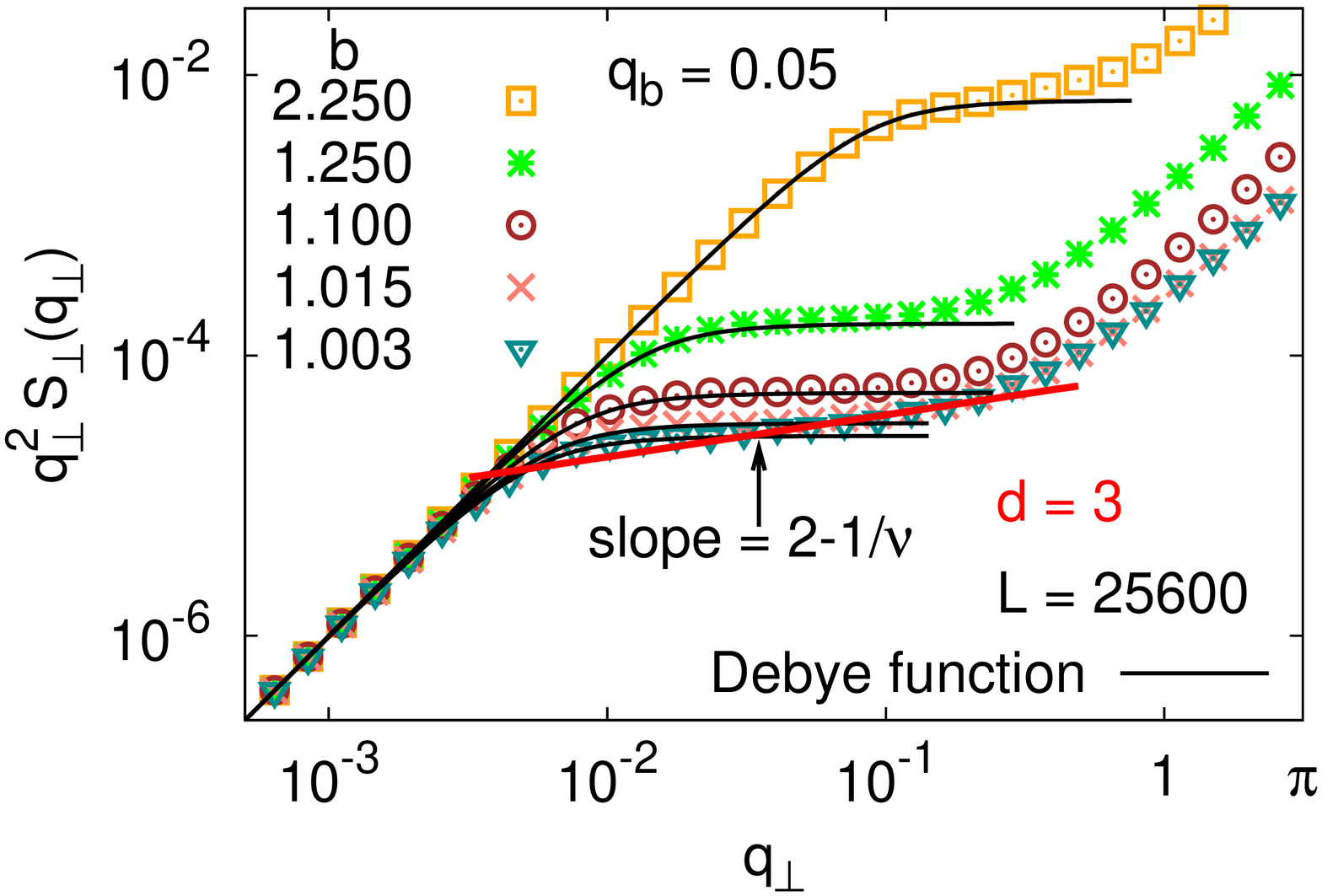}\\
(c)\includegraphics[scale=0.29,angle=270]{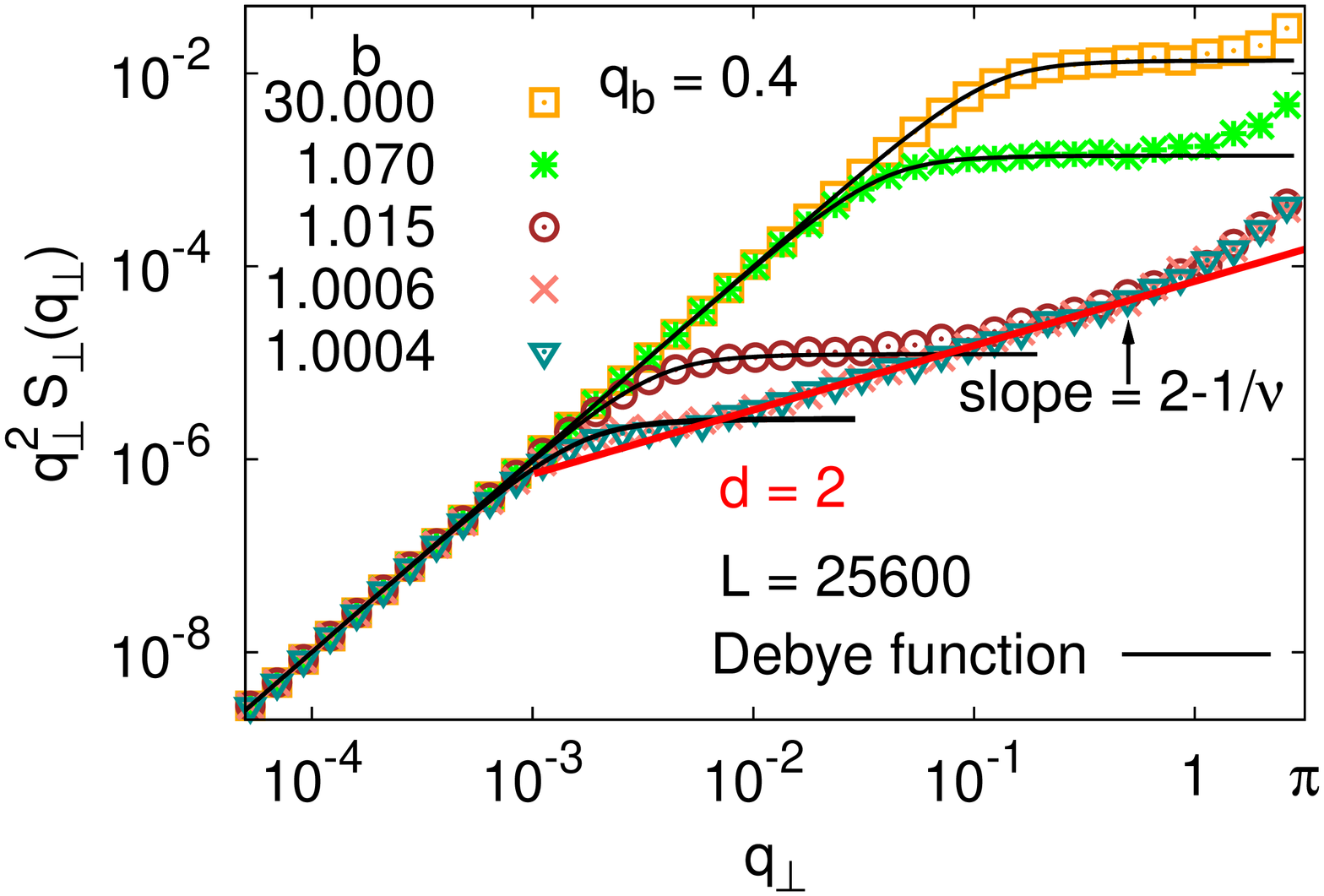}\hspace{0.4cm}
(d)\includegraphics[scale=0.29,angle=270]{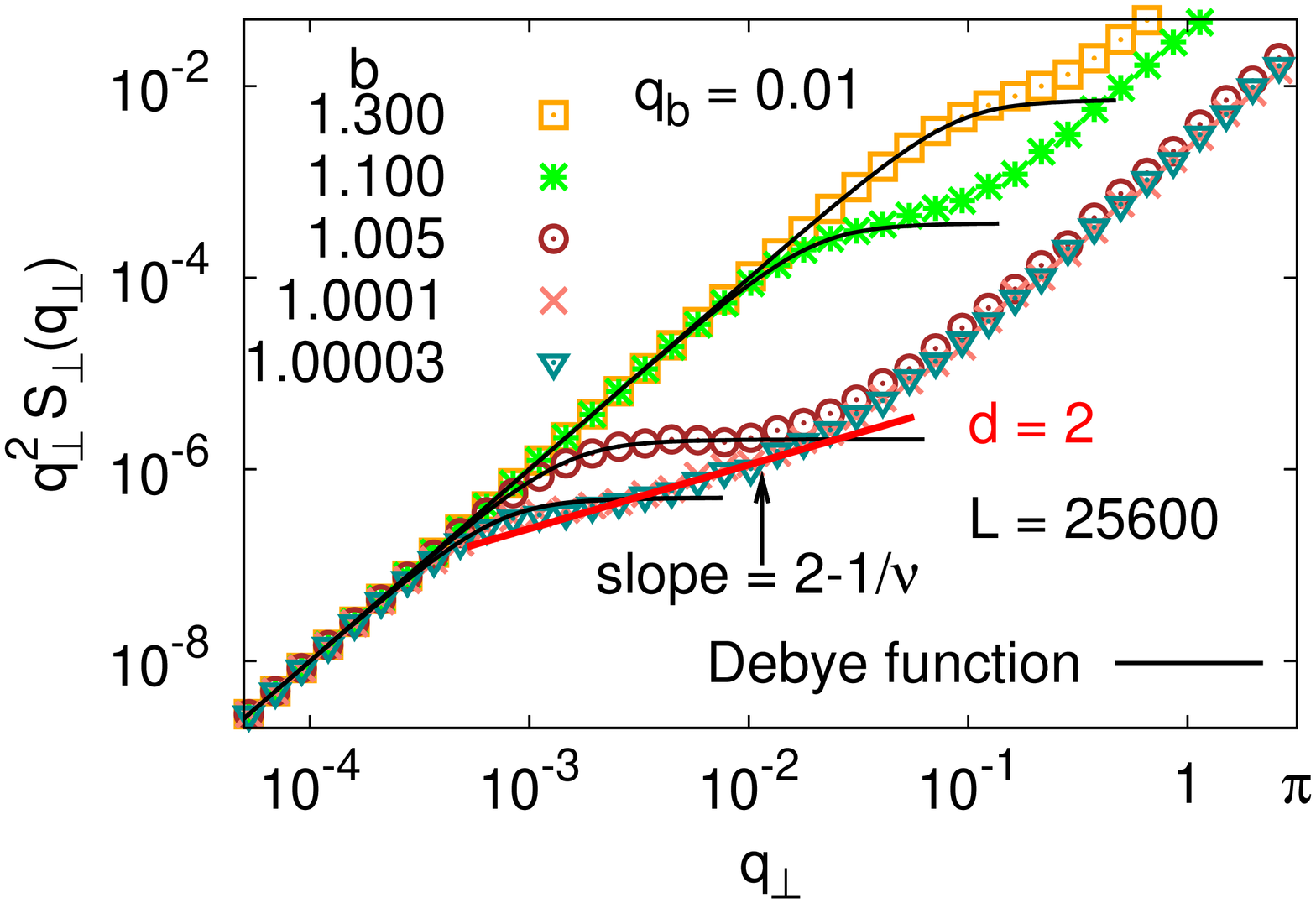}\\
\caption{\label{fig13}
Log-log plot of $q^2_\bot S_\bot(q_\bot)$ vs.~$q_\bot$ for $d=3$ and
two choices of the stiffness parameter $q_b, q_b = 0.4$ (a) and
$q_b =0.05$ (b), and also for $d=2$ and two choices of $q_b$,
namely $q_b=0.4$ (c) and $q_b=0.01$ (d).
The curves are the Debye function, Eq.~(\ref{eq47}) with
$X_\bot=\frac{3}{2}q_\bot^2 \langle R_{g,\bot}^2 \rangle$ in $d=3$,
and $X_\bot=3q_\bot^2 \langle R_{g,\bot}^2 \rangle$ in $d=2$.
In each case several choices
of $b = \exp(f/k_BT)$ are included. Chain length is $L = 25600$ throughout.}
\end{center}
\end{figure*}

In order to understand these results more quantitatively, we first 
report $S_{\bot}(q_\bot)$ in form $q^2_\bot S_\bot (q_\bot)$ versus 
$q_\bot$ and compare to the Debye function, Eq.~(\ref{eq47}), but using
the value $\langle R^2_{g,\bot}\rangle$ as observed in the simulation (rather than 
any theoretical prediction for it). Fig.~\ref{fig13} shows that the Debye 
function works surprisingly well: for $d=3$ the 
slope $2-1/\nu$ indicating non-Gaussian behavior is seen for $q_b=0.4$ only 
for weak stretching ($b=1.001$ and $1.003$), while for larger stretching forces 
a horizontal part in the plot
$q^2_\bot S_\bot (q)$ has developed. Also in $d=2$ the excluded volume 
regime, where the slope $2-1/\nu$ is compatible with the data, is pronounced 
only for rather flexible chains (such as
$q_b=0.4$, Fig.~\ref{fig13}c) while for stiff chains in $d=2$ (such as $q_b=0.01$) 
excluded volume effects show up in $S_\bot(q_\bot)$ only for extremely weak 
stretching (such as $b=1.0001$, i.e. $f/k_BT=10^{-4})$. for stronger stretching 
of semiflexible chains in $d=2$ the Debye function seems to describe the data 
for small $q_\bot$ $(q_\bot \leq 10^{-2})$, but then a crossover to a 
behavior $S_\bot(q_\bot)\approx const$ and hence 
$q_\bot^2S_\bot(q_\bot)\propto q_\bot^2$ sets in.

\begin{figure*}[htb]
\begin{center}
(a)\includegraphics[scale=0.29,angle=270]{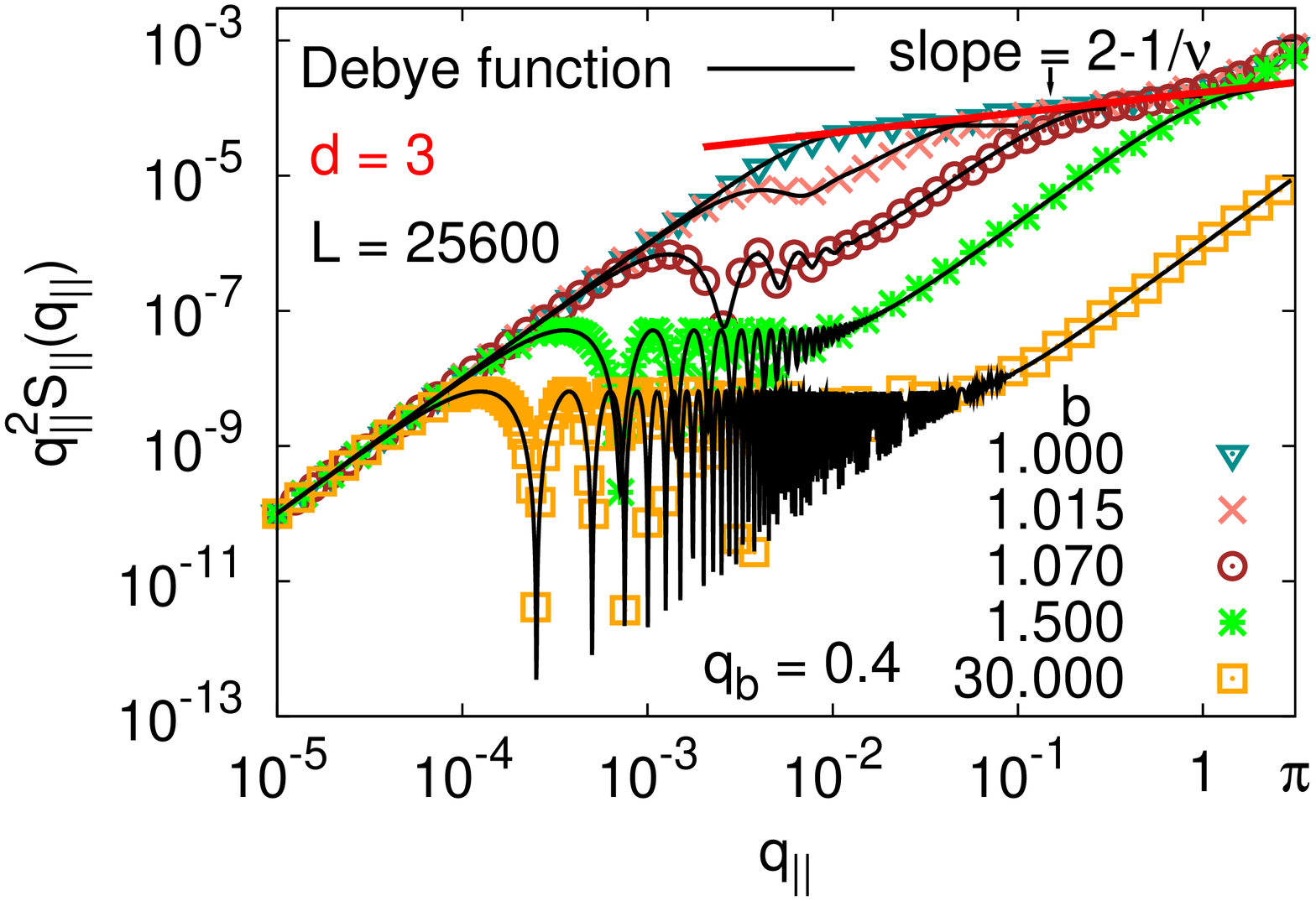}\hspace{0.4cm}
(b)\includegraphics[scale=0.29,angle=270]{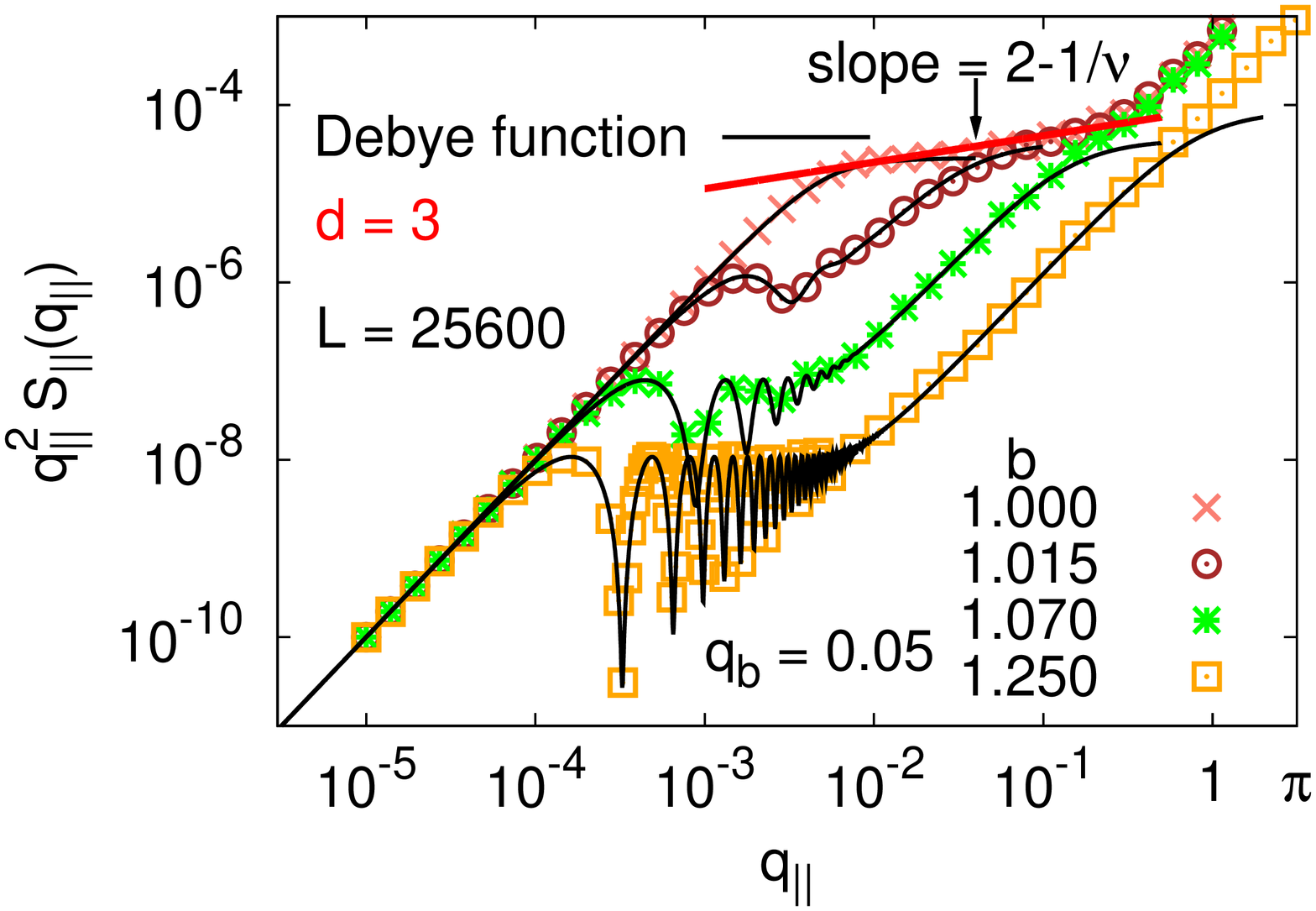}\\
(c)\includegraphics[scale=0.29,angle=270]{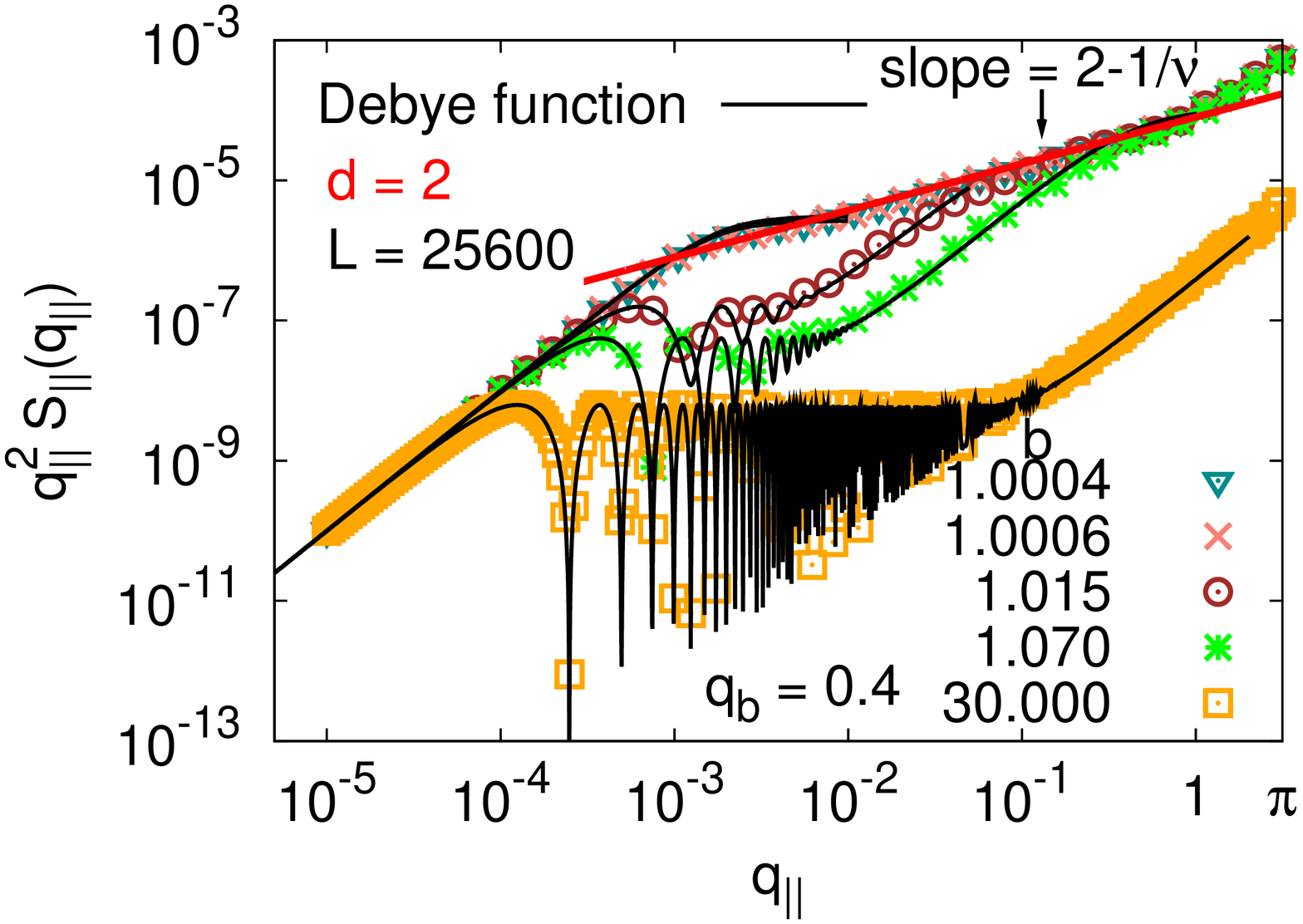}\hspace{0.4cm}
(d)\includegraphics[scale=0.29,angle=270]{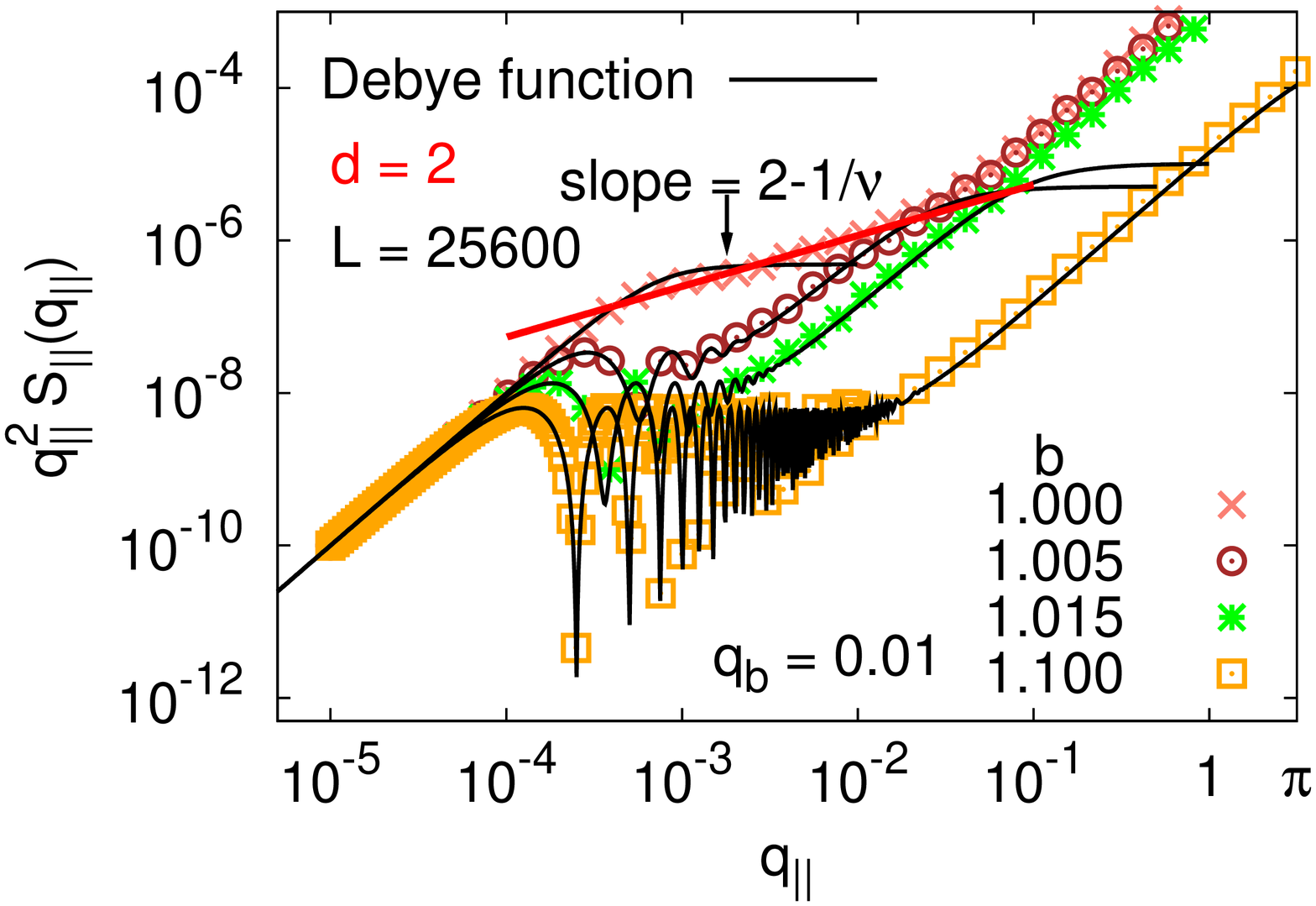}\\
\caption{\label{fig14}
Log-log plot of $q_{||}^2 S_{||}(q_{||})$ vs.~$q_{||}$ for $d=3$ and
two choices of the stiffness parameter $q_b$, namely $q_b = 0.4 $ (a)
and $q_B=0.05$ (b), and also for $d=2$ and two choices of $q_b=0.4$ (c)
and $q_b = 0.01 $ (d).
The curves are the
Debye function, Eq.~(\ref{eq58}) with complex
$X_{||}=q_{||}^2(\langle X^2 \rangle - \langle X \rangle^2)/2+iq_{||}<X>$,
and the straight line shows the excluded volume power law
(slope $=2 - 1/\nu$).
In each case several choices of $b= \exp(f/k_BT)$
are included. Chain length is $L = 25600$ throughout.}
\end{center}
\end{figure*}

The behavior of $S_{||}(q_{||})$ when plotted in the form 
$q^2_{||}S_{||}(q)$ vs.~$q_{||}$ is particularly striking (Fig.~14). 
Again excluded volume effects (described in this representation by a 
slope $2-1/\nu$ again) are pronounced only for very small forces, while 
for somewhat larger forces (such as in $d=3$ for $b\ge 1.070$ 
for $q_b=0.4$ and for $b \ge 1.015$ for $q_b=0.05$, and in $d=2$ for 
$b\ge 1.015$ for $q_b=0.4$ and for 
$b\ge 1.005$ for $q_b=0.01$), the oscillatory behavior of the structure factor, 
as described by the Debye function with complex $X_{||}$ \{Eq.~(\ref{eq46})\}, 
sets in. In order to interpret this behavior in more detail, 
we write $X_{||}=a+ic,$ where 
$a=q_{||}^2(\langle X^2\rangle - \langle X\rangle ^2)/2$ and 
$c=q_{||} \langle X \rangle$, to rewrite Eq.~(\ref{eq47}) as follows
\begin{widetext}
\begin{equation}\label{eq58}
S_{||}(q_{||}) = 2\frac{\exp (-a) \left[(a^2-c^2) \cos c-2ac 
\sin c \right] + a^3+ac^2+c^2-a^2}{(a^2+c^2)^2}
\end{equation}
\end{widetext}
We recognize that there are two rather distinct parts, an exponentially damped 
oscillatory part and a ``background part'' which survives when the oscillatory 
part has died out. For $a \gg 1$ this background part can be written as (note 
that for large stretching there is a regime where $a^2 \ll c^2$)
\begin{equation}\label{eq59}
S_{||} (q_{||}) \approx \frac{2a}{c^2} = \frac{\langle X ^2 \rangle }{\langle X \rangle^2}-1
\end{equation}
In the regime where the oscillations have died out again and there is hence a 
flat part of $S_{||}(q_{||})$, independent of $q_{||}$ again, the structure 
factor hence measures the relative fluctuation in the length of the strongly 
stretched polymer.

Let us now consider the oscillatory part of Eq.~(\ref{eq58}). Since we are in 
a regime where $a^2 \ll c^2$, the maxima are reached when $\cos c=-1$ and in the 
regime where $a \ll 1$ and hence $\exp(-a) \approx 1$ we hence have
\begin{eqnarray}\label{eq60}
S_{||}^{\rm max} (q_{||}) &\approx& \frac{4}{c^2} =
\frac{4}{q^2_{||}\langle X\rangle ^2} \, ,  \\ 
\quad q_{||} \langle X \rangle &=& (2m+1) \pi\,, \,  m = 0,1,\ldots \nonumber
\end{eqnarray}
and hence $q^2_{||} S_{||}^{\rm max}\approx \;const$, as observed from the full 
calculation of Eq.~(\ref{eq58}), and the simulation.

When we compare these results to the scattering from the rigid rods, however, 
Eq.~(\ref{eq30}) predicts maxima that are undamped and minima that are 
strictly zero, so for increasing $q_{||}$ the oscillations continue forever. 
However, this is a result for a rod that has a strictly fixed length 
$L_{\rm rod}$, while the polymer under strong stretch (with extension 
$\langle X \rangle$ such that $1-\langle X \rangle /L \ll 1$) still 
is only similar to a rod of fluctuating length, and this in fact is 
borne out by the structure factor at large $q$ \{Eq.~(\ref{eq59})\}.

In any case the success of the Debye function, Eqs.~(\ref{eq46}), 
(\ref{eq58}) 
for the description of the scattering from strongly stretched chains 
in both $d=2$ and $d=3$ dimensions is very remarkable, since it is 
derived from Gaussian chain statistics~\cite{30}, and we have seen that 
in $d=2$ in the absence of stretching forces Gaussian chain statistics 
does not work in $d=2$, irrespective of chain stiffness.

At the end of this section, we emphasize that the examples given for the 
success of the Debye function for stretched chains, as derived by 
Benoit et al.~\cite{30}, are not accidental, but this behavior is typical 
for a wide range of chain stiffnesses. As an example, we show further data 
for $S_\bot(q_\bot)$ in both $d=2$ and $d=3$ and various other choices of 
the stiffness parameter $q_b$ in Figs.~15, 16. Whenever the plots indicate a 
well-defined plateau, one can extract an estimate of $\langle R_{g, \bot}^2\rangle$ 
from it (note that the actual values of $\langle R_{g, \bot}^2\rangle$ that 
were independently estimated were used to predict the Debye functions as 
shown in Figs.~15, 16.). As has been shown already in Fig.~8, the radii 
$\langle R_{g, \bot}^2 \rangle$ do have a broad regime of forces $f/k_BT$ 
where excluded volume effects (``Pincus blob`'' - behavior) prevail, 
so the success of the Debye function must not be over-emphasized, it 
does not mean that the chain conformation follow Gaussian statistics.

\begin{figure*}[htb]
\begin{center}
(a)\includegraphics[scale=0.29,angle=270]{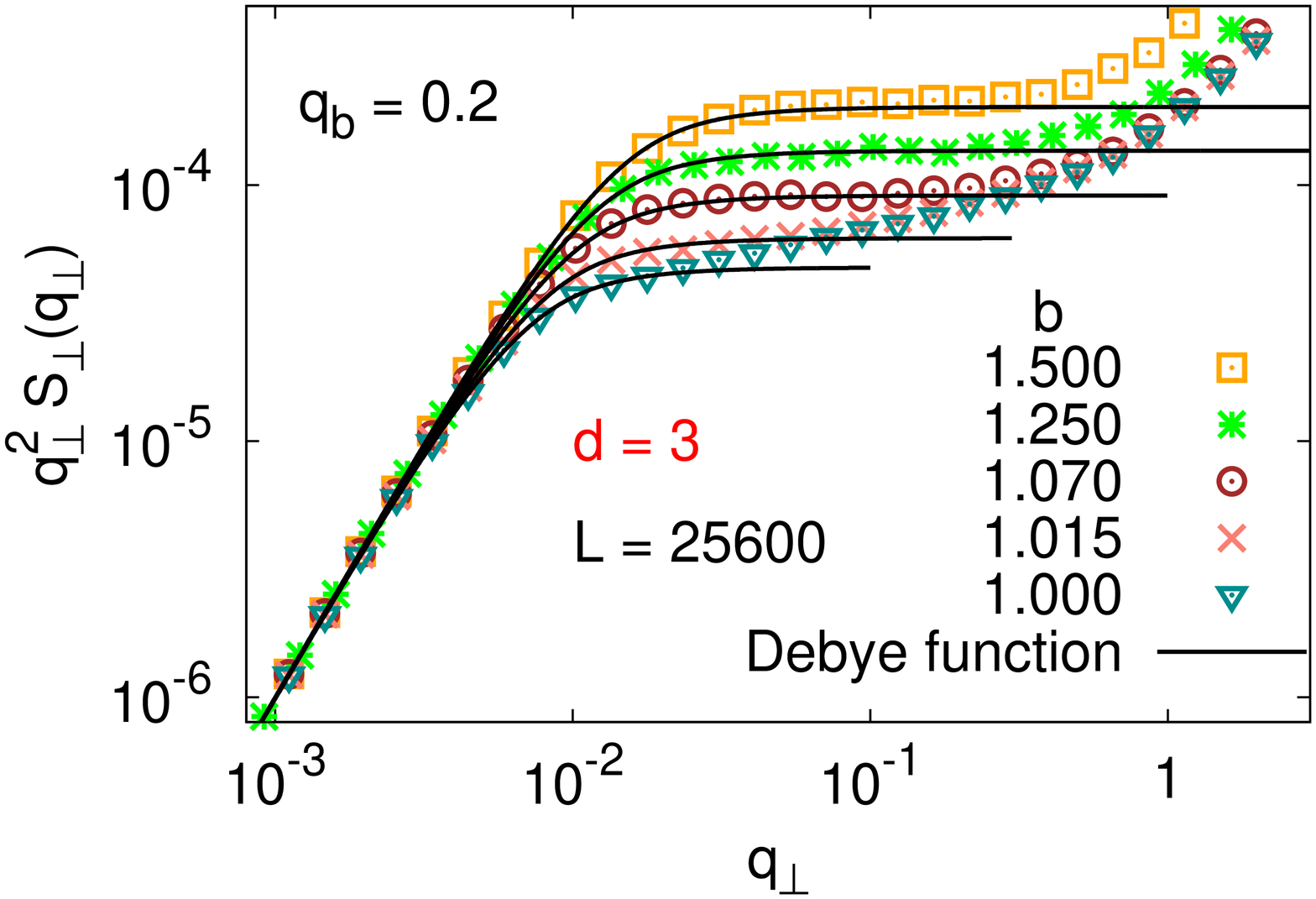}\hspace{0.4cm}
(b)\includegraphics[scale=0.29,angle=270]{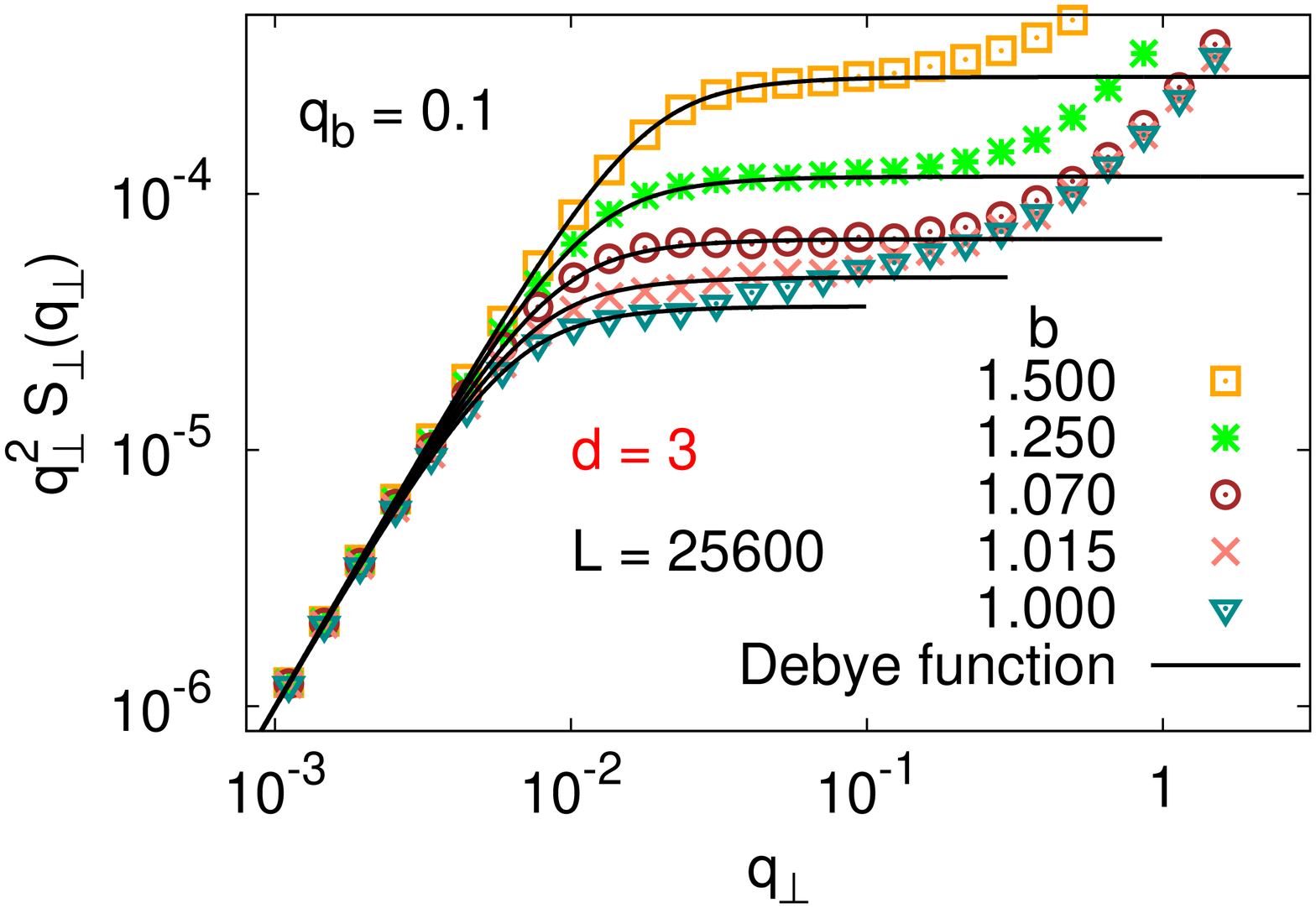}\\
(c)\includegraphics[scale=0.29,angle=270]{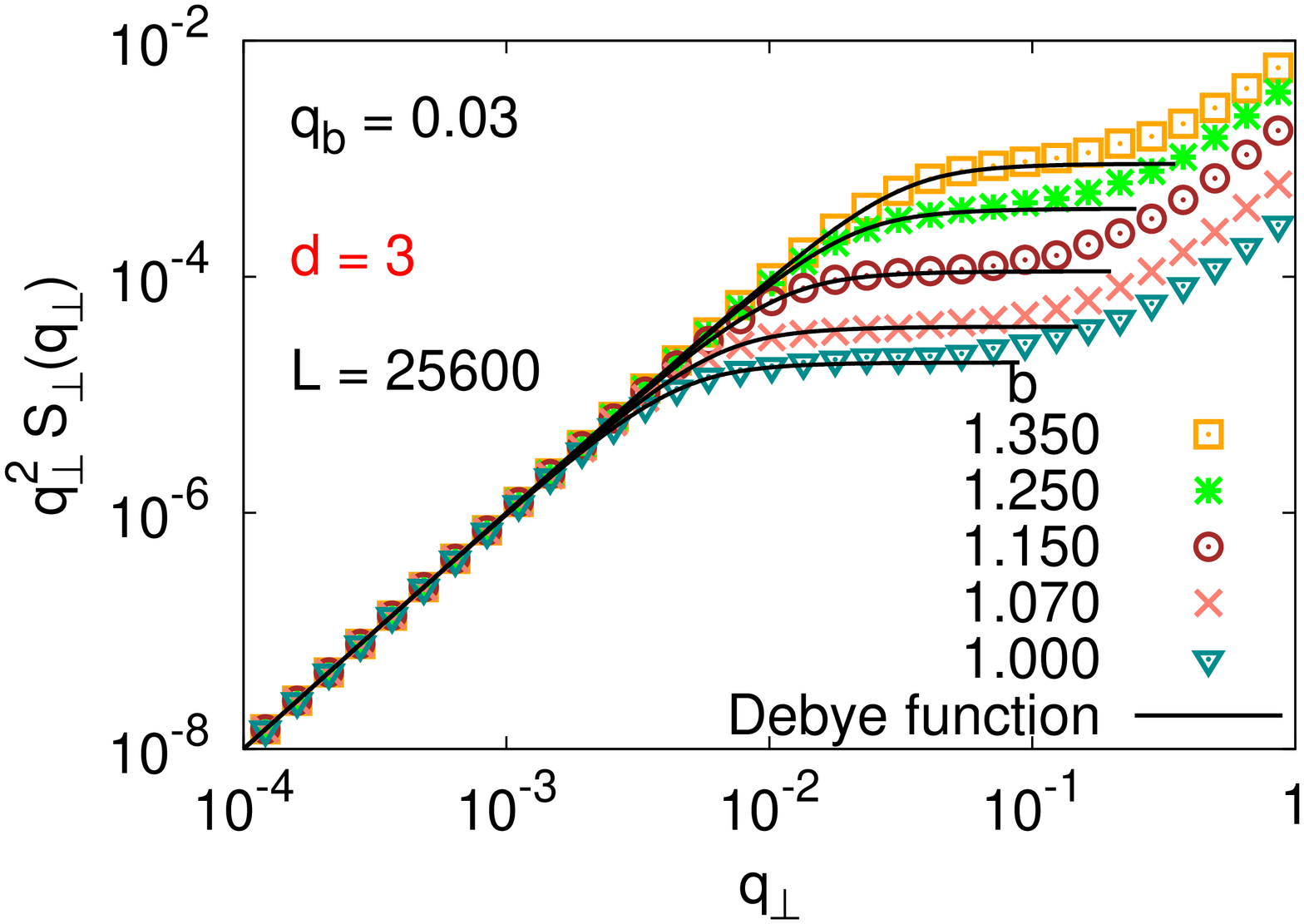}\hspace{0.4cm}
(d)\includegraphics[scale=0.29,angle=270]{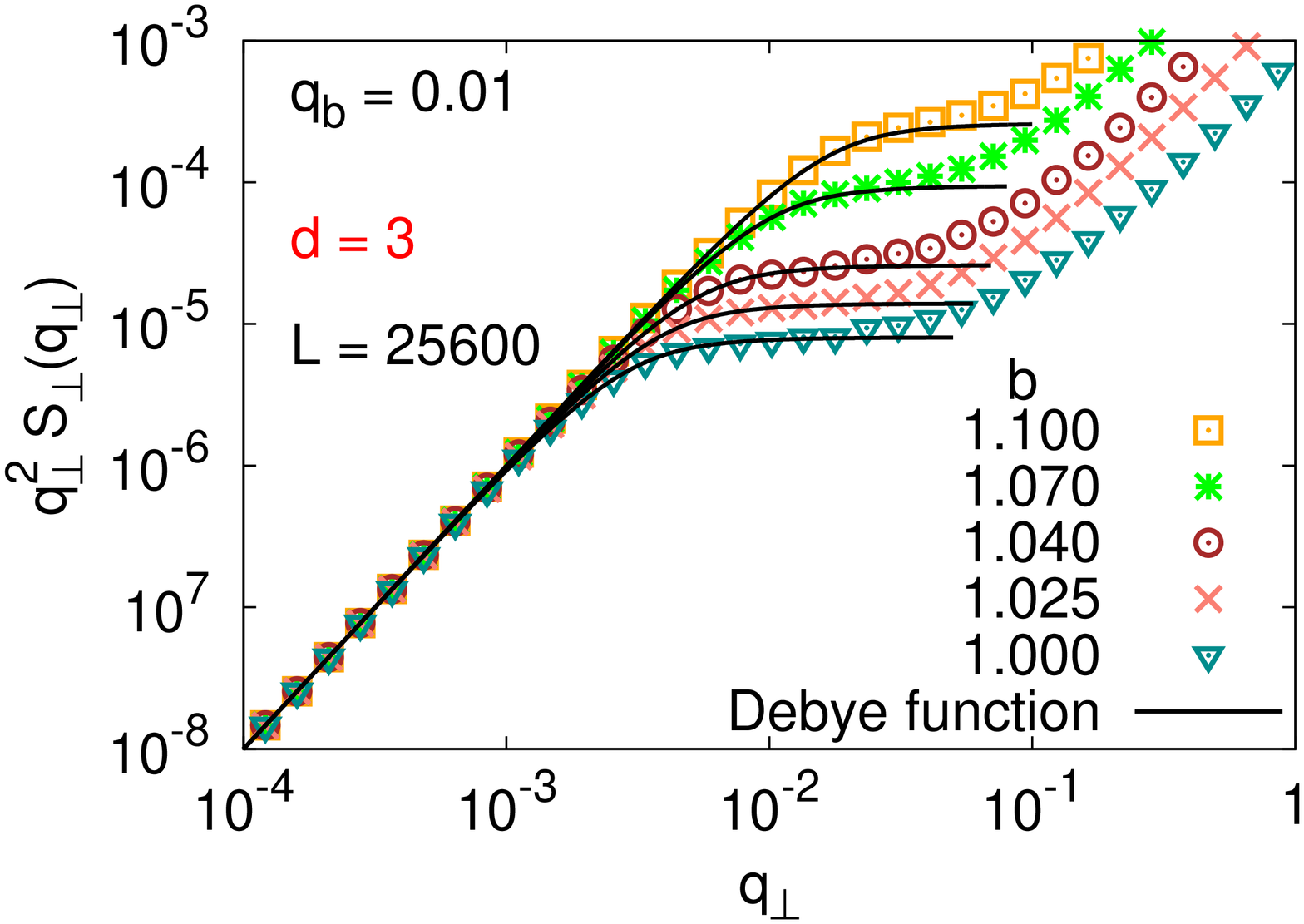}\\
\caption{\label{fig15}
Log-log plot of $q_\bot ^2 S_\bot(q_\bot)$ vs.~$q_\bot$ in $d=3$ dimensions
for $q_b=0.2$ (a), 0.1 (b), 0.03 (c) and 0.01 (d), for several choices of
the force parameter $b = \exp(f/k_BT)$, as indicated. The curves are the
Debye function, Eq.~(\ref{eq47}) with
$X_\bot=\frac{3}{2}q_\bot^2 \langle R_{g,\bot}^2 \rangle$.
When the data settle down at a horizontal
plateau, it yields an estimate of $4/(3 \langle R_{g,\bot}^2 \rangle)$.}
\end{center}
\end{figure*}

\begin{figure*}[htb]
\begin{center}
(a)\includegraphics[scale=0.29,angle=270]{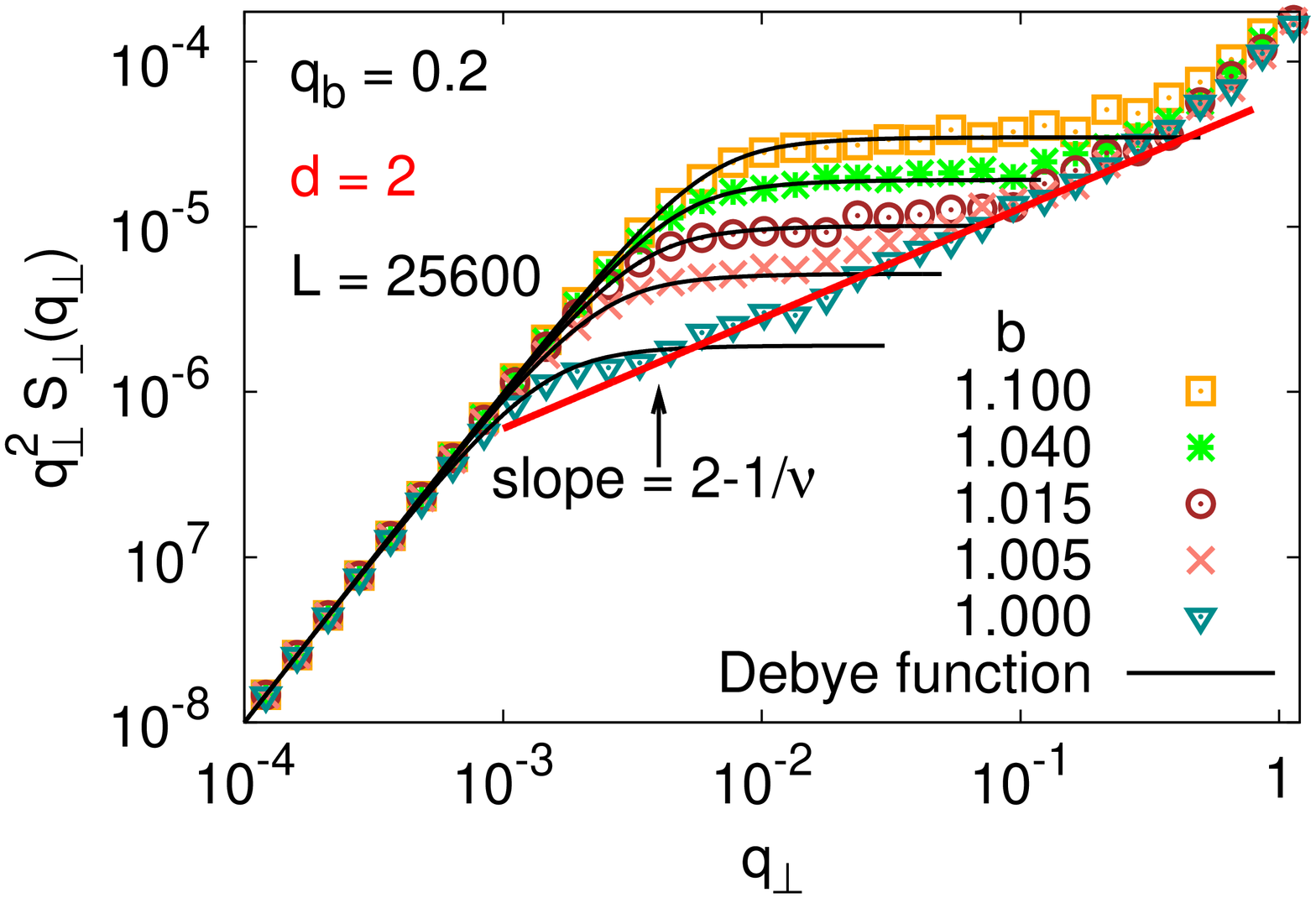}\hspace{0.4cm}
(b)\includegraphics[scale=0.29,angle=270]{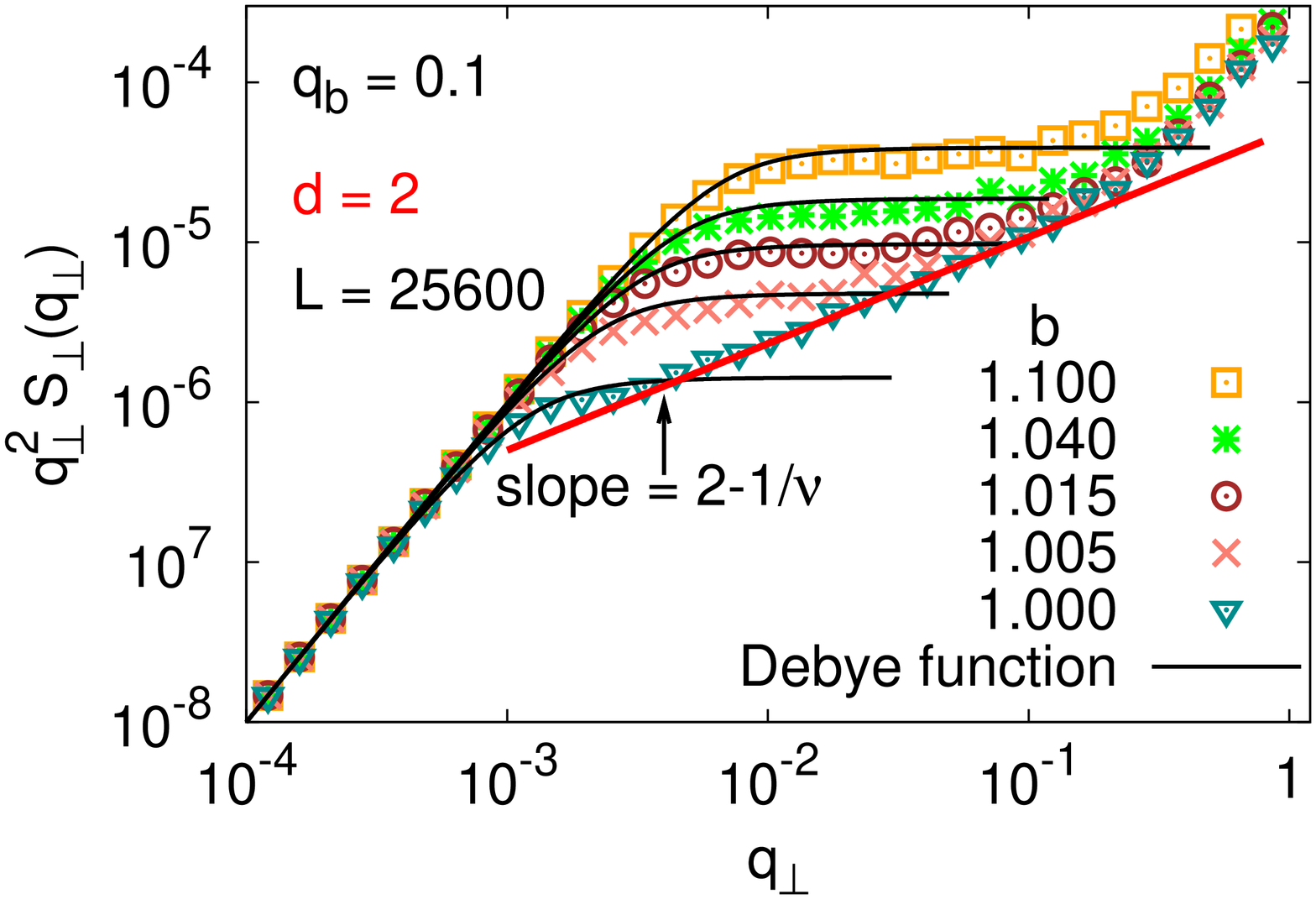}\\
(c)\includegraphics[scale=0.29,angle=270]{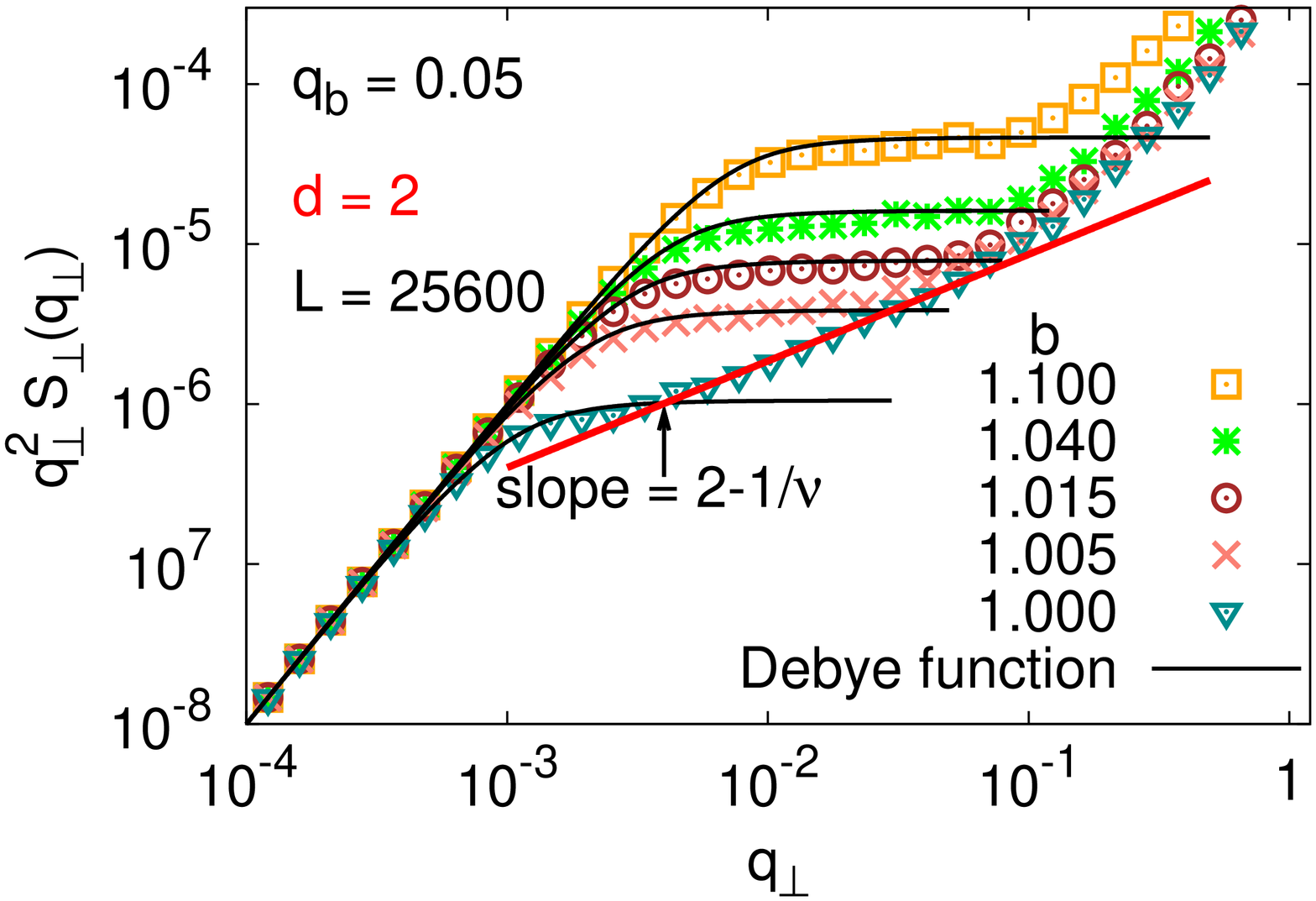}\hspace{0.4cm}
(d)\includegraphics[scale=0.29,angle=270]{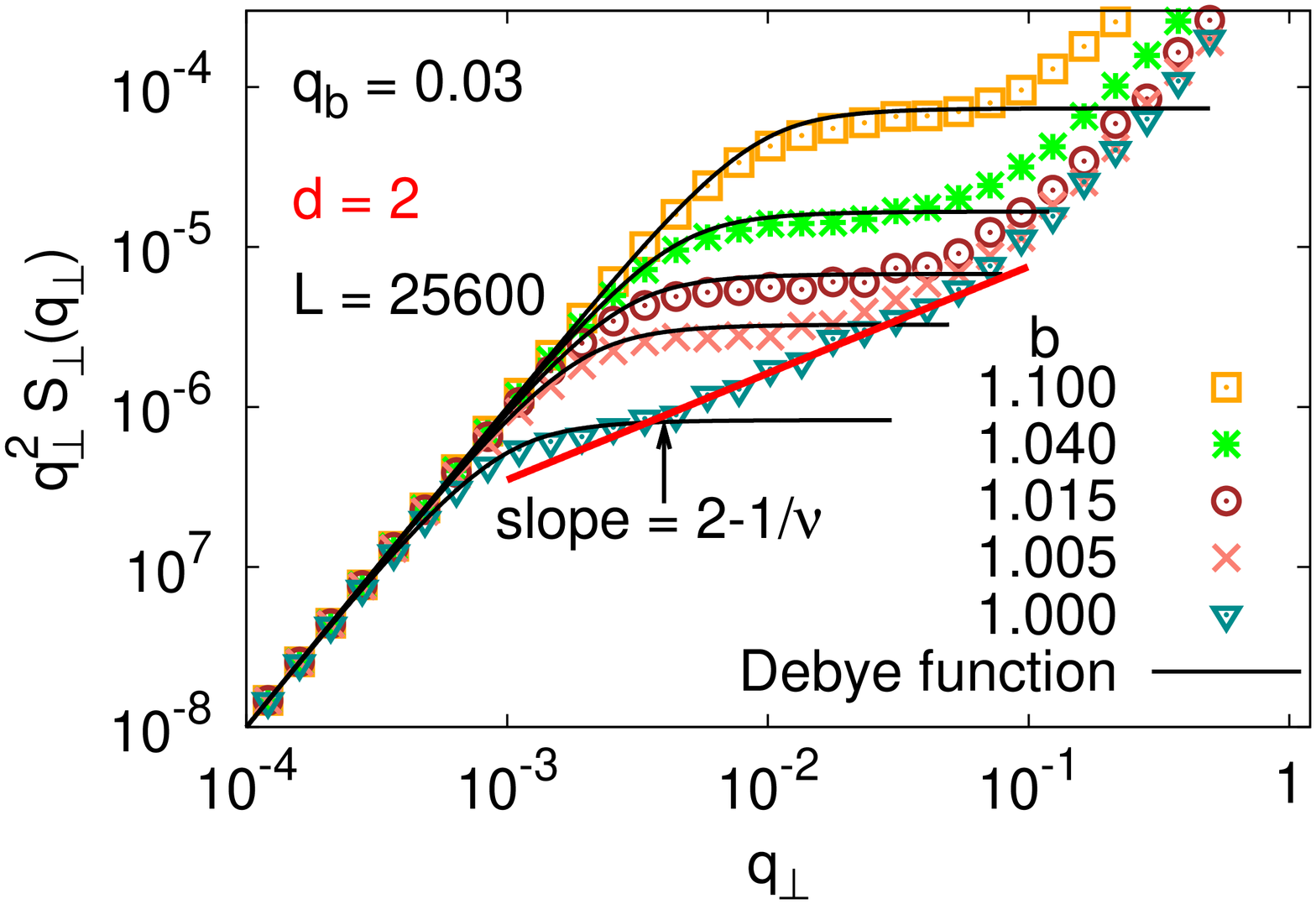}\\
\caption{\label{fig16}
Log-log plot of $q_\bot^2S_\bot(q_\bot)$ vs.~$q_\bot$ in $d=2$ dimensions
for $q_b=0.2$ (a), 0.1 (b), 0.05 (c) and 0.03 (d), for several choices of
the force parameter $b = \exp (f/k_BT)$, as indicated. The curves are the
Debye function, Eq.~(\ref{eq47}) with
$X_\bot=3q_\bot^2 \langle R_{g,\bot}^2 \rangle$,
and the straight line shows the excluded volume power law
(slope $=2 - 1/\nu$). From the Debye plateau
$2/(3\langle R_{g,\bot}^2\rangle)$ can be extracted.}
\end{center}
\end{figure*}

\section{Conclusions}

In this paper we have presented a comparative simulation study of the 
single-chain structure factor $S(q)$ for variable stiffness of the 
macromolecules in both $d=2$ and $d=3$ dimensions, both for coils in 
equilibrium in dilute solution under good solvent conditions, and for 
polymers under the influence of a stretching force. Characteristic linear 
dimensions of the macromolecules that are needed in the theoretical 
interpretation of $S(q)$, have in our Monte Carlo simulation always 
been estimated directly and hence independently, such as the mean square 
gyration radius $\langle R_g ^2 \rangle $ and the persistence length $\ell_p$. 
In the presence of stretching forces, the extension $\langle X \rangle $ 
in the direction of the force (as well as fluctuations 
$\langle X ^2 \rangle - \langle X \rangle ^2$, and components of the 
gyration radius $\langle R_{g \bot}^2\rangle$ and $\langle R_{g ||}^2 \rangle$ 
have been obtained as well. Of course, as usual the simulations are 
performed for the strictly monodisperse case, the number of bonds $N$ 
and hence also the contour length $L=N\ell_b$ of the chain molecules 
are known input parameters of the simulation. In this respect, a more 
definite interpretation of the outcome of the simulations can be expected, 
than would be expected for corresponding experiments (where polydispersity 
is a problem, and often the average contour length is not a priori known 
but must also be extracted from fitting suitable experimental data). 
Of course, the drawback of our Monte Carlo simulations on a lattice is 
the highly idealized character of such coarse-grained model as used here, 
the self-avoiding walk with additional energy penalty for making kinks. 
Nevertheless, the comparison between our simulation results for the mean 
square gyration radius versus the number of ``Kuhn segments'' $n_K$ with 
corresponding experimental data (Fig.~1) is very encouraging: one not only 
notes a striking similarity between simulation and experiment, but we 
stress that also the same range of dimensionless variables $(n_K$ and 
$\langle R_g^2\rangle / (2 \ell_p)^2$ ) is accessible. The simulation 
has the bonus that directly single-chain properties are accessible 
(no extrapolation as a function of the concentration $c$ of the solution 
towards $c \rightarrow 0$ is required and by changing the energy parameter 
$\epsilon_b/k_BT$, that describes the cost of making a kink along the walk, 
in units of the thermal energy, the stiffness is easily controlled. 
In experiment, stiffness can only be widely varied by combining data for 
polymers with different chemical structure.

From our data we have confirmed a conclusion drawn already from our previous 
study of mean square end-to-end distances, namely that in $d=2$ a direct 
crossover occurs from rod-like behavior to self-avoiding walks, with a 
scaling $\langle R_g^2 \rangle \propto \ell_p^{1/2} L^{3/2}$, without 
the existence of any intermediate regime with Gaussian behavior (Fig.~3). 
In $d=3$, however, such an  intermediate regime has been found, Fig.~2, 
for $1 \ll n_K \ll n_K^*\propto (\ell_p/D)^\zeta$, where $D$ is the local 
chain diameter and the exponent $\zeta$ is in the range $1.5 \leq \zeta \leq 2$. 
Thus, there is no universal value $n_K^*$ where excluded volume 
effects set in, but rather $n^*_K\rightarrow \infty$ for $\ell_p /D \rightarrow \infty$.

In the equilibrium structure factor $S(q)$, in the absence of 
stretching forces, correspondingly several regimes can be distinguished. 
For small enough $q$, the standard Guinier behavior always occurs, 
which contains the information on $\langle R _g^2\rangle$, of course. 
For $d=2$, one then always has the excluded volume regime 
(for long enough chains), $S(q) \propto q^{-4/3}$, and possibly 
(for rather stiff chains) a crossover to rod-like behavior
$(S(q)\propto q^{-1})$ sets in gradually. A Gaussian behavior 
$S(q) \propto q^{-2}$ is never seen, unlike the case $d=3$, where this 
behavior does become visible for very stiff chains (before for still 
larger $q$ the rod-like behavior starts). The excluded volume power law, 
$S(q) \propto q^{-1/\nu}$ with $\nu \approx 0.588$, is only visible for 
not very stiff chains (if chain lengths $L \leq 25600$ are analyzed, 
as done here: if $L \rightarrow \infty$, this power law would emerge 
for any finite value of the persistence length, of course). This pattern 
of behavior (Fig.~4), of course, could have been a priori expected, but 
we are also able to show via Kratky plots $(qLS(q) $ vs.~$qL$) that for 
semiflexible chains in $d=3$ the expressions derived by Kholodenko and 
by Stepanow provide a quantitatively accurate description. It is found 
that for large $q$ this quantity $qLS(q)$ settles down at $\pi$, unlike 
the behavior predicted for flexible chains (the Debye function predicts 
$qLS(q) \propto q^{-1}$ for large $q$). However, the onset of the plateau 
occurs gradually in the decade $1 < q \ell_p < 10$; thus the onset of the 
plateau allows an estimation of $\ell_p$ only somewhat roughly. 
The peak position of the Kratky plot (Figs.~5,6) reflects the 
theoretically expected scaling of the gyration radius with $L$ and 
$\ell_p$, even though in the Kratky plot (Fig.~5) direct evidence 
for excluded volume effects seem to be minor.

Des Cloizeaux~\cite{18} derived from the Kratky-Porod model that for 
$L \rightarrow \infty$ one should have 
$LqS(q) = \pi + \textrm{const} (q \ell_p)^{-1}$, with the constant 
being predicted to be 2/3 \{Eq.~(\ref{eq1})\}. Unfortunately, this 
result is at variance with our numerical results (Fig.~7). 
The reason for this problem is still not clear.

Turning to the behavior of chains under the influence of stretching 
forces, we have shown that for weak forces, where linear response holds, 
excluded volume effects invalidate the Kratky-Porod model completely 
in $d=2$ dimensions, and one typically observes a broad range of forces 
where the extension versus force relation is a power law, and also 
$\langle R_{g \bot}^2 \rangle , \langle X^2\rangle - \langle X \rangle ^2$ 
are found to scale like $(f \ell_p/k_BT)^{1/\nu-2}$ in this 
``Pincus blob'' regime. In $d=3$, dimensions, however, a Pincus blob 
regime also exists, but its observability also is restricted. 
For large $k_BT/f$, however, the (continuum) Kratky-Porod descriptions 
is not valid for our discrete lattice model either: it is 
found that then $\langle R_{g \bot}^2 \rangle $ and 
$\langle X^2 \rangle - \langle X \rangle ^2$ decrease 
like $\exp(-f/k_BT)$ for $f/k_BT \gg 1$.

Although the excluded volume effects show up rather clearly in the 
chain extensions and gyration radii components, it turns out that 
Benoit's extension of the Debye formula to stretched chains~\cite{30} 
is a surprisingly accurate description of both the transverse 
$(S_\bot(q_\bot))$ and parallel parts ($S_{||}(q_{||})$) of the 
structure factor. The oscillatory behavior of $S_{||}(q_{||}$) 
for strongly stretched chains shows that their conformations resemble
a string of elastically coupled particles. Thus, if measurable, the structure 
factor of stretched chains would add valuable details to the picture 
of their conformations.

As we have emphasized in our paper, the statistical mechanics of 
semiflexible polymers has been a longstanding and controversial 
problem of polymer science. The subject is of great relevance 
for biopolymers, but also of broad interest in material science. 
We expect that the present study will be useful both for the 
interpretation of experiments and stimulate further theoretical 
studies, such as of the interplay between solvent quality and chain stiffness.

\begin{acknowledgments}
We are grateful to the Deutsche Forschungsgemeinschaft (DFG) for support under 
grant No SFB 625/A3, and to the John von Neumann Institute for Computing 
(NIC J\"ulich) for a generous grant of computer time. We are particularly 
indebted to S. Stepanow for his help with the explicit calculation of his 
exact formula for the structure factor of the Kratky-Porod model. We are also 
indebted to Hyuk Yu for pointing out Ref.~\cite{12} to us, and 
to C. Pierleoni for drawing our attention to Ref.~\cite{34}.
H.-P. Hsu thanks K. Ch. Daoulas for stimulating discussions.
\end{acknowledgments}

\appendix
\section{Scattering function of random walk chains under
constant pulling force}

The paper by Benoit et al.~\cite{30} 
uses the distribution function
of the end-to-end vector of a Gaussian chain of $\mid i-j \mid$
repeat units for the calculation of the single chain scattering function. 
The result is the well-known Debye function.

Following an idea from Doi's book~\cite{Doibook}
one can derive the diffusion equation
yielding the end-to-end vector distribution in the following way. Assume a
step-wise Markov growth of the chain
\begin{equation}
    P(\vec{R},N)=\sum_i P(\vec{R}-\vec{l}_i,N-1)p(\vec{l}_i) \, .
\end{equation}
The sum over ``i" goes over an isotropic bond vector set, i.e., 
both $\vec{l}_i$ and $-\vec{l}_i$ are members of the set.

Expanding the right side to first order in $N$ and to second order in 
$\vec{R}$ yields
\begin{widetext}
\begin{equation}
   P(\vec{R}-\vec{l}_i,N-1)=P(\vec{R},N)-\frac{\partial P}{\partial N}
-\sum_{\alpha=1}^3 \frac{\partial P}{\partial R_\alpha} l_{i,\alpha}
+\frac{1}{2}\sum_{\alpha=1}^3 \sum_{\beta=1}^3 
\frac{\partial^2 P}{\partial R_\alpha \partial R_\beta} l_{i,\alpha}
l_{i,\beta}+\ldots \, .
\label{eqA2}
\end{equation}
\end{widetext}
Averaging the derivatives with respect to $R_\alpha$ with a symmetric 
bond probability $p(\vec{l}_i)$ (in the simplest case this is just one 
over the number of bonds) gives zero for the first derivative and 
$\delta_{\alpha \beta} l^2/3$ for the second term, resulting
in the diffusion equation
\begin{equation}
    \frac{\partial P}{\partial N} = \frac{l^2}{6} \nabla^2 P
\end{equation}
Solving this for a bulk chain gives the well known result for the end-to-end
vector distribution
\begin{equation}
    P(\vec{R},N) = (2 \pi Nl^2/3)^{-3/2} 
\exp \left(-\frac{3R^2}{2Nl^2} \right) \, ,
\end{equation}
The Debye function can than be derived by averaging with this probability,
as is done in the Benoit et al. paper.

The derivation above is useful as a starting point for a calculation 
of the scattering function for a chain that is pulled. 
In this case the bond probabilities are not symmetric. For our model we have
\begin{eqnarray}
   p_0 &=& \frac{b}{b^2+4b+1} \\
   p_+ &=& \frac{b^2}{b^2+4b+1} \nonumber \\
   p_- &=& \frac{1}{b^2+4b+1}
\end{eqnarray}
for moves perpendicular to the pulling direction, in $+X$ direction and 
in $-X$ direction, respectively, where $b = \exp(fl/k_BT)$ with 
$l = 1$ is used as in the main text. When we now perform the expansion of 
Eq.~(\ref{eqA2}) and perform the average
over the bond probabilities we obtain
\begin{widetext}
\begin{equation}
  \frac{\partial P}{\partial N} = -(p_+ - p_-)l \frac{\partial P}{\partial X}
+p_0 l^2 \left(\frac{\partial^2 P}{\partial Y^2}+
\frac{\partial^2P}{\partial Z^2}\right)+\frac{p_+ + p_-}{2} l^2 
\frac{\partial^2 P}{\partial X^2}
\end{equation}
\end{widetext}
For $p_+=p_-=p_0$ this reduces to the normal diffusion equation. This
equation has to be solved with the boundary conditions
\begin{eqnarray}
  P(\vec{R},0) &=& \delta(\vec{R}) \\
  P(\vec{R},N) &\rightarrow& 0 \qquad {\rm for} \qquad
R \rightarrow \infty \nonumber
\end{eqnarray}
Let us define $D_{\bot}=2p_0l^2$, $D_{||}=(p_++p_-)l^2$, and 
$v=(p_+ - p_-)l$, so we have
\begin{equation}
  \frac{\partial P}{\partial N}=-v \frac{\partial P}{\partial X}
+\frac{1}{2}D_\bot\left(\frac{\partial^2 P}{\partial Y^2}
+\frac{\partial^2 P}{\partial Z^2} \right)+\frac{1}{2}
D_{||} \frac{\partial^2 P}{\partial X^2} \, .
\label{eqA9}
\end{equation}
These are three diffusion processes in the three Cartesian directions, $X$ is
parallel to the force, $Y$ and $Z$ are perpendicular. The solutions for the
perpendicular directions are the same as for the force-free case. For the
parallel direction we have an altered diffusion coefficient and a drift part
to the process, i.e., a Gaussian diffusion around a deterministic drift. The
complete solution to Eq.~(\ref{eqA9}) is therefore given by
\begin{equation}
  P(X,Y,Z,N)=\frac{1}{2 \pi N D_\bot} \frac{1}{\sqrt{2 \pi N D_{||}}}
e^{-\frac{Y^2+Z^2}{2ND_\bot}} e^{-\frac{(x-vN)^2}{2ND_{||}}} \, .
\label{eqA10}
\end{equation}
For $p_0=p_+=p_-=1/6$ we obtain back the force free solution. To calculate
the scattering function we follow the procedure employed in the calculation
of the Debye function in the force free case.
\begin{equation}
     S(\vec{q}) = \frac{1}{N^2} \sum_{i,j} \langle 
e^{i\vec{q} \cdot \vec{r}_{ij}} \rangle
\end{equation}
can be calculated when we assume a continuous chain model (i.e. only look
at distances much larger than the lattice constant) so that the distribution
for the $\vec{r}_{ij}$ is given by the Gaussian distribution we just calculated.
\begin{equation}
   S(\vec{q})=\frac{1}{N^2} \sum_{i,j} \int d^3 \vec{r}_{ij}
P(\vec{r}_{ij}, \mid i-j \mid) e^{i\vec{q} \cdot \vec{r}_{ij}}
\end{equation}
where we have $P(\vec{r}_{ij},\mid i-j \mid) = P_Y(Y, \mid i-j \mid)
P_Z(Z,\mid i-j \mid) P_X(X, \mid i-j \mid)$ and $P_Y$ and $P_Z$ have
the same functional form and all $P_i$ are normalized to 
one individually.
\vskip 1.0truecm
\noindent \textbf{Scattering in the perpendicular direction}

\begin{eqnarray}
S(\vec{q}_\bot) 
= \frac{1}{N^2} && \sum_{i,j}  \int dY \int dZ  \\
&& P_Y(Y,\mid i-j \mid) P_Z(Z, \mid i-j \mid)
e^{i\vec{q}_\bot \cdot \vec{\rho}_{ij}} \nonumber 
\end{eqnarray}
where $\rho_{ij}=Y \hat{e}_Y + Z \hat{e}_Z$. So we have to evaluate
\begin{eqnarray}
&&  S(q_{_Y},q_{_Z}) \nonumber \\
&& =\frac{1}{N^2} \sum_{i,j} \int  d Y
\frac{1}{\sqrt{2\pi \mid i-j \mid D_\bot}} 
e^{-\frac{Y^2}{2\mid i-j \mid D_\bot}} e^{iq_{_Y}Y} \nonumber \\
&& \int dZ 
\frac{1}{\sqrt{2\pi \mid i-j \mid D_\bot}} 
e^{-\frac{Z^2}{2\mid i-j \mid D_\bot}} e^{iq_{_Z}Z} 
\end{eqnarray}
resulting in
\begin{equation}
       S(q_\bot) = \frac{1}{N^2} \sum_{i,j} 
e^{-\frac{q^2_\bot D_\bot \mid i-j \mid}{2}}
\end{equation}
This is evaluated by a continuum approximation for the two sums,
$\sum_i \rightarrow \int_0^N du$ and $\sum_j \rightarrow \int_0^N dv$
which finally yields:
\begin{equation}
    S(q_\bot) = \frac{4}{Nq_\bot^2 D_\bot}+\frac{8}{N^2q_\bot^4D_\bot^2}
\left(e^{-\frac{q_\bot^2 D_\bot N}{2}} -1 \right) \, .
\label{eqA16}
\end{equation}

\noindent \textbf{Scattering in the parallel direction}

\begin{equation}
      S(q_{||})=\frac{1}{N^2} \sum_{i,j} \int dX 
\frac{1}{\sqrt{2 \pi \mid i-j \mid D_{||}}} 
e^{-\frac{(x-v\mid i-j \mid)^2}{2 \mid i-j \mid D_{||}}}
e^{iq_{||}X}
\end{equation}
which now results in
\begin{equation}
    S(q_{||}) = \frac{1}{N^2} \sum_{i,j}
e^{-\frac{q_{||}D_{||} \mid i-j \mid}{2}} \cos(q_{||}v \mid i-j \mid) \, .
\end{equation}
Performing the final calculation again in the continuum approximation gives
\begin{eqnarray}\label{eqA19}
&& S(q_{||}) \nonumber  \\
&& = \frac{4}{N} \frac{D_{||}q_{||}^2}
{q_{||}^4D_{||}^2+4v^2q_{||}^2} \nonumber \\
&& + \frac{8}{N^2}\frac{q^4_{||}D^2_{||}-4v^2q_{||}^2}
{(q^4_{||}D_{||}^2+4 v^2q_{||}^2)^2} \left(
e^{-\frac{q_{||}^2D_{||}N}{2}}\cos(Nvq_{||})-1 \right) \nonumber \\
&& - \frac{32}{N^2}\frac{q_{||}^3D_{||}v}
{(q^4_{||}D^2_{||}+4v^2q_{||}^2)^2}e^{-\frac{q_{||}^2D_{||}N}{2}}
\sin(Nvq_{||}) 
\end{eqnarray}
This result determines our scattering functions with parameters depending
on the applied force $f$, so these equations contain no free parameters. Note
that both functions reduce to the Debye function for the force free isotropic
case ($v=0$, $D_\bot = D_{||} = l^2/3$) as it should be, 
because the scattering function
does not depend on the direction of the scattering vector in this case. 

   For fitting purposes it might yield better results to replace some of the
quantities by average values determined in the simulation. For the $Y$ and
the $Z$ components simple Gaussian statistics holds
\begin{eqnarray}
       \langle Y \rangle &=& \langle Z \rangle = 0 \\
      \langle Y^2 \rangle &=& \langle Z^2 \rangle = N D_\bot \\
      \langle R_{g,Y}^2 \rangle &=& \langle R^2_{g,Z} \rangle 
= \frac{ND_\bot}{6}
\end{eqnarray}
but in the force direction we can calculate from Eq.~(\ref{eqA10}) 
for the moments of the end-to-end distance
\begin{eqnarray}
          \langle X \rangle &=& Nv \\
    \langle X^2 \rangle &=& N^2 v^2 +N D_{||} \\
    \langle \Delta X^2 \rangle &=& ND_{||} \\
  \langle R^2_{g,X} \rangle &=& \frac{ND_{||}}{6} +\frac{D_{||}N^2 v^2}{12} \, .
\end{eqnarray}
When we now rewrite the scattering functions in terms of moments of the
end-to-end vector or gyration tensor, we obtain
\begin{eqnarray}
&&    S(q_\bot)  \\
&&=\frac{4}{3q_\bot^2 \langle R^2_{g,\bot}\rangle}
+\frac{8}{\left[3q_\bot^2 \langle R^2_{g,\bot} \rangle \right]^2}
\left(e^{-\frac{3q_\bot^2 \langle R^2_{g,\bot} \rangle}{2}}-1 \right) \, .
\nonumber 
\label{eqA27}
\end{eqnarray}
This is the Debye function with the appropriate prefactors, because for
$\langle R^2_{g,\bot} \rangle=2/3 \langle R_g^2 \rangle$
it reduces to the standard function, Eq.~(\ref{eq22}). For the scattering
parallel to the pulling direction we can write

\begin{eqnarray}\label{eqA28}
&&    S(q_{||})  \\
&&= 4 \frac{q_{||}\langle \Delta X^2 \rangle}
{q_{||}^4 \langle \Delta X^2 \rangle^2 +4 q_{||}^2 \langle X \rangle^2 } \nonumber \\
&& + 8 
 \frac{q_{||}^4 \langle \Delta X^2 \rangle^2 - 4q_{||}^2 \langle X \rangle^2}
{(q_{||}^4\langle \Delta X^2 \rangle^2+4q^2_{||} \langle X \rangle^2)^2}
\left(e^{-\frac{q_{||}^2\langle \Delta X^2 \rangle}{2}}
\cos(q_{||}\langle X \rangle )-1 \right) \nonumber \\
&&- 32 \frac{q^3_{||} \langle \Delta X^2 \rangle \langle X \rangle}
{(q_{||}^4 \langle \Delta X^2 \rangle^2 +4q_{||}^2 \langle X \rangle^2)^2}
e^{-\frac{q_{||}^2 \langle \Delta X^2 \rangle}{2}} 
\sin (q_{||}\langle X \rangle ) \, . \nonumber 
\end{eqnarray}
This gives the same formula as Eq.~(\ref{eq58}) when $X_{||}$ 
in Eq.~(\ref{eq46}) is written by $X_{||}=a+ic$ with 
$a=q_{||}^2 \langle \Delta X^2 \rangle/2$, and $c=q_{||}\langle X \rangle$.
For $\langle X \rangle=0$ and $\langle \Delta X^2 \rangle = 
\langle X^2 \rangle=6 \langle R_{g,X}^2 \rangle =2 \langle R_g^2 \rangle$
this again reduces to the Debye function, Eq.~(\ref{eq22}).
\vskip 1.0 truecm
\noindent \textbf{Scattering in $d=2$}

Both scattering functions as calculated in Eqs.~(\ref{eqA16}) and 
(\ref{eqA19}) remain formally
unchanged. However, the probabilities for the single steps change to
\begin{eqnarray}
    p_0 &=& \frac{b}{b^2+2b+1} \nonumber \\
    p_+ &=& \frac{b^2}{b^2+2b+1} \\
   p_- &=& \frac{1}{b^2+2b+1} \nonumber 
\end{eqnarray}
The parallel and perpendicular diffusion coefficients as well as the drift 
velocity still have the same functional dependence on these probabilities. 
However, introducing the chain extensions into Eqs.~(\ref{eqA16}) and 
(\ref{eqA19}) for $d = 2$ changes the prediction Eq.~(\ref{eqA27}) for 
the perpendicular scattering to
\begin{eqnarray}\label{eqA30}
&&    S(q_\bot)  \\
&&= \frac{2}{3q_\bot^2 \langle R_{g,\bot}^2 \rangle}
+ \frac{2}{(3q_\bot^2 \langle R^2_{g,\bot} \rangle)^2}
\left(e^{-3q_\bot^2 \langle R^2_{g,\bot} \rangle} -1 \right) \, , \nonumber 
\end{eqnarray}
whereas it leaves Eq.~(\ref{eqA28}) unchanged.
Taking $X_{\bot}=3 q_\bot^2 \langle R^2_{g,\bot} \rangle$, 
the expression of Eq.~(\ref{eqA30}) has the same formula as
Eq.~(\ref{eq47}).


\begin{thebibliography}{99}
\bibitem{1} P. G. de Gennes, \textit{Scaling Concepts in Polymer Physics} (Cornell University Press, Ithaca, NY, 1979).
\bibitem{2} J. Des Cloizeaux and G. Jannink, \textit{Polymers in Solution: Their Modeling and Structure} (Clarendon Press, Oxford, UK, 1990).
\bibitem{3} L. Sch\"afer, \textit{Excluded Volume Effects in Polymer Solutions as Explained by the Renormalization Group} (Springer, Berlin, 1999).
\bibitem{4} J. S. Higgins and H. C. Benoit, \textit{Polymers and Neutron Scattering} (Clarendon Press, Oxford, 1994).
\bibitem{5} P. J. Flory, \textit{Principles of Polymer Chemistry} (Cornell University Press, NY, 1953).
\bibitem{6} M. Rubinstein and R. H. Colby, \textit{Polymer Physics} (Oxford Universit Press, Oxford, UK, 2003).
\bibitem{7} J. P. Cotton, D. Decker, H. Benoit, B. Farnoux, J. Higgins,
G. Jannink, R. Ober, C. Picot, and J. des Cloizeaux, Macromolecules \textbf{7}, 863 (1974).
\bibitem{8} B. Farnoux, F. Boue, J. P. Cotton, M. Daoud, G. Jannink, M. Nierlich, and P. G. de Gennes, J. Phys. \textbf{39}, 77 (1978).
\bibitem{9} J. C. Le Guillou and J. Zinn-Justin, Phys. Rev. B \textbf{21}, 3976 (1980).
\bibitem{10} H.-P. Hsu, W. Paul and K. Binder, Macromolecules \textbf{43}, 3094 (2010).
\bibitem{11} H.-P. Hsu, W. Paul and K. Binder, 
Europhys. Lett. \textbf{92}, 28003 (2010).
\bibitem{12} T. Norisuye and H. Fujita, Polymer J. \textbf{14}, 143 (1982).
\bibitem{13} H.-P. Hsu and K. Binder, J. Chem. Phys. \textbf{136}, 024901 (2012).
\bibitem{14} O. Kratky and G. Porod, J. Colloid Sci. \textbf{4}, 35 (1949).
\bibitem{15} D. W. Schaefer, J. F. Joanny, and P. Pincus, Macromolecules \textbf{13}, 1280 (1980).
\bibitem{16} R. R. Netz and D. Andelman, Phys. Rep. \textbf{380}, 1 (2003).
\bibitem{161} H. Yamakawa and W. H. Stockmayer, J. Chem. Phys. \textbf{57}, 2843 (1972).
\bibitem{162} H. Yamakawa and J. Shimada, J. Chem. Phys. \textbf{83}, 2607 (1985).
\bibitem{163} J. Shimada and H. Yamakawa, J. Chem. Phys. \textbf{85}, 591 (1986).
\bibitem{164} H. Fujita, Macromolecules \textbf{21}, 179 (1988).
\bibitem{165} A. Tsuboi, T. Norisuye, and A. Teramoto, 
Macromolecules \textbf{29}, 3597 (1996).
\bibitem{17} H. P. Hsu, W. Paul and K. Binder, Europhys. Lett. \textbf{95} 68004 (2011).
\bibitem{18} J. des Cloizeaux, Macromolecules \textbf{6}, 403 (1973).
\bibitem{19} G. Porod, Z. Naturforschung \textbf{4a}, 401 (1949).
\bibitem{20} G. Porod, J. Polym. Sci. \textbf{10}, 157 (1953).
\bibitem{21} A. Peterlin, J. Polym. Sci. \textbf{47}, 403 (1960).
\bibitem{22} W. R. Krigbaum and S. Sasaki, J. Polym. Sci.: Polym. Phys. Ed. 
\textbf{19}, 1339 (1981).
\bibitem{23} M. G. Bawendi and K. F. Freed, J. Chem. Phys. \textbf{83}, 2491 (1985).
\bibitem{231} M. Rawiso, R. Duplessix and C. Picot, 
Macromolecules \textbf{20}, 630 (1987).
\bibitem{24} A. L. Kholodenko, Ann. Phys. \textbf{202}, 186 (1990).
\bibitem{25} A. L. Kholodenko, J. Chem. Phys. \textbf{96}, 700 (1992).
\bibitem{26} A. L. Kholodenko, Macromolecules \textbf{26}, 4179 (1993).
\bibitem{27} A. J. Spakowitz and Z.-G. Wang, Macromolecules \textbf{37}, 5814 (2004).
\bibitem{271} S. Stepanow, Eur. Phys. J. B \textbf{39}, 499 (2004).
\bibitem{28} S. Stepanow, J. Phys.: Condens. Matter \textbf{17}, S1799 (2005).
\bibitem{281} J. S. Pedersen, M. Laso and P. Schurtenberger, Phys. Rev. E \textbf{54}, R5917 (1996).
\bibitem{29} J. S. Pedersen and P. Schurtenberger, Macromolecules \textbf{29}, 7602 (1996).
\bibitem{291} J. S. Pedersen and P. Schurtenberger, Europhys. Lett. \textbf{45}, 666 (1999).
\bibitem{292} D. P\"otschke, P. Hickl, M. Ballauff, P.-O. Astrand, and J. S. Pedersen, Macromol. Theory Simul. \textbf{9}, 345 (2000).
\bibitem{30} H. Benoit, R. Duplessix, R. Ober, M. Daoud, J. P. Cotton, B. Farnoux, and G. Jannink, Macromolecules \textbf{8}, 451 (1975).
\bibitem{31} P. Pincus, Macromolecules \textbf{9}, 386 (1976).
\bibitem{32} S. B. Smith, L. Finzi, and C. Bustamante, Science \textbf{258}, 1122 (1992).
\bibitem{33} J. F. Marko and E. D. Siggia, Macromolecules \textbf{28}, 8759 (1995).
\bibitem{34} C. Pierleoni, G. Arialdi, and J.-P. Ryckaert, Phys. Rev. Lett. \textbf{79}, 2990 (1997).
\bibitem{35} R. R. Netz, Macromolecules \textbf{34}, 7522 (2001).
\bibitem{36} M.-N. Dessinges, B. Maier, Y. Zhang, M. Peliti, D. Bensimon, and V. Croquette, Phys. Rev. Lett. \textbf{89}, 248102 (2002).
\bibitem{37} L. Livadaru, R. R. Netz, and H. J. Kreuzer, Macromolecules \textbf{36}, 3732 (2003).
\bibitem{38} T. Hugel, M. Rief, M. Seitz, H. E. Gaub, and R. R. Netz, Phys. Rev. Lett. \textbf{94}, 048301 (2005).
\bibitem{39} Y. Seol, G. M. Skinner, and K. Visscher, Phys. Rev. Lett. \textbf{93}, 118102 (2004).
\bibitem{40} A. Prasad, Y. Hori, and J. Kondev, Phys. Rev. E \textbf{72}, 041918 (2005).
\bibitem{41} G. Morrison, C. Hyeon, N. M. Toan, B.-Y. Ha, and D. Thirumalai, Macromolecules \textbf{40}, 7343 (2007).
\bibitem{42} O. A. Saleh, D. B. McIntosh, P. Pincus and N. Ribeck, Phys. Rev. Lett. \textbf{102}, 068301 (2009).
\bibitem{43} D. B. McIntosh, N. Ribeck and O. A. Saleh, Phys. Rev. E \textbf{80}, 041803 (2009).
\bibitem{44} N. M. Toan and D. Thirumalai, Macromolecules \textbf{43}, 4394 (2010).
\bibitem{45} A. Dittmore, D. B. McIntosh, S. Halliday, and O. A. Saleh, Phys. Rev. Lett. \textbf{107}, 148301 (2011).
\bibitem{46} A. Onuki, J. Phys. Soc. Japan \textbf{54}, 3656 (1985).
\bibitem{461}P. Lindner and R. Oberth\"ur, Physica B \textbf{156} \& \textbf{157}, 410 (1989).
\bibitem{462} C. Pierleoni and J.-P. Ryckaert, Macromolecules \textbf{28}, 5097 (1995).
\bibitem{463} T. Q. Nguyen and H.-H. Kausch (eds.) \textit{Flexible Polymer Chains in Elongational Flow} (Springer, Berlin, 1999).
\bibitem{47} W. Reisner, K. J. Morton, R. Riehn, Y. M. Wang, Z. Yu, M. Rosen, J. C. Sturm, S. Y. Chou, E. Frey, and R. H. Austin, Phys. Rev. Lett. \textbf{94}, 196101 (2005).
\bibitem{48} P. Cifra, J. Chem. Phys. \textbf{131}, 224903 (2009).
\bibitem{49} F. Th\"uroff, B. Obermayer, and E. Frey, Phys. Rev. E \textbf{83}, 021802 (2011).
\bibitem{50} A. Yu. Grosberg and A. R. Khokhlov, \textit{Statistical Physics of Macromolecules} (AIP Press, New York, 1994).
\bibitem{51} J. P. Wittmer, H. Meyer, J. Baschnagel, A. Johner, S. Obukhov, L. Mattioni, M. M\"uller and A. N. Semenov, Phys. Rev. Lett. \textbf{93}, 147801 (2004).
\bibitem{52} J. P. Wittmer, P. Beckrich, H. Meyer, A. Cavallo, A. Johner, and J. Baschnagel, Phys. Rev. E\textbf{76}, 011803 (2007).
\bibitem{53} D. Shirvanyants, S. Panyukov, Q. Liao, and M. Rubinstein, Macromolecules \textbf{41}, 1475 (2008).
\bibitem{54} L. Sch\"afer, A. Ostendorf, and J. Hager, J. Phys. A: Math. Gen. \textbf{32}, 7875 (1999).
\bibitem{55} H. Benoit and P. Doty, J. Phys. Chem. \textbf{57}, 958 (1953).
\bibitem{56} T. Neugebauer, Ann. Phys. \textbf{434}, 509 (1943).
\bibitem{57} H.-P. Hsu, W. Paul, and K. Binder, J. Chem. Phys. \textbf{129}, 204904 (2008).
\bibitem{58} S. Stepanow (private communication).
\bibitem{Emery1978} V. J. Emery and J. D. Axe, Phys. Rev. Lett. \textbf{40},
1507 (1978).
\bibitem{Ricci2007} A. Ricci, P. Nielaba, S. Sengupta, and K. Binder,
Phys. Rev. E \textbf{75}, 011405 (2007).
\bibitem{Doibook} M. Doi and s. F. Edwards, 
\textit{The Theory of Polymer Dynamics} (Clarendon Press, Oxford, 1986).
\end{thebibliography}
\end{document}